\documentclass[aps,prl,reprint,amssymb,showpacs,superscriptaddress,notitlepage,onecolumn]{revtex4-2}
\usepackage{bm}
\usepackage{graphicx}
\usepackage[dvipsnames]{xcolor}
\usepackage{hyperref}
\usepackage{dcolumn}
\usepackage{braket}
\usepackage{natbib}
\usepackage{amsmath}
\usepackage{color}
\usepackage{mathtools,amssymb}
\usepackage{soul}
\usepackage{float}
\usepackage{amssymb}
\usepackage{esint}
\usepackage{mathtools}
\usepackage{physics}
\usepackage{makecell}
\usepackage{multirow}
\usepackage{array}
\usepackage{geometry}
\usepackage{url}
\usepackage{adjustbox}

\begin{document}
	\title{Quantum simulation of interacting bosons with propagating waveguide photons}
	
	\author{Xinyuan Zheng}\email{
		xzheng16@terpmail.umd.edu}
	\altaffiliation{Equally contributing authors}
	\affiliation{Department of Electrical and Computer Engineering,
		Institute for Research in Electronics and Applied Physics,
		and Joint Quantum Institute, University of Maryland, College Park, MD 20742, USA}
	
	\author{Mahmoud Jalali Mehrabad*}\email{mjalalim@umd.edu}
	\affiliation{Joint Quantum Institute and Quantum Technology Center, University of Maryland,
		College Park, MD 20742, USA}
	
	\author{Avik Dutt}
	\affiliation{Department of Mechanical Engineering, and Institute for Physical Science and Technology, University of Maryland, College Park, Maryland 20742, USA}
	
	\author{Nathan Schine}
	\affiliation{Joint Quantum Institute and Department of Physics, University of Maryland, College Park, MD 20742, USA}
	
	
	\author{Edo Waks}\email{edowaks@umd.edu}
	\affiliation{Department of Electrical and Computer Engineering,
		Institute for Research in Electronics and Applied Physics,
		and Joint Quantum Institute, University of Maryland, College Park, MD 20742, USA}
	
	
	\begin{abstract}
		
		\section{Abstract}
		\textbf{Optical networks composed of interconnected waveguides are a versatile platform to simulate bosonic physical phenomena. Significant work in the non-interacting regime has demonstrated the capabilities of this platform to simulate many exotic effects such as photon transport in the presence of gauge fields, dynamics of quantum walks, and topological transition and dissipation phenomena. However, the extension of these concepts to simulating interacting quantum many-body phenomena such as the Bose-Hubbard and the fractional quantum Hall (FQH) physics has remained elusive. In this work, we address this problem and demonstrate a framework for quantum many-body simulation as well as drive and dissipation in photonic waveguides. Specifically, we show that for waveguide photons, a tunable on-site interaction can be simulated using a photon-number-selective phase gate. We propose an implementation of such a phase gate based on a three-level-atom-mediated photon subtraction and addition. We apply this approach to bosonic lattice models and propose circuits that can accurately simulate the Bose-Hubbard and FQH Hamiltonian as benchmarking examples. Moreover, we show how to simulate the Lindbladian evolution with engineered dissipators such that the steady state of the Lindbadlian corresponds to the ground state of desired Hamiltonians. Our scheme extends the time-multiplexed waveguide photonic simulation platform to the strongly interacting quantum many-body regime while retaining all of its crucial advantages, such as single-site addressability, Hamiltonian parameter controllability, and hardware efficiency.
		}
	\end{abstract}
	
	\maketitle

	
	\section{Introduction}

	Photonics provides an attractive platform for the simulation of quantum many-body phenomena due to its many advantageous properties~\cite{noh2016quantum,hartmann2016quantum}. In particular, light can efficiently be generated, detected, and manipulated over broad spectral bandwidths and at the single photon regime. Moreover, photons provide many useful control knobs and degrees of freedom such as frequency, mode, time bin, polarization, and orbital angular momentum~\cite{yuan2018synthetic}, which can enable the engineering of versatile synthetic dimensions, opening up pathways for photonic implementation of quantum simulation and universal quantum computation~\cite{aspuru2012photonic,underwood2012bose,childs2014bose,knill2001scheme}.
	
	Recent decades have witnessed a surge of theory and experimental research on the implementation of many-body photonic systems~\cite{umucalilar2012fractional,cho_fractional_2008}. Examples include optical cavities containing atoms~\cite{birnbaum2005photon}, photonic crystal cavities~\cite{reinhard2012strongly}, and superconducting circuits~\cite{wang2024realization,karamlou2024probing,kapit2014induced}. Photons propagating in a 1D continuum offer another versatile platform for quantum simulation~\cite{schreiber20122d,marques2023observation,knill2001scheme}, with many desirable properties such as interconnectivity~\cite{bogaerts2020programmable}, dynamic tunability~\cite{monika2024quantum,zheng2024dynamic}, synthetic gauge fields~\cite{chalabi2019synthetic,chalabi2020guiding}, and scalable synthetic dimensions~\cite{yuan2018synthetic}. However, to-date, the waveguide photon toolbox has been mainly utilized for simulating non-interacting phenomena~\cite{schreiber20122d,leefmans2022topological,parto2023non,weidemann2020topological,weidemann2022topological,chalabi2019synthetic,chalabi2020guiding,lin2023manipulating,zheng2024dynamic} as well as classical nonlinear effects~\cite{marques2023observation,jurgensen2021quantized,leefmans2024topological,flower2024observation,jalali2024strain}. The extension of this platform 
	to quantum many-body simulation of strongly interacting bosons is extremely challenging and has remained elusive. 
	One method to create interactions for propagating waveguide photons is to use atomic systems coupled to waveguide modes which mediate interaction through nonlinear photon scattering~\cite{yang2022deterministic,lund2024subtraction}. These effects have been demonstrated in a broad class of quantum emitters including quantum dots~\cite{tomm2023photon}, color centers~\cite{babin2022fabrication}, and circuit QED~\cite{ferreira2024deterministic}. However, these nonlinearities distort the form of the wavefunction~\cite{shen2007strongly,mahmoodian2020dynamics,peyronel2012quantum,chang2014quantum} or do not apply to photon numbers greater than two~\cite{yang2022deterministic}, leading to imperfect photon-photon interactions. Moreover, practical schemes for drive and dissipation in this interacting regime are not available.
	
	In this work, we address this problem and demonstrate a framework for quantum many-body simulation in photonic waveguides. Specifically, we demonstrate that the required strong and tunable on-site interaction term can be simulated using a photon-number-selective phase gate, while engineered dissipation can be implemented with beamsplitters and ancilla bosonic modes. We propose a physical implementation of the phase gate that preserves the temporal mode and indistinguishability of the photon states using a three-level atom. Importantly, this photon nonlinearity works for arbitrary numbers of photons, and when cascaded, allows a fully tunable phase control. Therefore it can be used as a fundamental interaction term to simulate many collective quantum phenomena such as the FQH physics and Bose Hubbard models. Using this nonlinearity, we propose resource-efficient optical circuits that accurately simulate both the Bose-Hubbard and fractional quantum Hall models as benchmarking examples. Moreover, we show how to simulate the Lindbladian evolution with engineered dissipators such that the steady state of the Lindbadian corresponds to the ground state of desired Hamiltonians.
	Our architecture inherits the advantages of the waveguide photon toolbox, combined with the crucial on-site interaction, opening up new avenues for the simulation of many-body phenomena.
	
	
	\section{Basics}
\begin{figure*}[t]
		\centering
		\includegraphics[width=0.85\textwidth]{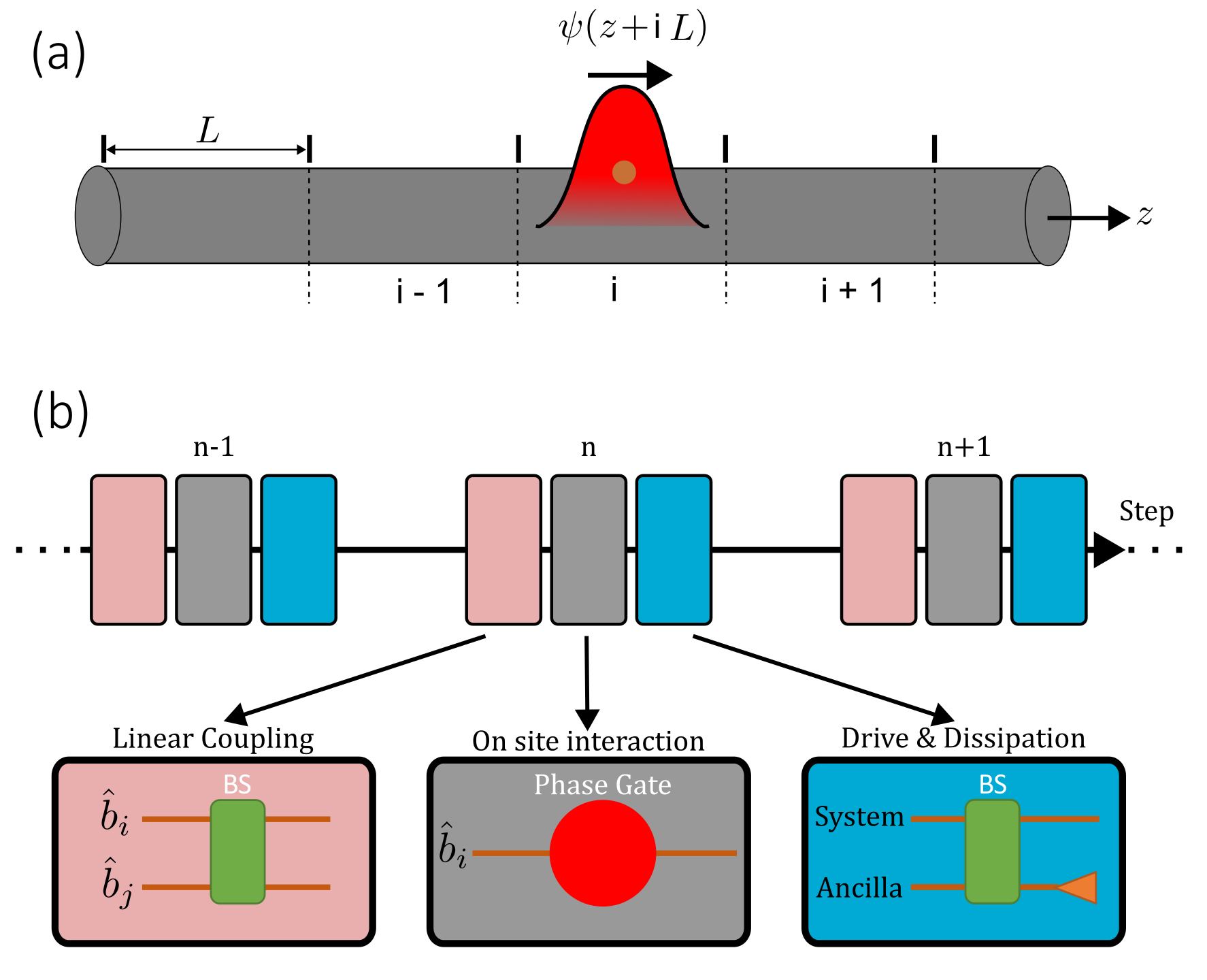}
		\caption{(a) Encoding discrete bosonic modes into a waveguide by separating the continuum into discrete time bins of length $L$, labeled as $...(i-1),i,(i+1)...$. The pulse shown here represents the state $\hat{b}_{i}^{\dagger}\ket{vac}$. $z$ is the real-space coordinate. (b) Trotterizing the Lindbladian evolution into a sequence of local quantum channels. In such a simulation framework, the linear coupling terms $(J_{ij}\hat{b}_{i}\hat{b}_{j}^{\dagger}+h.c.)$ are simulated by linear beams splitters (BS) while the on-site interaction terms $f(\hat{n}_{i})$ can be simulated by photon number selective phase gates. Further, by introducing ancilla degrees of freedoms, we can simulate various types of dissipation.}
		\label{Fig:intro}
	\end{figure*}

	We first introduce our scheme for encoding discrete bosonic modes in waveguides. 
	A waveguide has a 1D continuum of bosonic modes. We can encode discrete bosonic modes in a waveguide by separating the continuum into discrete time bins of length $L$, as shown in Figure~\ref{Fig:intro}(a). 
	This encoding scheme can also be formalized mathematically, which
	we present here by considering only the right propagating modes of a waveguide. These modes can be modeled by a continuum of bosonic operators $\hat{a}_{z}$ for $z\in R$. Here $z$ is the real-space coordinate of the waveguide and $\hat{a}_{z}(\hat{a}_{z}^{\dagger})$ annihilates (creates) a right propagation photon at location $z$ in the waveguide. The annihilation and creation operators satisfies the canonical relation $[\hat{a}_{z},\hat{a}_{z'}^{\dagger}]=\delta(z-z')$ where $\delta(z)$ is the Dirac delta function. We now choose a normalized function $\psi(z)$ that acquires non-zero values only in the time bin interval $[0,L]$. We define a set of discrete bosonic operators $\hat{b}_{n}^{\dagger}=\int_{-\infty}^{\infty}dz \,\psi(z+nL)\,\hat{a}_{z}^{\dagger}$. Since the function $\psi(z)$ is normalized, i.e. $\int_{-\infty}^{\infty}dz\, |\psi(z)|^2=1$,  the operators $\hat{b}_{n}$ satisfy the canonical relations $[\hat{b}_{n},\hat{b}_{m}^{\dagger}]=\delta_{m,n}$. Here $\delta_{m,n}$ is the Kronecker delta function, thus the operators $\hat{b}_{n}$ satisfy the canonical relations for discrete bosonic modes. As schematically shown in Figure~\ref{Fig:intro}(a), the operator $\hat{b}_{n}^{\dagger}$ creates a waveguide photon with pulse shape $\psi(z)$ in the $n-th$ time bin. Also, since any photon state in the waveguide evolves according to the waveguide Hamiltonian $\hat{H}_{wg}=-i\int_{-\infty}^{\infty}dz \ v_{g} \hat{a}_{z}^{\dagger}\partial_{z}\hat{a}_{z}$~\cite{shen2007strongly}, the time-bins are propagating with a group velocity $v_g$.
	
	Next, we introduce the general master equation that describes the dynamics that we wish to simulate.
	We focus on multimode bosonic systems, whose density matrix $\hat{\rho}$ evolves according to the Lindblad master equation, 
	$d\hat{\rho}/dt=-i[\hat{H},\hat{\rho}]+\sum_{\nu}\gamma_{\nu}(\hat{c}_{\nu}\hat{\rho}\hat{c}_{\nu}^{\dagger}-\frac{1}{2}\{\hat{c}_{\nu}^{\dagger}\hat{c}_{\nu},\hat{\rho}\})$. Here $\hat{H}$ is the system Hamiltonian that describes coherent evolution and $\hat{c}_{\nu}$s are quantum jump operators that characterize the dissipation with rate $\gamma_{\nu}$. 
	A broad family of multimode bosonic Hamiltonians, including the Bose-Hubbard~\cite{karamlou2024probing,rosen2024synthetic}, FQH~\cite{wang2024realization}, and non-Abelian gauge theories~\cite{cheng2023artificial}, can be written in the form of:
	
	\begin{equation}
		\hat{H} = \sum_{\langle i,j\rangle}(J_{ij}\hat{b}_{i}\hat{b}_{j}^{\dagger}+h.c.) + \sum_{i}f(\hat{n}_{i})
		\label{general hamiltonian}
	\end{equation}
	
	Here $J_{ij}\hat{b}_{i}\hat{b}_{j}^{\dagger}+h.c.$ is the linear coupling term and $J_{ij}$ is the tunneling rate between bosonic modes $\hat{b}_{i}$ and $\hat{b}_{j}$. The function $f$, which is a formal function of the number operator $\hat{n}_{i}=\hat{b}_{i}^{\dagger}\hat{b}_{i}$, is the on-site interaction for bosonic mode $\hat{b}_i$. In this work, we primarily focus on this family of Hamiltonians~\cite{umucalilar2012fractional,weidemann2020topological,weidemann2022topological}.
	
	Apart from the Hamiltonian, engineering the dissipation is also of crucial importance. 
	Firstly, the dissipation combined with the coherent Hamiltonian evolution can drive the state of the system into exotic quantum many-body states with novel properties~\cite{kapit2014induced}. Moreover, it is possible to prepare the ground state of a target Hamiltonian by properly engineering the dissipation. Such state preparation can be made robust against certain types of noise, which are known as autonomous quantum error corrections~\cite{albert2016geometry}.
	
	To simulate the master equation discussed above, we use the Suzuki-Trotter formula to break down the evolution of the system into a sequence of simple local quantum operations. This strategy, which is known as Trotterization~\cite{lloyd1996universal}, has been widely utilized in the non-interacting regime to simulate Hamiltonian evolution or even Lindbladian evolution with waveguide photons~\cite{schreiber20122d,leefmans2022topological,parto2023non,weidemann2020topological,weidemann2022topological,ferreira2024deterministic,chalabi2019synthetic,chalabi2020guiding,lin2023manipulating,zheng2024dynamic,leefmans2022topological,monika2024quantum}. The full master equation can be written as $d\hat{\rho}/dt=\mathbb{L}[\hat{\rho}]$, where the superoperator $\mathbb{L}[*]$ is known as the quantum dynamic generator. An initial state $\hat{\rho}_{init}$ thus evolves into the state $e^{\mathbb{L}T}[\hat{\rho}_{init}]$ after evolution time $T$. By properly splitting the generator $\mathbb{L}$ into a sum of generators $\mathbb{L}=\sum_{\mu}\mathbb{L}_{\mu}$ and splitting the full evolution time $T$ into $N$ Trotter steps with duration $\delta t = T/N$, we can approximate the evolution within time $T$ as:
	\begin{equation}
		e^{\mathbb{L}T}=(\prod_{\mu}e^{\mathbb{L}_{\mu}\delta t})^{N}
		\label{Suzuki Trotter}
	\end{equation}
	With a careful choice of splitting, each quantum channel $e^{\mathbb{L}_{\mu}\delta t}$ may become spatially local, so that they can be easily implemented on the encoded bosonic modes, as discussed below and more concretely in sections 3 and 4.
	
	Following equation~\ref{Suzuki Trotter}, we show that the on-site interaction term and the linear coupling term in equation~\ref{general hamiltonian} can be simulated by photon-number-selective phase gate and linear beamsplitters, respectively.
	For simplicity, let $\mathbb{L}[*]=-i[\hat{H},*]$, where $\hat{H}$ is given in equation~\ref{general hamiltonian}. We split the Hamiltonian into several terms, such that $\hat{H}=\sum_{\mu=1}^{\mu_{max}}\hat{H}_{\mu}$. We thus have $\mathbb{L}=\sum_{\mu=1}^{\mu_{max}}\mathbb{L}_{\mu}$, where $\mathbb{L}_{\mu}[*]=-i[\hat{H}_{\mu},*]$. 
	For $\mu=1,...,\mu_{max}-1$, we group all the linear coupling terms $J_{ij}\hat{b}_{i}\hat{b}_{j}^{\dagger}+h.c.$ into $\mu_{max}-1$ disjoint groups so that any two terms in each group commute with each other. Then the channel $e^{\mathbb{L}_{\mu}\delta t}$ can be obtained by applying unitary gates $e^{-i\delta t(J_{ij}\hat{b}_{i}\hat{b}_{j}^{\dagger}+h.c.)}$ between all nearest neighbor pairs $\langle i,j \rangle $ in the same group, and by definition these unitaries are linear beamsplitter operations between two bosonic modes. For $\mu=\mu_{max}$, we set $\hat{H}_{\mu_{max}}=\sum_{i}f(\hat{n}_i)$. The channel $e^{\mathbb{L}_{\mu}\delta t}$ can then be obtained by applying a unitary gate $e^{-i\delta t f(\hat{n}_{i})}$ for each bosonic mode $\hat{b}_{i}$. By definition, these gates are photon number selective phase gates that apply a phase $\phi(n)=-\delta t f(n)$ when there is $n$ photon in the mode $\hat{b}_{i}$. These linear beamsplitter gates and photon number selective phase gates are shown as pink and gray boxes in Figure~\ref{Fig:intro}(b). For a full Lindbladian, there could be other terms in $\mathbb{L}$ that describe dissipation, and we can implement $e^{\mathbb{L}_{diss}\delta t}$ by properly introducing ancilla degrees of freedom. These will be demonstrated explicitly for certain types of dissipation such as the one in Figure~\ref{Fig:FQH coherent drive}(a).

	
	\section{Photon phase gate via three-level atoms}
	
	Implementing a photon-number-selective phase gate for waveguide photons using waveguide QED systems poses significant challenges. 
	Ideally, the phase gate should map the $k$-photon state $\frac{1}{\sqrt{k!}}(\hat{b}^{\dagger})^{k}\ket{vac}$ to the state $e^{i\phi(k)}\frac{1}{\sqrt{k!}}(\hat{b}^{\dagger})^{k}\ket{vac}$ where $\hat{b}^{\dagger}=\int dz \psi(z)\hat{a}_{z}^{\dagger}$. This means that the gate should apply a controllable phase for arbitrary photon number $k$ in a single time bin without distorting the shape of the wavefunction, and should be distinguished from \textbf{\textit{control}} phase gates where there's as most one photon in each time bin~\cite{duan2004scalable,hacker2016photon}. The ``distortion free" requirement for our phase gate was considered hard to satisfy for most realistic nonlinearities in waveguide QED, such as two-level atoms coupled to waveguides. This is in part because few-photon pulse propagation through these systems suffer from momentum mixing~\cite{shen2007strongly,hafezi2012quantum}, which leads to distortion of the few-photon wavefunction~\cite{yang2022deterministic}. Existing protocols~\cite{yang2022deterministic} that provide a solution to this challenge do not apply to photon numbers greater than two. Moreover, such protocols are limited to a specific wavefunction that optimizes the fidelity. Any bias from the optimized wavefunction deteriorates the gate fidelity. 
	
	\begin{figure*}[t]
		\centering
		\includegraphics[width=0.99\textwidth]{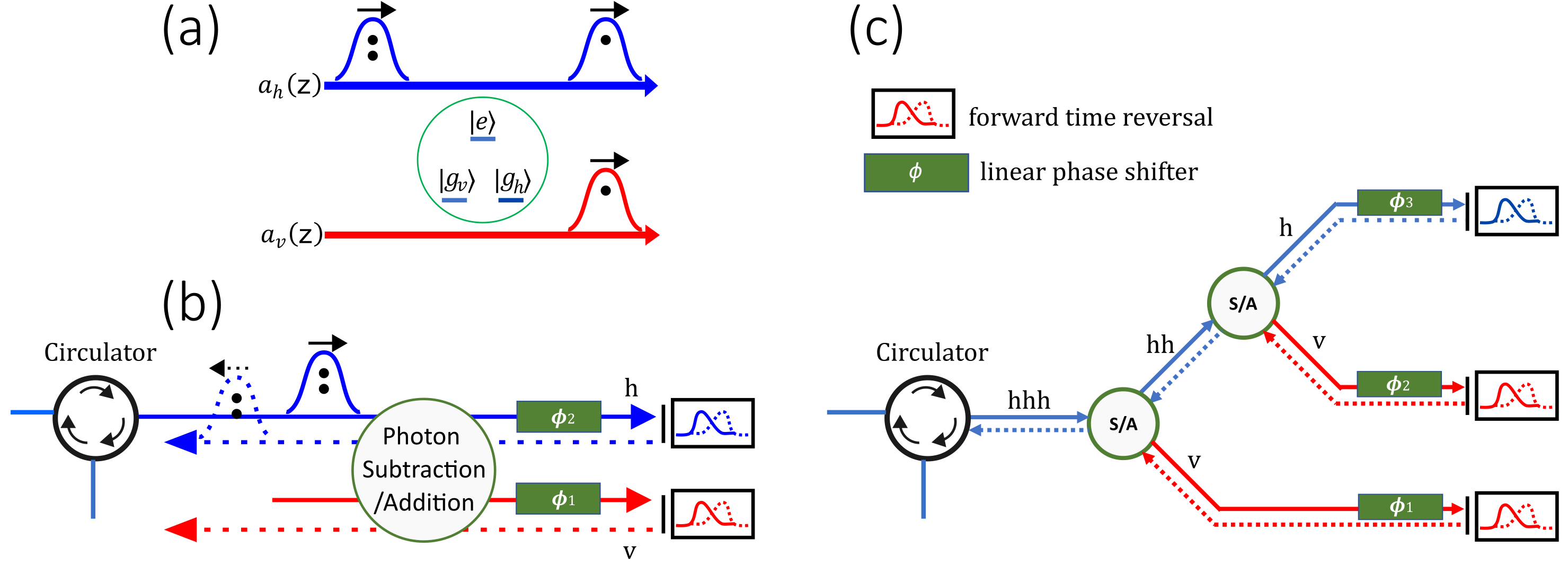}
		\caption{Implementation of photon number selective phase gate for waveguide-encoded bosonic modes using $\Lambda$-systems. (a) A $\Lambda$-system coupled to two unidirectional waveguides (labeled $``h"$ and $``v"$) can serve as an asymptotically-ideal deterministic photon subtractor/adder for Fock states. (b) The architecture of the nonlinear phase gate, which applies a phase $e^{i(\phi_{1}+(k-1)\phi_{2})}$ for any photon number $k$ greater than 1. (c) A cascading version of the system in (b), which enables a more sophisticated nonlinear function f(n) for the photon number dependent phase.}
		\label{Fig:phase gate}
	\end{figure*}
	
	Here we propose a phase gate based on deterministic photon subtraction for Fock state via $\Lambda$-systems~\cite{shomroni2014all,rosenblum2011photon}, which overcomes the above challenges. The photon subtractor is sketched in Figure~\ref{Fig:phase gate}(a), containing a $\Lambda$-system with two ground states $\ket{g_h},\ket{g_v}$ and an excited state $\ket{e}$. Here chiral coupling between the atom and the waveguide is considered to avoid decay in the backward direction. The $\ket{g_h}\leftrightarrow\ket{e}$ transition is only coupled to the upper ``h"-waveguide, and the $\ket{g_v}\leftrightarrow\ket{e}$ transition is only coupled to the lower ``v''-waveguide, with the same transition energy. The photon subtraction process is a quantum scattering process, where the atom is initialized in the $\ket{g_h}$ ground state and a $k$-photon Fock state $\frac{1}{\sqrt{k!}}(\hat{b}^{\dagger})^{k}\ket{vac}$ is injected towards the atom from the ``h''-waveguide from $-\infty$. The central frequency of the injected pulse is resonant with the atom's transition energy. 
	As shown in the SI section I-III, as the waveguide atom coupling rate $\gamma\rightarrow\infty$, the $k$-photon state that scatters off the atom converges to a state with $(k-1)$ photons in ``h"-waveguide and $1$ photon in the ``v" waveguide, and the atom state is switched to $\ket{g_v}$ after the scattering process. For any finite $\gamma$, the scattering output state is in a superposition of $\psi$($k-1$ ``h" and $1$ ``v" photons) and $\psi$($k$ ``h" photon). In SI section I-III, we show that the norm of the latter component, i.e., the subtraction failure probability $p_{fail}$, converges to $0$ as $\gamma\rightarrow\infty$. We focus on this limit thereafter. 
	
	Making use of photon subtraction, we propose the phase gate in Figure~\ref{Fig:phase gate}(b). Immediately after the subtraction, we attach two linear phase shifters to the $``h"$ and $``v"$-waveguides. By picking a phase $\phi_2$ for the linear phase shifter in $``h"$-waveguide and $\phi_1$ for the one in the $``v"$-waveguide, the state after the subtraction picks up a phase $(\phi_1+(k-1)\phi_2)$. Similar to~\cite{yang2022deterministic}, we then reflect the phase-shifted state back to the atom using two phase conjugating mirrors~\cite{sivan2011time,chumak2010all,yanik2004time} to time-inverse the subtraction process so that the final output state of the phase gate is the original Fock state with a phase $(\phi_1+(k-1)\phi_2)$. 
	We note that for an asymmetric wavefunction $\psi(z)$, this transformation will also generate a time reversal which can be undone with a second phase-conjugating mirror. But for a symmetric $\psi(z)$ this additional correction is not required. The definition of gate infidelity $INF_{gate}$ and its relation with $p_{fail}$ is analyzed in SI section I-III, and is shown to vanish for $\gamma\rightarrow\infty$.
	
	One can achieve an even more controllable photon-number selective phase using the extended scheme shown in Figure~\ref{Fig:phase gate}(c). Here another layer of photon subtraction is implemented after the first layer. As proved in SI section IV, the scattering output state has exactly one photon in each $``v"$-waveguide and $(k-2)$ photons in the $``h"$-waveguide after all subtractions, for $\gamma\rightarrow\infty$. Thus, using the linear phase shifters in Figure~\ref{Fig:phase gate}(c), a $k$-photon state picks up a phase $\phi_{1}$, $(\phi_{1}+\phi_{2})$ and $(\phi_{1}+\phi_{2}+\phi_{3})$ for $k=1,2,3$, respectively. For $k>3$, the phase is $(\phi_{1}+\phi_{2}+(k-2)\phi_{3})$. Generally, additional photon subtraction layers can implement the desired photon-number selective phase for any number of input photons fewer or equal to the number of photon subtraction layers $n_{layer}$. Namely, using the more general scheme, one can implement a phase shift of the form $\sum_{l=1}^{k}\phi_{l}$ for all $k<n_{layer}$, and thus the phase shift for any photon number is uniquely programmable by choosing the $\phi_{l}$s. We note that to integrate this phase gate in photonic circuits, we also included an optical circulator to avoid back-reflection.

	
	\section{Simulating the Bose-Hubbard model and Fermionization}
	
	We first apply our Trotterization scheme to simulate the 1D Bose-Hubbard Hamiltonian. The 1D Bose-Hubbard Hamiltonian is given by $H = -J\sum_{i}(\hat{b}_{i}^{\dagger}\hat{b}_{i+1}+h.c.)+\sum_{i}\frac{U}{2}(\hat{n}_{i}-1)\hat{n}_{i}$. Here $\hat{b}_{i}$ is the bosonic annihilation operator for the $i$-th lattice site, with $J$ being the tunneling rate. The operator $\hat{n}_{i}=\hat{b}_{i}^{\dagger}\hat{b}_{i}$ is the photon number operator and $U$ represents the strength of on-site interaction. 
	
\begin{figure*}[t]
		\centering
		\includegraphics[width=0.95\textwidth]{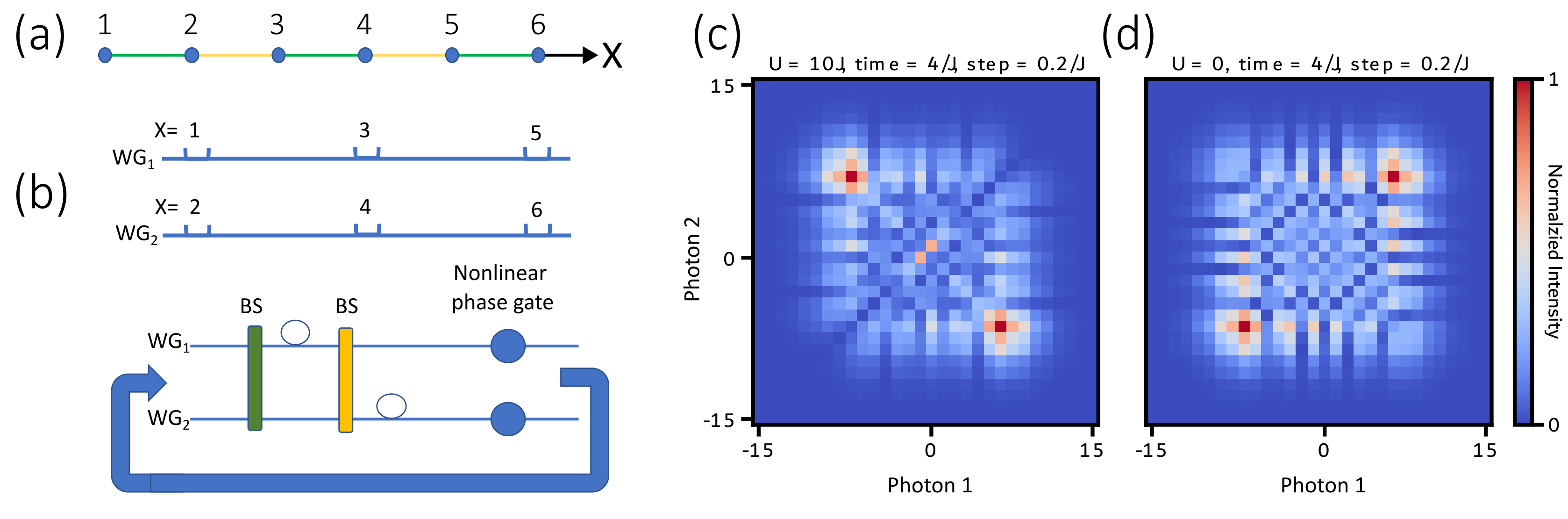}
		\caption{Schematic and results for Bose-Hubbard Simulation. (a) We decompose the Full 1D Bose-Hubbard Hamiltonian into three parts, shown in three different colors, and implement the exponentiation of each part sequentially. (b) Proposed circuit for Bose-Hubbard Hamiltonian simulation. (c-d) two photon correlator $\langle\hat{b}_{i}^{\dagger}\hat{b}_{j}^{\dagger}\hat{b}_{j}\hat{b}_{i}\rangle$ for a quantum quench experiment with strong (c) and no interaction (d). The initial state is $\ket{init}=\hat{b}_{N_{x}/2}^{\dagger}\hat{b}_{N_{x}/2+1}^{\dagger}\ket{vac}$, which we evolve using the circuit in (b).  }
		\label{Fig:bose hubbard}
	\end{figure*}

	
	To Trotterize the Hamiltonian evolution, we split the Hamiltonian into three terms, as shown in Figure~\ref{Fig:bose hubbard}(a). We sweep all the linear coupling terms 
	$-(J\hat{b}_{i}^{\dagger}\hat{b}_{i+1}+h.c.)$ between pairs of nearest neighbors denoted as green edges into the first term. The second term thus includes all the linear coupling terms denoted as yellow edges. The third term is the on-site interaction term, denoted as blue vertices. After exponentiation, the first two terms can be implemented by linear beamsplitters between corresponding edges, and the on-site interaction term can be implemented by the nonlinear phase gates, as discussed in Section 1.
	
	A compact and hardware-efficient experimental realization of our Trotterized Hamiltonian evolution is shown in Figure~\ref{Fig:bose hubbard}(b).
	To encode all the bosonic modes, we make use of equally spaced time bins in two parallel waveguides. The time bins in waveguide 1 (WG1) are used to encode all bosonic modes with odd $x$, and in waveguide 2 (WG2) we encode all bosonic modes with even $x$. The spacing between two adjacent time bins is $l_x$ for both waveguides, and in the initial configuration the $x=1$ time bin is aligned with the $x=2$ time bin. It can then be verified that the waveguide circuit in Figure~\ref{Fig:bose hubbard}(b) implements all the desired unitary gates in a single Trotter step. Particularly, the green beamsplitter implements all the beamsplitter operations denoted as green edges in Figure~\ref{Fig:bose hubbard}(a) as the time bins fly through it, and similarly for the yellow beamsplitter. This is enabled by a fiber delay of $l_x$ in WG1 which creates a new time bin configuration with $x=3$ aligned with $x=2$. After re-aligning the time bins to the original configuration with a second time fiber delay, we use two phase gates coupled to the WG1 and WG2 for on-site interaction simulation.
	
	
	We numerically examine the validity of the Bose-Hubbard simulator by demonstrating interaction-induced fermionization for bosons with a quantum quench experiment.
	Following previous works~\cite{preiss2015strongly}, we initialize the system into the state $\ket{init}=\hat{b}_{N_{x}/2}^{\dagger}\hat{b}_{N_{x}/2+1}^{\dagger}\ket{vac}$ i.e. two photons are located in two adjacent sites at the center of the lattice. Periodic boundary condition is used here for all calculations (see SI section IX). 
	We take Trotter step size $\delta t=0.2/J$ and simulate Hamiltonian evolution with a total evolution time $4/J$. After evolving the initial state with a Hamiltonian with no on-site interactions, we observe bunching behavior originating from bosonic HBT (Hanbury Brown-Twiss) interference, as shown in Figure~\ref{Fig:bose hubbard}(d). This means that the two particles are more likely to be observed on the same side of the chain. However, when we turn on the repulsive on-site interaction $U=10J$, we observe fermionization as expected, which means that the two particles are more likely to be spotted on the opposite sides of the chain~\cite{preiss2015strongly}, as shown in Figure~\ref{Fig:bose hubbard}(c).

	
	\section{the FQH Hamiltonian}
	
\begin{figure*}[t]
		\centering
		\includegraphics[width=0.99\textwidth]{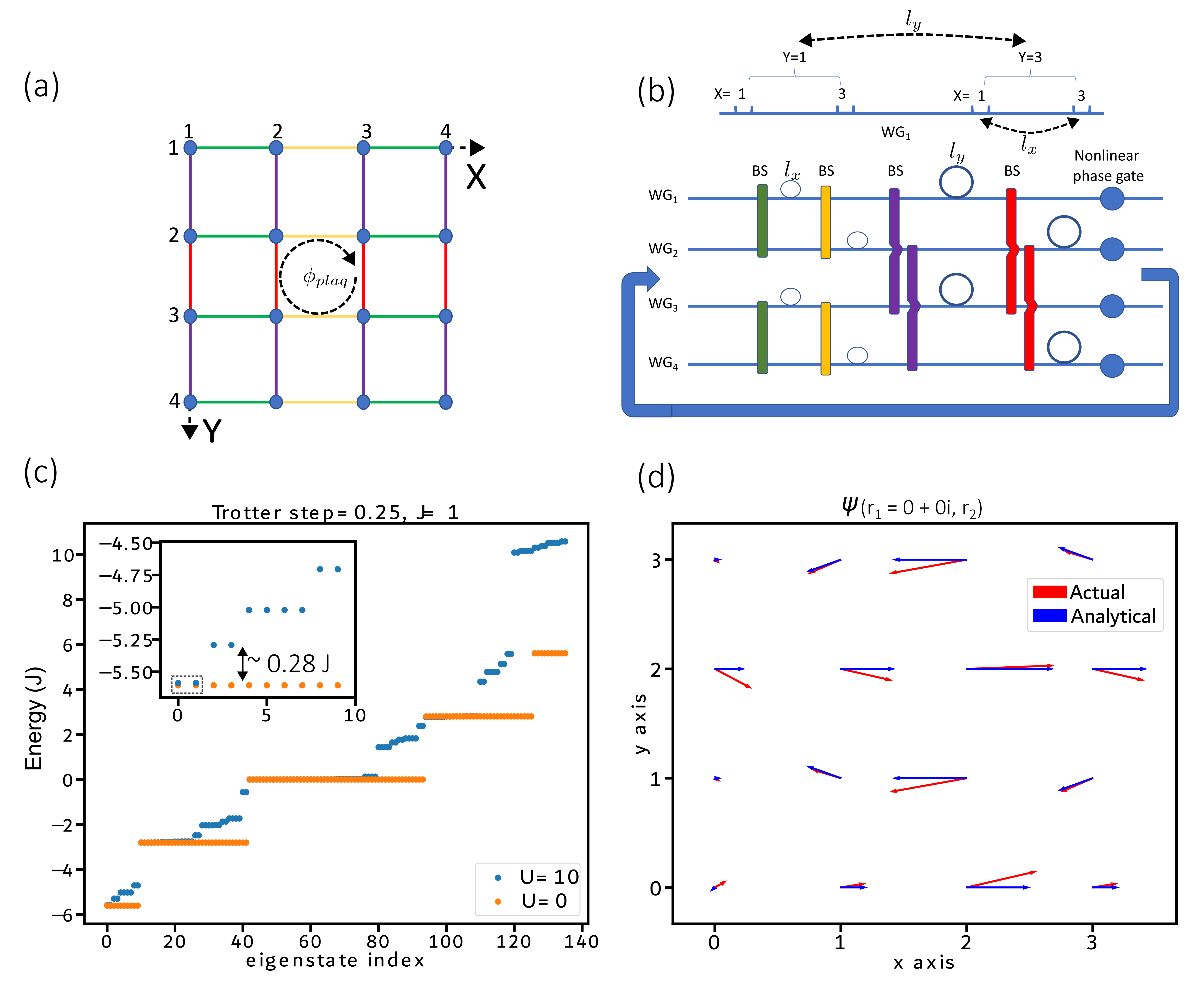}
		\caption{Schematic for FQH Hamiltonian simulation. (a) The FQH Hamiltonian defined on a 4x4 square lattice. The accumulated hopping phase along an oriented look around a plaquette is indicated by the $\phi_{plaq}$. The blue circles indicate the on-site interaction term. The lines indicated the linear coupling terms between nearest neighbors. (b)  The circuit used to simulate the FQH Hamiltonian. BS: beamsplitter. WG: Waveguide. Top inset: the encoding scheme of bosonic modes. (c) The calculated effective eigenenergies for the circuit, without (orange) and with (blue) on-site interaction. The Inset of the Figure shows the energy gap opened by on-site interaction. (d) Comparison between the analytic (blue) ground state wavefunction $\psi_{ana}^{(1)}(r_1,r_2)$ and the optimized linear combination of the actual solutions for ground state wavefunction (red). }
		\label{Fig:FQH Hamiltonian simulation}
	\end{figure*}

	We further consider the task of simulating the FQH Hamiltonian.
	The fractional quantum Hall Hamiltonian is defined on a 2D square lattice with nearest neighbor coupling, as shown in Figure~\ref{Fig:FQH Hamiltonian simulation}(a). The Hamiltonian is given by:
	
	\begin{equation}
		H_{FQH} = -J\sum_{\langle i,j\rangle}(\hat{b}_{i}^{\dagger}\hat{b}_{j}e^{i\phi_{ij}}+h.c.)+\sum_{i}\frac{U}{2}\hat{n}_{i}(\hat{n}_{i}-1)
	\end{equation}
	Here $\langle i,j \rangle$ denotes pairs of nearest neighbor lattice sites, shown as edges in Figure~\ref{Fig:FQH Hamiltonian simulation}(a). The hopping phases $\phi_{ij}$s are chosen such that the integration along an oriented loop around a single $1\times 1$ lattice plaquette, given by $\phi_{plaq}=\frac{1}{2\pi}\sum_{loop}\phi_{ij}$, is a fixed value. This value is equivalent to the effective magnetic flux through each plaquette. 
	
	To simulate this Hamiltonian evolution, we break the Hamiltonian into five terms. 
	We sweep all the nearest neighbor pairs in the Hamiltonian into four groups, denoted by edges with different colors in Figure~\ref{Fig:FQH Hamiltonian simulation}(a). The first four Hamiltonian terms are obtained by summing over linear coupling terms $-J(\hat{b}_{i}^{\dagger}\hat{b}_{j}e^{i\phi_{ij}}+h.c.)$ in the same group (i.e. edges with the same color). The fifth term is the on-site interaction term, i.e. $H_{on-site}=\sum_{i}U/2(\hat{n}_{i}-1)\hat{n}_{i}$. 
	
	Figure~\ref{Fig:FQH Hamiltonian simulation}(b) shows the waveguide circuit that simulates the Trotterized FQH Hamiltonian evolution.
	We use four waveguides to encode all the bosonic modes in the 2D lattice. The sites with odd $x$ and odd $y$ are encoded in time bins in WG1, and those with even $x$ and odd $y$ are encoded in time bins in WG2. Similarly, sites with even $y$ are encoded in WG3 and WG4. For each WG, sites with different $x$s but the same $y$ are encoded in a cluster of equally spaced time bins with spacing $l_x$, and two adjacent clusters that represent two different $y$s are separated by a larger distance $l_y$. Similar to the BH example, the circuit in Figure~\ref{Fig:FQH Hamiltonian simulation}(b) implements all the unitary gates in one Trotter step. The beamsplitters with different colors are used to simulate the linear coupling terms denoted by the same color, and the fiber delays are used to create the correct time bin alignment. The nonlinear phase gates are used for on-site interaction simulation. 

	To verify that our scheme simulates the correct Hamiltonian, we compute the effective eigenenergies for the circuit shown in Figure~\ref{Fig:FQH Hamiltonian simulation}(b). We pick a minimal model for FQH, in which we have a $4\times4$ square lattice and the flux per plaquette is $\phi_{plaq}=\frac{1}{4}$. We fix $J=1$ and $U=10J$ with Trotter step size $\delta t = 0.25$. The circuit shown in Figure~\ref{Fig:FQH Hamiltonian simulation}(b) implements a unitary $U_{\delta t}$ which conserves the total photon number in the lattice. Thus, one can diagonalize the unitary operator $U_{\delta t}$ within the $k$-photon sector, 
	and obtain an ensemble of eigenvalues $u_{j}^{(k)}$ and corresponding eigenvectors. These eigenvalues for such a unitary operator are complex numbers with radius $1$, and we define the effective eigenenergy as $\epsilon_{j}^{(k)}=-ilog(u_{j}^{(k)})/\delta t$.
	
	Figure~\ref{Fig:FQH Hamiltonian simulation}(c) shows all the eigenenergies with photon number $k=2$, and compares the case with $U=0$  (no interaction) v.s. $U=10J$  (strong interactions). We note that here we used periodic boundary conditions (see SI section IX). For no interaction, as shown in orange, the two photons can occupy the three possible Landau levels with no extra contact energy, and thus we have five possible values of eigenenergies. When the interaction is turned on, however, the contact interaction between two photons occupying the lowest landau level opens up a gap between the ground state and the lowest excited state, shown as an inset of Figure~\ref{Fig:FQH Hamiltonian simulation}(c). We thus have a pair of degenerate ground states in the strongly interacting case, consistent with the original Hamiltonian.
	
	Having identified the pair of ground states from the effective energies (i.e. the two blue circles in the dashed box in the inset of Figure~\ref{Fig:FQH Hamiltonian simulation}(c)), we now compare their wavefunction to that of the original Hamiltonian. For the original FQH model defined on a torus, the analytic expression of the ground state wavefunctions can be expressed using Jacobi theta functions, as shown in SI section V. For our configuration, there are two orthonormal analytic wavefunctions that span the ground space, which we denote as $\Psi_{ana}^{(1)}(r_1,r_2)$ and $\Psi_{ana}^{(2)}(r_1,r_2)$, where $r=x + iy$. The two ground state wavefunctions for our actual circuit model are denoted as $\Psi_{act}^{(1)}(r_1,r_2)$ and $\Psi_{act}^{(2)}(r_1,r_2)$. We then pick $\Psi_{ana}^{(1)}(r_1,r_2)$ and maximize its overlap between the optimal linear combination $\alpha\Psi_{act}^{(2)}(r_1,r_2)+\beta\Psi_{act}^{(1)}(r_1,r_2)$. Namely, we solve the optimization problem
	
	\begin{equation}
		\begin{aligned}
			& \max_{\alpha,\beta}|\sum_{r_1,r_2}\Psi_{ana}^{(1)*}(r_1,r_2)\times(\alpha\Psi_{act}^{(2)}(r_1,r_2)+\beta\Psi_{act}^{(1)}(r_1,r_2))|^2\\
			& s.t.\ \ |\alpha|^2+|\beta|^2=1
		\end{aligned}
	\end{equation}
	
	We obtain the optimizer $\tilde{\alpha}$ and $\tilde{\beta}$ and hence the optimal wavefunction is $(\tilde{\alpha}\Psi_{act}^{(2)}(r_1,r_2)+\tilde{\beta}\Psi_{act}^{(1)}(r_1,r_2))$. We observe the optimal value for the overlap to be $94.5\%$. Further, we plot the comparison between the analytic wavefunction $\Psi_{ana}^{(1)}(r_1,r_2)$ in our model and the optimized linear combination of the actual solution, shown as blue and red arrows in Figure~\ref{Fig:FQH Hamiltonian simulation}(d). Note that to simplify the presentation, we have fixed $r_1= 0 + 0i$ and vary $r_2$ across the whole $4\times4$ lattice. Here the length and angle of the arrows represent the amplitude and phase of the wavefunction. A good agreement can be observed between the actual ground state wavefunction for our circuit and the analytic wavefunction. 

	
	\section{the FQH Hamiltonian with coherent drive}
	
	In the previous section, we demonstrated the simulation of the FQH Hamiltonian. In this section, we further demonstrate how to probe the ground state shown in Figure~\ref{Fig:FQH Hamiltonian simulation}(c-d) by simulating a Lindbladian describing drive and dissipation.
	The Lindbladian master equation is given by:
	\begin{equation}
		\frac{d\rho}{dt}=\mathbb{L}[\rho]=\mathbb{L}_{FQH}[\rho]+\sum_{i}\mathbb{L}_{i,drive}[\rho]+\sum_{i}\mathbb{L}_{i,diss}[\rho]
	\end{equation}
	where $\mathbb{L}_{FQH}[\rho]=-i[H_{FQH},\rho]$. We include coherent pump for each lattice bosonic mode $\hat{b}_{i}$, i.e. $\mathbb{L}_{i,drive}[\rho]=-i[H_{i,drive},\rho]$ with $H_{i,drive}=Fe^{i\Omega_{drive}t}\hat{b}_{i}+h.c.$. The parameters $\Omega_{drive}$ and $F$ are the driving frequency and amplitude, respectively. The dissipation is single photon loss with rate $\gamma$, given by $\mathbb{L}_{i,diss}[\rho]=\gamma( \hat{b}_{i}\rho\hat{b}_{i}^{\dagger}-\frac{1}{2}\{\hat{b}_{i}^{\dagger}\hat{b}_{i},\rho\} )$.
	
\begin{figure*}[t]
		\centering
		\includegraphics[width=0.7\textwidth]{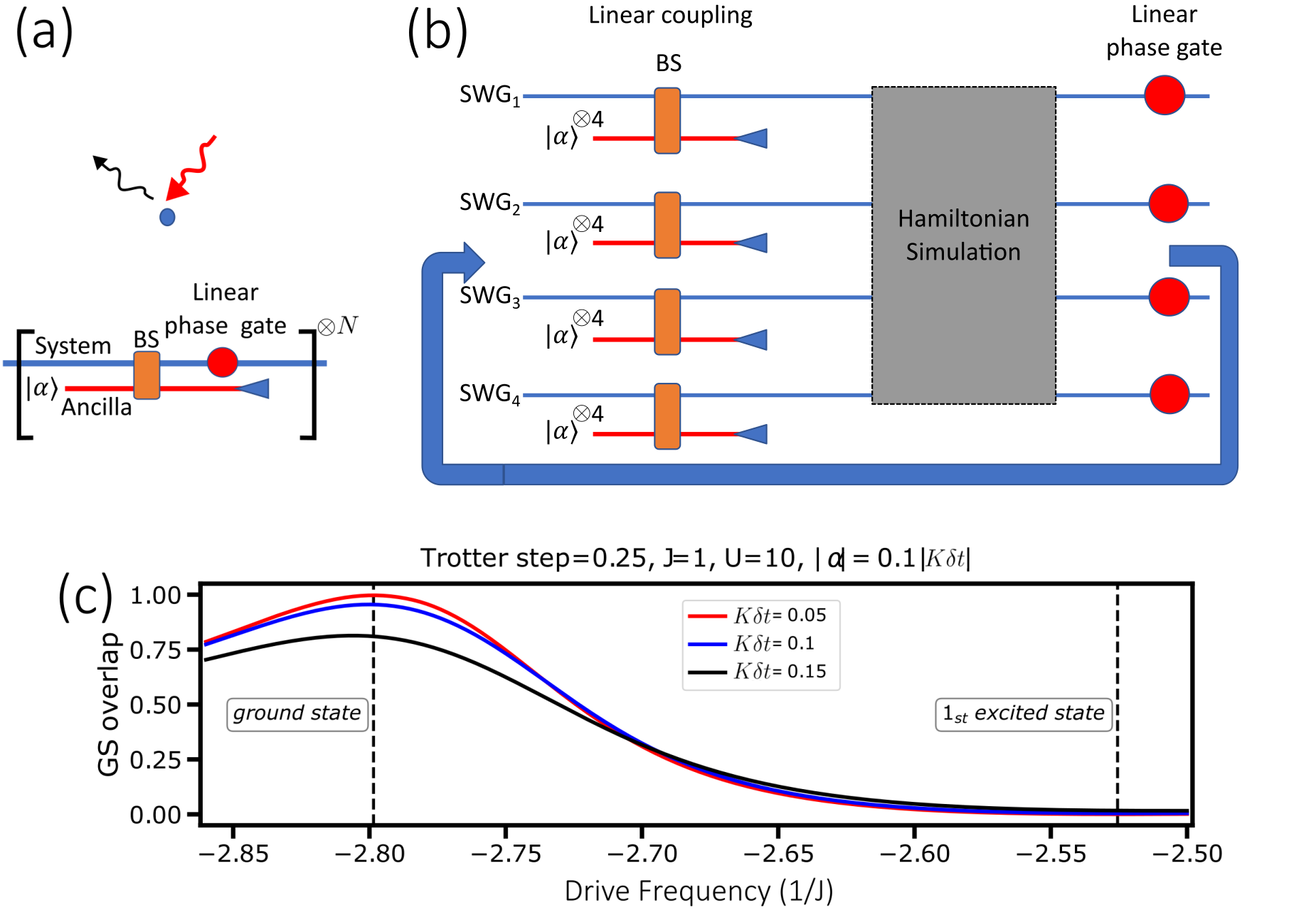}
		\caption{Simulating the FQH model with coherent drive and dissipation. (a) The effect of coherent drive and single photon loss combined can be simulated by interacting the system bosonic mode with an ancilla bosonic mode initialized in the coherent state $\ket{\alpha}$ before dumping the ancilla. (b) The complete circuit for the implementation of the FQH Hamiltonian with coherent drive and dissipation. BS: beamsplitter. WG: waveguide. The gray box indicates the Hamiltonian simulation described in Figure 4. (c) the overlap of the fixed point $\rho_{fix}$ with the FQH ground state subspace after post-selection. The (c) panel is plotted as a function of the drive frequency.}
		\label{Fig:FQH coherent drive}
	\end{figure*}

	In Figure~\ref{Fig:FQH coherent drive}(a) we show how to simulate drive and dissipation for only one bosonic mode $\hat{b}$. We thus need to apply the channel $\epsilon_{DS} = e^{(\mathbb{L}_{drive}+\mathbb{L}_{diss})\delta t}$ in each Trotter step. To implement this channel we can introduce an ancillary bosonic mode $\hat{c}$ initialized in the coherent state $\ket{\alpha}$, and apply a beamsplitter unitary $e^{-iK \delta t(\hat{b}^{\dagger}\hat{c}+h.c.)}$ between the system and ancilla, where $K$ is the beamsplitter coupling rate. We then trace out the ancilla and apply a linear phase gate $e^{i\Phi\hat{n}}$ to the system bosonic mode. The above operations are summarized in Figure~\ref{Fig:FQH coherent drive}(a). To simulate drive and dissipation with parameter $(F,\Omega,\gamma)$, we should take $\alpha^{*} K = F$, $\gamma \delta t = (K\delta t)^2$ and $\Phi=\Omega \delta t$, as shown in the SI section VI. 
	
	Generalizing Figure~\ref{Fig:FQH coherent drive}(a) to the $4\times 4$ FQH lattice, we propose the circuit in Figure~\ref{Fig:FQH coherent drive}(b). We use the same four-waveguide encoding scheme for the ancilla as for the system lattice, and each ancilla waveguide is initialized with four coherent laser pulses aligned with the system time bins. The four linear linear beamsplitters coupling the system and ancilla waveguides applies the linear beamsplitter operations between the encoded system and ancilla bosonic modes as the time bins fly through them. The gray box represents the simulation circuit in Figure~\ref{Fig:FQH Hamiltonian simulation}(b), and the linear phase gates shown as red circles are used to simulate different coherent drive frequencies.

	The circuit shown in Figure~\ref{Fig:FQH coherent drive}(b) applies a quantum channel $\epsilon^{full}$ on the system at each circulation, and drive any initial state to the steady state $\rho_{fix}$ defined by the equation $\epsilon^{full}[\rho_{fix}]=\rho_{fix}$.
	We fix the parameter $J=1, U=10, \delta t=0.25$ and pick several values of $K\delta t=0.05, 0.1, 0.15$, which represents different rates of dissipation. We also fix the ratio between the drive and dissipation, i.e. $\alpha^{*}/(K\delta t)=0.1$. The most important parameter is the driving frequency $\Omega_{drive}$, and we vary this parameter continuously from $-2.85$ to $-2.5$, forming the x-axis in Figure~\ref{Fig:FQH coherent drive}(c). 
	Since we are only probing the Laughlin state, we focus on the post-selected state $\rho^{(2)}_{fix}=\Pi_{2}\rho_{fix}\Pi_{2}/Tr[\Pi_{2}\rho_{fix}\Pi_{2}]$, obtained by projecting the fixed point $\rho_{fix}$ onto the $2$-photon configuration subspace. We examine the overlap of this post-selected state with the ground state subspace, which is spanned by the two ground states in Figure~\ref{Fig:FQH Hamiltonian simulation}(c). For a driving frequency that is resonant with the two photon transition energy from the vacuum to the FQH ground state~\cite{umucalilar2012fractional}, i.e. $\Omega_{drive}=\frac{1}{2}\epsilon_{FQH}$, this overlap is maximized, as shown in Figure~\ref{Fig:FQH coherent drive}(c). For $K\delta t=0.1$, the maximized overlap is beyond $95\%$. For more observables regarding the fixed point $\rho_{fix}$ see SI section VII.   
	
	\section{Summary and outlook}
	
	In conclusion, we show how strongly interacting bosonic quantum many-body models can be simulated using propagating waveguide photons. 
	Importantly, we showed that the crucial on-site interaction term can be simulated with a photon-number-selective phase gate based on three-level atom mediated photon subtraction. Notably, the photon subtraction subroutine has been experimentally demonstrated~\cite{shomroni2014all}. As benchmarking examples, we accurately simulated the Bose-Hubbard~\cite{karamlou2024probing} and the FQH~\cite{wang2024realization} Hamiltonian. Moreover, we demonstrate how to simulate an engineered dissipation to probe the ground state of the Hamiltonian. All examples in this work inherit the site controllability, hardware efficiency and readout techniques of the simulation platform.
	
	Looking forward, the on-site interaction demonstrated in this work can be integrated with all the previous works which simulate non-interacting models using propagating waveguide photons, enabling exploration of a diverse family of quantum collective phenomena~\cite{suarez2025chiral,jalali2023topological}. Moreover, going beyond Lindbladian simulation with Trotterization, our approach enables exploring general bosonic circuit models containing beamsplitters, nonlinear phase gates and measurements. This may allow the exploration of the bosonic version of measurement-induced phase transitions~\cite{skinner2019measurement} or to construct quantum neural networks~\cite{steinbrecher2019quantum}. Overall, our work opens the pathway towards the simulation of an enormous family of strongly interacting quantum many body models using waveguide photons.

	
	
	
	
	

	
	\section{Supplementary Information}
	
	\subsection{I. Definition of gate fidelity: Single subtraction case}
	We consider the phase gate architecture shown in Figure~\ref{Fig:phase gate}(a-b). As an example, we consider a two-photon Fock input state. 
	The input state can be written as:
	\begin{equation}
		\ket{\psi} = \frac{1}{\sqrt{2}}\int_{-\infty}^{\infty}dz_{h,1}dz_{h,2} \psi(z_{h,1})\psi(z_{h,2})\hat{a}_{h,z_{h,1}}^{\dagger}\hat{a}_{h,z_{h,2}}^{\dagger}\ket{vac,g_h}
		\label{gate input state}
	\end{equation}
	The photon subtractor shown in Figure~\ref{Fig:phase gate}(a) contains a three level atom coupled to two waveguides. The Hamiltonian is given by: $H=H_{wg}+H_{atom}+H_{couple}$. Here $\hat{H}_{wg}=-i\int dz v_{g}\hat{a}_{h,z}^{\dagger}\partial_{z}\hat{a}_{h,z}-i\int dz v_{g}\hat{a}_{v,z}^{\dagger}\partial_{z}\hat{a}_{v,z}$ describes the waveguide part of the Hamiltonian, whereas $\hat{H}_{atom}=E_e\ket{e}\bra{e}+E_g\ket{g_h}\bra{g_h}+E_g\ket{g_v}\bra{g_v}$ describes the atomic part. The atom-waveguide interaction is described by $\hat{H}_{couple} = \sqrt{2\pi \gamma}\int dz (\hat{a}_{v,z}\delta(z)\ket{e}\bra{g_v} + \hat{a}_{h,z}\delta(z)\ket{e}\bra{g_h} + h.c.)$ where $\gamma$ is the decay rate of the atom. Here chiral coupling between the atom and the waveguide is considered to avoid decay in the backward direction. Then, after the subtraction, the output state is given by:
	\begin{equation}
		\begin{aligned}
			\ket{\psi'} = &\int_{-\infty}^{\infty}dz_{h,1}dz_{h,2} \psi'^{(1)}(z_{h,1},z_{h,2})\hat{a}_{h,z_{h,1}}^{\dagger}\hat{a}_{h,z_{h,2}}^{\dagger}\ket{vac,g_h}\\
			+&\int_{-\infty}^{\infty}dz_{v}dz_{h,1} \psi'^{(2)}(z_{v};z_{h,1})\hat{a}_{v,z_{v}}^{\dagger}\hat{a}_{h,z_{h,1}}^{\dagger}\ket{vac,g_v}
			\label{general state after single subtraction}
		\end{aligned}
	\end{equation}
	
	For the phase gate architecture shown in Figure~\ref{Fig:phase gate}(b), the above output state then flies through a pair of linear phase shifters. Following that, we then send it to a phase conjugating mirror that reflects the state back while reversing the waveguide photon wavefunction. Finally, the state undergoes another atom-light scattering process before leaving the whole system. The final state that fully leaves the phase gate can be written as:
	\begin{equation}
		\begin{aligned}
			\ket{\psi''} = &\int_{-\infty}^{\infty}dz_{h,1}dz_{h,2} \psi''^{(1)}(z_{h,1},z_{h,2})\hat{a}_{h,z_{h,1}}^{\dagger}\hat{a}_{h,z_{h,2}}^{\dagger}\ket{vac,g_h}\\
			+&\int_{-\infty}^{\infty}dz_{v}dz_{h,1} \psi''^{(2)}(z_{v};z_{h,1})\hat{a}_{v,z_{v}}^{\dagger}\hat{a}_{h,z_{h,1}}^{\dagger}\ket{vac,g_v}
		\end{aligned}
	\end{equation}
	
	For the ideal phase gate, the output state should be:
	\begin{equation}
		\ket{\psi''_{ideal}} = \frac{e^{i(\phi_1+\phi_2)}}{\sqrt{2}}\int_{-\infty}^{\infty}dz_{h,1}dz_{h,2} \psi(z_{h,1}) \psi(z_{h,2}) \hat{a}_{h,z_{h,1}}^{\dagger}\hat{a}_{h,z_{h,2}}^{\dagger}\ket{vac,g_h}
	\end{equation}
	
	Then, the gate fidelity is defined by:
	\begin{equation}
		\begin{aligned}
			F_{gate} &= |\bra{\psi''}\ket{\psi''_{ideal}}|^{2}\\
			\bra{\psi''}\ket{\psi''_{ideal}} & = \sqrt{2}\int_{-\infty}^{\infty}dz_{h,1}dz_{h,2}(\psi''^{(1)}(z_{h,1},z_{h,2}))^{*} \psi(z_{h,1}) \psi(z_{h,2})
		\end{aligned}
	\end{equation}
	
	Generally, it is hard to find the analytic solution of $\psi''^{(1)}(z_{h,1},z_{h,2})$. Nevertheless, we can obtain the analytic solution for gate fidelity $F_{gate}$ using the following trick. Consider $\ket{\psi'}$. After the two linear phase shifter in Figure~\ref{Fig:phase gate}(b), it becomes:
	\begin{equation}
		\begin{aligned}
			\ket{\tilde{\psi}'} &= e^{i 2\phi_2}\int_{-\infty}^{\infty}dz_{h,1}dz_{h,2} \psi'^{(1)}(z_{h,1},z_{h,2})\hat{a}_{h,z_{h,1}}^{\dagger}\hat{a}_{h,z_{h,2}}^{\dagger}\ket{vac,g_h}\\
			& + e^{i (\phi_1+\phi_2)} \int_{-\infty}^{\infty}dz_{v}dz_{h,1} \psi'^{(2)}(z_{v};z_{h,1})\hat{a}_{v,z_{v}}^{\dagger}\hat{a}_{h,z_{h,1}}^{\dagger}\ket{vac,g_v}\\
		\end{aligned}
	\end{equation}
	
	On the other hand, we define $\ket{\tilde{\psi}'_{ideal}}$ as 
	\begin{equation}
		\begin{aligned}
			\ket{\tilde{\psi}'_{ideal}} &= e^{i (\phi_1+\phi_2)}\int_{-\infty}^{\infty}dz_{h,1}dz_{h,2} \psi'^{(1)}(z_{h,1},z_{h,2})\hat{a}_{h,z_{h,1}}^{\dagger}\hat{a}_{h,z_{h,2}}^{\dagger}\ket{vac,g_h}\\
			& + e^{i (\phi_1+\phi_2)} \int_{-\infty}^{\infty}dz_{v}dz_{h,1} \psi'^{(2)}(z_{v};z_{h,1})\hat{a}_{v,z_{v}}^{\dagger}\hat{a}_{h,z_{h,1}}^{\dagger}\ket{vac,g_v}\\
		\end{aligned}
	\end{equation}
	
	Now the state $\ket{\tilde{\psi}'_{ideal}}$ would \textbf{\textit{exactly}} evolve into $\ket{\psi''_{ideal}}$ after the phase conjugating mirror and another round of atom-light scattering. This is simply because it picks up an overall phase (i.e. the same phase $e^{i (\phi_1+\phi_2)}$ for the two components $\psi'^{(1)}$ and $\psi'^{(2)}$). Also, since the ``phase conjugating mirror reflection" and ``atom-light scattering process" are unitary operations, the inner product between any two states is conserved. Therefore, we can alternatively write the gate fidelity $F_{gate}$ as:
	
	\begin{equation}
		\begin{aligned}
			F_{gate} &= |\bra{\tilde{\psi}'}\ket{\tilde{\psi}'_{ideal}}|^{2}\\
			\bra{\tilde{\psi}'}\ket{\tilde{\psi}'_{ideal}} & = 2e^{i (\phi_1-\phi_2)}\int_{-\infty}^{\infty}dz_{h,1}dz_{h,2}|\psi'^{(1)}(z_{h,1},z_{h,2})|^{2} \\
			& + \int_{-\infty}^{\infty}dz_{v}dz_{h,1}|\psi'^{(2)}(z_{v};z_{h,1})|^{2} 
		\end{aligned}
	\end{equation}
	
	From the above expression, we can define the ``successful subtraction probability" $p_{succ}$ and ``subtraction failure probability" $p_{fail}$ as:
	\begin{equation}
		\begin{aligned}
			p_{fail} & = 2\int_{-\infty}^{\infty}dz_{h,1}dz_{h,2}|\psi'^{(1)}(z_{h,1},z_{h,2})|^{2} \\
			p_{succ} & = \int_{-\infty}^{\infty}dz_{v}dz_{h,1}|\psi'^{(2)}(z_{v};z_{h,1})|^{2}  \\
		\end{aligned}
	\end{equation}
	Note that we have $p_{fail}+p_{succ}=1$ since $\bra{\tilde{\psi}'}\ket{\tilde{\psi}'}=1$. For general photon number $k$, these probabilities are defined by:
	\begin{equation}
		\begin{aligned}
			p_{fail} & = k!\int_{-\infty}^{\infty}dz_{h,1}dz_{h,2}\cdots dz_{h,k}|\psi'^{(1)}|^{2} \\
			p_{succ} & = (k-1)!\int_{-\infty}^{\infty}dz_{v}dz_{h,1}\cdots dz_{h,k-1}|\psi'^{(2)}|^{2}  \\
			\label{eq: p_fail_p_succ_def}
		\end{aligned}
	\end{equation}
	
	Given $p_{fail}$ and $p_{succ}$, we can then simplify the expression of $F_{gate}$ to be:
	\begin{equation}
		\begin{aligned}
			F_{gate} & = (p_{fail} \ e^{i\delta\phi} + 1 - p_{fail})(p_{fail} \ e^{-i\delta\phi} + 1 - p_{fail})\\
			& = 1 - (2p_{fail}-2p_{fail}^{2})(1-cos(\delta\phi)) 
		\end{aligned}
	\end{equation}
	where $\delta\phi = \phi_1 - \phi_2$ is the phase difference between the two branches ``h" and ``v" in Figure~\ref{Fig:phase gate}(b). The gate infidelity is thus:
	\begin{equation}
		INF_{gate} = 1 - F_{gate} = 2p_{fail}(1-p_{fail})(1 - cos(\delta\phi))
		\label{eq: infidel subtract failure relation}
	\end{equation}
	
	Hence, we see that the infidelity is determined by two factors: (i) The probability of a failed photon subtraction $p_{fail}$, in which all the input photons still remain in the ``h"-waveguide, and (ii) The fact that we are using the phase gate as a \textbf{\textit{nonlinear}} phase gate, for which we need to choose $\phi_1\neq\phi_2$. In the worst case, we could have $INF_{gate}=4p_{fail}(1-p_{fail})$. To find the analytic solution for $F_{gate}$ and to show that it converges to $1$ in the long-pulse limit, we just need to find the analytic solution for $p_{fail}$. In fact, we can even find the analytic solution of $\psi'^{(1)}(z_{h,1},z_{h,2})$, for any input waveform $\psi(z)$, using the virtual cavity method. This will be shown in the following subsection.

	\subsection{II. Solving and bounding gate fidelity: Single subtraction case}
	
	Here, we present a detailed, quantitative analysis of few-photon propagation in the system shown in Figure~\ref{Fig:phase gate}(a). We will solve for the analytic expression of $\psi'^{(1)}$ in equation~\ref{general state after single subtraction}, for photon number $k=1,2$. Integrating $\psi'^{(1)}$ over all the $z_{h}$ variables gives analytic solution for $p_{fail}$. For photon number $k=1$, we directly show from the expression of $p_{fail}$ that for any positive $\epsilon$, we can make $p_{fail}<4\epsilon$ by choosing sufficiently large $\gamma$. For $k\geq 2$, we obtain an upper bound for $p_{fail}$ by investigating $p_{succ}$ instead, which can be obtained from $\psi'^{(2)}$. From this upper bound we can again show that $p_{fail}$ can be made arbitrarily small for sufficiently large $\gamma$. Numerically, for $k=1$ we plot $p_{fail}$ as a function of $\gamma$ for different input waveform $\psi(z)$. For $k\geq2$, we only plot the upper bound for $p_{fail}$ against $\gamma$. These numerical results then offer an estimate for the sufficiently large $\gamma$ that is required for some target gate infidelity.
	
\begin{figure*}[h]
		\centering
		\includegraphics[width=0.5\textwidth]{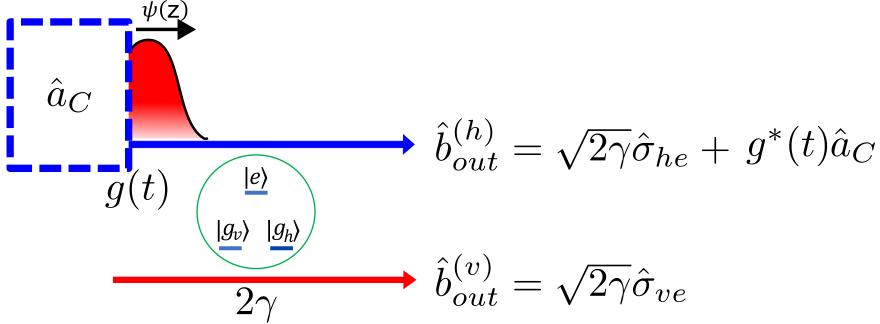}
		\caption{The virtual cavity method for a single layer of photon subtraction. The dashed box is the virtual cavity that injects a quantum pulse in the ``h"-waveguide towards the three-level atom. One can solve for the time-ordered correlation functions and hence the wavefunctions of the output state by applying the output operators indicated on the right.}
		\label{Fig:virtual cavity single}
	\end{figure*}

	To obtain $\psi'^{(1)}$ or $\psi'^{(2)}$, we make use of the virtual cavity method demonstrated in~\cite{kiilerich2019input,rosenblum2011photon}. For a $k$-photon state injected towards the atom in $\ket{g_h}$ from the left through the ``h”-waveguide, we can imagine that it is injected from a ``virtual cavity" coupled to the ``h"-waveguide, as shown in Figure~\ref{Fig:virtual cavity single}. To be precise, the virtual cavity $\hat{a}_{C}$ has time-dependent complex coupling $i[g^{*}(t)\hat{a}_{C}\hat{a}_{h,0-}^{\dagger}-h.c.]$ to the waveguide field $\hat{a}_{h,0-}$, which is the input field of the three-level atom. In particular, if one takes the coupling coefficient to be $g(t)=\frac{\psi^{*}(-t)}{\sqrt{1 - \int_{0}^{t}dt|\psi(-t')|^2}}$ and initializes the virtual cavity into a $k$-photon Fock state $\frac{1}{\sqrt{k!}}(\hat{a}_{C}^{\dagger})^{k}\ket{0}$, the state emitted into the waveguide is exactly the input state $\ket{\psi}$ in equation~\ref{gate input state}. For convenience, we define the temporal profile $u(t)=\psi(-t)$ where $u(t)$ only takes non-zero values on $[0,L]$ since $\psi(z)$ is only supported on the interval $[-L,0]$. We have the relation: 
	\begin{equation}
		g(t)=\frac{u^{*}(t)}{\sqrt{1 - \int_{0}^{t}dt' |u(t')|^2}}
		\label{decay rate waveform relation}
	\end{equation}
	
	
	
	The evolution of the reduced density matrix of the virtual-cavity-and-atom system shown in Figure~\ref{Fig:virtual cavity single} is given by the Lindbladian~\cite{kiilerich2019input}:
	
	\begin{equation}
		\begin{aligned}
			\frac{d\rho}{dt} = -i [\hat{H}, \rho] + \sum_{p=h,v}(\hat{b}^{(p)}_{out}\rho\hat{b}^{(p)\dagger}_{out}-\frac{1}{2}\{\hat{b}^{(p)\dagger}_{out}\hat{b}^{(p)}_{out},\rho \})\\
		\end{aligned}
	\end{equation}
	
	In the above equation, we are already working in the rotating frame of the atom so that the resonant frequency of the atom $\Omega=0$. The Hamiltonian $\hat{H}$ and the collapse operators are given by:
	
	\begin{equation}
		\begin{aligned}
			\hat{H} &= \frac{i}{2}(\sqrt{2\gamma}g(t)\hat{a}^{\dagger}_{C}\hat{\sigma}_{he}-\sqrt{2\gamma}g^{*}(t)\hat{a}_{C}\hat{\sigma}_{eh})\\
			\hat{b}^{(h)}_{out} &= g^{*}(t) \hat{a}_{C} + \sqrt{2\gamma}\hat{\sigma}_{he}\\
			\hat{b}^{(v)}_{out} &= \sqrt{2\gamma}\hat{\sigma}_{ve}
		\end{aligned}
	\end{equation}
	we now solve the promised results one by one.
	
	\subsubsection{One photon input state}
	We start with a single photon input state. Now the wavefunction $\psi'^{(1,2)}$ in equation~\ref{general state after single subtraction} can now be obtained by computing correlation functions from the above Lindbladian. In particular:
	\begin{equation}
		\begin{aligned}
			\bra{0;g_h}\hat{b}^{(h)}_{out}(t)\ket{1;g_h} & = \psi'^{(1)}(z_{h,1}=-t)\\
			\bra{0;g_v}\hat{b}^{(v)}_{out}(t)\ket{1;g_h} & = \psi'^{(2)}(z_{v,1}=-t)\\
			& t \geq 0
		\end{aligned}
	\end{equation}
	where the state $\ket{1;g_h}$ denotes the initial state at time $T=0$, with one photons in the virtual cavity and the atom in $\ket{g_h}$. To solve for these correlation functions, we can unravel the Lindbladian evolution into non-unitary evolution under a non-Hermitian Hamiltonian interleaved by a series of quantum jumps induced by $(\hat{b}_{out})$s, as in~\cite{rosenblum2011photon}. The effective non-Hermitian Hamiltonian is given by:
	
	\begin{equation}
		\begin{aligned}
			\hat{H}_{eff} &= -i\frac{|g(t)|^2}{2} \hat{a}^{\dagger}_{C} \hat{a}_{C} - 2i\gamma \hat{\sigma}_{ee}\\
			&-i\sqrt{2\gamma}g^{*}(t)\hat{a}_{C}\hat{\sigma}_{eh}
			\label{non hermitian single}
		\end{aligned}
	\end{equation}
	
	Starting with the state $\ket{1;g_h}$, the state at $T=t$ can be written as $\alpha_{0}(t)\ket{1;g_h}+\alpha_{1}(t)\ket{0;e}$, with $\alpha_{0,1}$ satisfying the following differential equations:
	\begin{equation}
		\begin{aligned}
			i\frac{d}{dT}\alpha_{0}(T) &= \left( -i  \frac{|g(T)|^2}{2} \right)\,\alpha_{0}(T)\\
			i\frac{d}{dT}\alpha_{1}(T) &= \left(-i\sqrt{2\gamma}g^{*}(T)\right)\,\alpha_{0}(T)+\left(-2i\gamma \right)\,\alpha_{1}(T)\\
		\end{aligned}
	\end{equation}
	From the above equations, we obtain a closed-form solution for $\alpha_{0}$, given by:
	\begin{equation}
		\frac{\alpha_{0}(t)}{\alpha_{0}(0)}=\alpha_{0}(t) = e^{ - \int_{0}^{t} \frac{|g(T)|^2}{2} dT}
	\end{equation}
	We can now examine the integral $\int (-|g(T)|^2) dT$, and by plugging in the relation in~\ref{decay rate waveform relation}, we get:
	\begin{equation}
		\int_{0}^{t} (-|g(T)|^2) dT = \int \frac{-|u(T)|^2}{1-\int_{0}^{T}|u(T')|^2 dT'} dT
	\end{equation}
	But note that:
	\begin{equation}
		\frac{d}{dT}\ln(1-\int_{0}^{T}|u(T')|^2 dT'|)=-\frac{|u(T)|^2}{1-\int_{0}^{T}|u(T')|^2 dT'}
	\end{equation}
	with $\ln(*)=\log_{e}(*)$ being the natural logarithm function. Then we have:
	\begin{equation}
		\int_{t_1}^{t_2} \frac{-|u(T)|^2}{1-\int_{0}^{T}|u(T')|^2 dT'} dT  = \ln( 1-\int_{0}^{t_2}|u(T')|^2 dT' ) - \ln(1-\int_{0}^{t_1}|u(T')|^2 dT'|)
		= \ln\left(\frac{G(t_2)}{G(t_1)}\right)
	\end{equation}
	where we have defined
	\begin{equation}
		G(t) = 1-\int_{0}^{t}|u(T')|^2 dT' = \int_{t}^{\infty}|u(T')|^2 dT'
		\label{eq: G define}
	\end{equation}
	for simplicity.
	Plugging in the above equations, we have:
	\begin{equation}
		\alpha_{0}(t) = (G(t))^{\frac{1}{2}}
	\end{equation}
	As for $\alpha_{1}$, we can write the differential equation as:
	\begin{equation}
		\frac{d}{d T}\alpha_{1}(T)+ 2\gamma \alpha_{1}(T)= (-\sqrt{2\gamma}g^{*}(T))\alpha_{0}(T)
	\end{equation}
	so we have:
	\begin{equation}
		\frac{d}{d T}(e^{2\gamma T} \alpha_{1}(T))= (-\sqrt{2\gamma}g^{*}(T))\alpha_{0}(T)e^{2\gamma T}
	\end{equation}
	By plugging in equation~\ref{decay rate waveform relation} and $\alpha_{0}(T)$ and integrate on both sides from $0$ to $t$, we get:
	\begin{equation}
		\alpha_{1}(t) = -\frac{1}{\sqrt{2\gamma}}\int_{0}^{t}2\gamma e^{-2\gamma(t-T)}u(T)dT 
	\end{equation}
	Therefore, by applying $\hat{b}^{(h)}_{out}=\sqrt{2\gamma}\hat{\sigma}_{he}+g^{*}(t)\hat{a}_{C}$ to the state $\alpha_{0}(t)\ket{1;g_h}+\alpha_{1}(t)\ket{0;e}$, we get:
	\begin{equation}
		\begin{aligned}
			&\hat{b}^{(h)}_{out}(\alpha_{0}(t)\ket{1;g_h}+\alpha_{1}(t)\ket{0;e})\\
			&=(g^{*}(t)\alpha_{0}(t) + \sqrt{2\gamma}\alpha_{1}(t))\ket{0;g_h}\\
			&= (u(t) - \int_{0}^{t}2\gamma e^{-2\gamma(t-T)}u(T)dT) \ket{0;g_h}
		\end{aligned}
	\end{equation}
	Therefore, we can obtain $\psi'^{(1)}(z_{h,1})$ by plugging in the correlation-wavefunction correspondence relation at the beginning of this subsection. We can also find $p_{fail}$, which is now:
	\begin{equation}
		p_{fail} = \int_{0}^{\infty}|u(t) - \int_{0}^{t}2\gamma e^{-2\gamma(t-T)}u(T)dT)|^{2} dt
	\end{equation}
	
	We will now prove that, for any target $0<\epsilon<\frac{1}{4}$, there exist some sufficiently large $\gamma$ such that $p_{fail}<4\epsilon$. For all the examples, we consider smooth, positive and real function $u(t)$. For simplicity, we set the size of the time bin to be $1$, and thus $u(t)\geq 0$ only for $t\in[0,1]$ and is $0$ otherwise. Due to these properties of $u(t)$, we know that $u(t)$ is bounded on $[0,1]$, i.e. $|u(t)|\leq M$ for some fixed $M$. Also, due to its smoothness, the function $u(t)$ is a Lipschitz continuous function. That is, there exist a constant $\eta$ such that for any $t_1 , t_2 \in [0,1]$ with $|t_1-t_2|<\eta \epsilon$, we have $|u(t_1)-u(t_2)|<\epsilon$.
	
	We start by estimating the integral on $[1,\infty]$. For $t>1$, $u(t)=0$. Therefore:
	\begin{equation}
		\begin{aligned}
			&|u(t) - \int_{0}^{t}2\gamma e^{-2\gamma(t-T)}u(T)dT|^{2} \\
			&=|\int_{0}^{t}2\gamma e^{-2\gamma(t-T)}u(T)dT|^{2} \\
			&=|\int_{0}^{1}2\gamma e^{-2\gamma(t-T)}u(T)dT|^{2} \\
			&\leq M^2 |\int_{0}^{1}2\gamma e^{-2\gamma(t-T)}dT|^{2} = M^2 (e^{2\gamma}-1)^2 e^{-4\gamma t}\\
			&\leq M^2 e^{4\gamma}e^{-4\gamma t}
		\end{aligned}
	\end{equation}
	where we use the bounded property of $u(t)$ for the first inequality. Hence:
	\begin{equation}
		\begin{aligned}
			&\int_{1}^{\infty}|u(t) - \int_{0}^{t}2\gamma e^{-2\gamma(t-T)}u(T)dT|^{2} dt\\
			&\leq \int_{1}^{\infty} M^2 e^{4\gamma}e^{-4\gamma t} dt = \frac{M^2}{4\gamma}\\
		\end{aligned}
	\end{equation}
	So taking $\gamma>\frac{M^2}{4\epsilon}$ renders $\int_{1}^{\infty}|u(t) - \int_{0}^{t}2\gamma e^{-2\gamma(t-T)}u(T)dT|^{2} dt<\epsilon$.
	
	Next, we consider the same integral on $[0,1]$. For simplicity, we introduce the following definition:
	\begin{equation}
		\begin{aligned}
			\tilde{u}(t) &= \int_{0}^{t}2\gamma e^{-2\gamma(t-T)}u(T)dT\\
			u^{(0)}(t) &= u(t)\int_{0}^{t}2\gamma e^{-2\gamma(t-T)}dT=u(t)(1-e^{-2\gamma t})
			\label{eq: u_tilde_u_0_define}
		\end{aligned}
	\end{equation}
	
	We now examine $|u(t)-\tilde{u}(t)|$ on the interval $[0,1]$. We first have:
	\begin{equation}
		|u(t)-\tilde{u}(t)|\leq|u(t)-u^{(0)}(t)|+|u^{(0)}(t)-\tilde{u}(t)|=u(t)e^{-2\gamma t} + |u^{(0)}(t)-\tilde{u}(t)|
	\end{equation}
	Notably, if we managed to show that the second term $|u^{(0)}(t)-\tilde{u}(t)|\leq 2\epsilon$ for any $t\in[0,1]$ for some sufficiently large $\gamma$, we would have:
	\begin{equation}
		\begin{aligned}
			\int_{0}^{1}|u(t)-\tilde{u}(t)|^2dt &\leq \int_{0}^{1} dt \left(4\epsilon^{2} + 4\epsilon u(t)e^{-2\gamma t} + |u(t)|^2e^{-4\gamma t}\right)\\
			&\leq 4\epsilon^{2} + 4\epsilon \frac{M}{2\gamma} + \frac{M^2}{4\gamma}
		\end{aligned}
	\end{equation}
	Hence, since we have assume $\epsilon<\frac{1}{4}$, by taking $\gamma>max\{2M,\frac{M^2}{4\epsilon}\}$, the above integration is less than $3\epsilon$. Combined with the integration on $[1,\infty]$, which is less than $\epsilon$, we obtain $p_{fail}<4\epsilon$ as desired. 
	
	We now prove $|u^{(0)}(t)-\tilde{u}(t)|\leq 2\epsilon$ for any $t\in[0,1]$ for some large enough $\gamma$. We have:
	\begin{equation}
		\begin{aligned}
			|u^{(0)}(t)-\tilde{u}(t)| &= |\int_{0}^{t}(u(t)-u(T))2\gamma e^{-2\gamma(t-T)} dT|\\
			&\leq \int_{0}^{t}|u(t)-u(T)|2\gamma e^{-2\gamma(t-T)} dT\\
		\end{aligned}
	\end{equation}
	Since $u(t)$ is Lipschitz continuous, we can find $\delta=\eta\epsilon$ such that for $T\in [t-\delta,t]$, $|u(t)-u(T)|<\epsilon$. Then, we have:
	\begin{equation}
		\begin{aligned}
			\int_{0}^{t}|u(t)-u(T)|2\gamma e^{-2\gamma(t-T)} dT &= (\int_{0}^{t-\delta}+\int_{t-\delta}^{t})|u(t)-u(T)|2\gamma e^{-2\gamma(t-T)} dT\\
			&\leq \int_{0}^{t-\delta}|u(t)-u(T)|2\gamma e^{-2\gamma(t-T)} dT + \epsilon\\
			&\leq M \int_{0}^{t-\delta}2\gamma e^{-2\gamma(t-T)} dT + \epsilon\\
			&=M(e^{-2\gamma \delta}-e^{-2\gamma t})+\epsilon \leq Me^{-2\gamma \delta} + \epsilon
		\end{aligned}
	\end{equation}
	where the second inequality again uses the boundedness of $u(t)$. If $t-\delta<0$ already, then the first term can be dumped and already the above is less than $\epsilon$.
	Finally, in order to make $Me^{-2\gamma \delta}<\epsilon$, we just need to take $\gamma>\frac{1}{2\eta \epsilon}ln(\frac{M}{\epsilon})$. Note that the parameters $\eta$ is the Lipschitz constant associate with the function $u(t)$ and does \textbf{\textit{not}} depend on $\epsilon$ or $\gamma$. To sum up, we need to choose 
	$\gamma>max\{\frac{1}{2\eta \epsilon}ln(\frac{M}{\epsilon}),2M,\frac{M^2}{4\epsilon}\}$ so that $p_{fail}<4\epsilon$. Numerically, we also plugged in different values of $\gamma$ and choose different temporal profiles $u(t)$ to see the vanishing $p_{fail}$, as shown in Figure~\ref{Fig:infidelity numeric single subtract}(a)(c).
	

	\subsubsection{Two photon input state}
	Here we show how to find $\psi'^{(1)}$ and $p_{fail}$ for photon number $k=2$. Already, the exact analytic expression for $p_{fail}$ becomes complicated, which motivates the need to find an upper bound for $p_{fail}$ whose analytic expression is much simpler. This upper bound will be presented in the next part.
	
	To obtain $\psi'^{(1)}(z_{h,1},z_{h,2})$, we make use of the following correlation-wavefunction correspondence relation:
	\begin{equation}
		\begin{aligned}
			\bra{0;g_h}\hat{b}^{(h)}_{out}(t+\tau)\hat{b}^{(h)}_{out}(t)\ket{2;g_h} & = 2!\psi'^{(1)}(z_{h,1}=-t;z_{h,2}=-t-\tau)\\
			t+\tau &\geq t \geq 0
		\end{aligned}
	\end{equation}
	Note that from boson symmetry, we only need to find the wavefunction $\psi'^{(1)}$ on the domain $z_{h,2}\leq z_{h,1}$, so the above time-ordered correlation function is enough to identify $\psi'^{(1)}$. For the above equation $\ket{2;g_h}$ denotes the initial state with two photons in the virtual cavity and the atom in $\ket{g_h}$. The initial state then undergoes two rounds of ``evolve and collapse" process, before it ends up in a state proportional to $\ket{0;g_h}$. Here, each collapse is induced by $\hat{b}^{(h)}_{out}=g^{*}(t)\hat{a}_{C}+\sqrt{2\gamma}\hat{\sigma}_{he}$.
	
	For the first evolve process from time $T=0$ to $T=t$, the initial state $\ket{2;g_h}$ evolves into $\alpha_{0}(t)\ket{2;g_h}+\alpha_{1}(t)\ket{1;e}$ under the effective Hamiltonian~\ref{non hermitian single}, where $\alpha_{0},\alpha_{1}$ satisfies:
	\begin{equation}
		\begin{aligned}
			i\frac{d}{dT}\alpha_{0}(T) &= \left( -i 2 \frac{|g(T)|^2}{2} \right)\,\alpha_{0}(T)\\
			i\frac{d}{dT}\alpha_{1}(T) &= \left(-i\sqrt{2}\sqrt{2\gamma}g^{*}(T)\right)\,\alpha_{0}(T)+\left(-i \frac{|g(T)|^2}{2}-2i\gamma \right)\,\alpha_{1}(T)\\
		\end{aligned}
	\end{equation}
	
	From which we find:
	\begin{equation}
		\begin{aligned}
			\alpha_{0}(t) &= G(t)\\
			\alpha_{1}(t) &= (G(t))^{\frac{1}{2}}\int_{0}^{t}(-\sqrt{2}\sqrt{2\gamma}u(t')e^{-2\gamma (t-t')})dt'
		\end{aligned}
	\end{equation}
	using the same trick as in the single photon case. For convenience, we define:
	\begin{equation}
		\tilde{u}(t_2,t_1) = \int^{t_2}_{t_1}2\gamma u(T)e^{-2\gamma(t_2-T)} dT
		\label{eq: general u tilde define}
	\end{equation}
	So that we can write $\alpha_{1}(t) = (G(t))^{\frac{1}{2}}  \frac{(-\sqrt{2})}{\sqrt{2\gamma}}\tilde{u}(t,0)$.
	After applying $\hat{b}^{(h)}_{out}$ at time $t$, the state $\alpha_{0}(t)\ket{2;g_h}+\alpha_{1}(t)\ket{1;e}$ collapse to:
	\begin{equation}
		\begin{aligned}
			&\hat{b}^{(h)}_{out}(\alpha_{0}(t)\ket{2;g_h}+\alpha_{1}(t)\ket{1;e})\\
			&=(\sqrt{2\gamma}\alpha_{1}(t)+\sqrt{2}g^{*}(t)\alpha_{0}(t))\ket{1;g_h} + g^{*}(t)\alpha_{1}(t)\ket{0;e}\\
			&=(G(t)^{\frac{1}{2}}\sqrt{2}(u(t)-\tilde{u}(t,0))\ket{1;g_h} + \frac{(-\sqrt{2})}{\sqrt{2\gamma}}u(t)\tilde{u}(t,0)\ket{0;e}
		\end{aligned}
	\end{equation}
	
	For the state $\ket{1;g_h}$, within the time interval $t\leq T\leq t+\tau$, it evolves into $\beta_{0}(t+\tau,t)\ket{1;g_h}+\beta_{1}(t+\tau,t)\ket{0;e}$. We can write down the differential equations for $\beta_{0},\beta_{1}$ and solve for them. The differential equations are:
	\begin{equation}
		\begin{aligned}
			i\frac{d}{dT}\beta_{0}(T,t) &= \left( -i  \frac{|g(T)|^2}{2} \right)\,\beta_{0}(T,t)\\
			i\frac{d}{dT}\beta_{1}(T,t) &= \left(-i\sqrt{2\gamma}g^{*}(T)\right)\,\beta_{0}(T,t)+\left(-2i\gamma \right)\,\beta_{1}(T,t)\\
		\end{aligned}
	\end{equation}
	from which we find
	\begin{equation}
		\begin{aligned}
			\beta_{0}(t+\tau,t) &= \left( \frac{G(t+\tau)}{G(t)} \right)^{\frac{1}{2}}\\
			\beta_{1}(t+\tau,t) &= -\frac{1}{\sqrt{2\gamma}}\tilde{u}(t+\tau,t)\left(\frac{1}{G(t)}\right)^{\frac{1}{2}}
		\end{aligned}
	\end{equation}
	For the state $\ket{0;e}$, within the time interval $t\leq T\leq t+\tau$ it just evolves into $e^{-2\gamma \tau}\ket{0;e}$ under the Hamiltonian in equation~\ref{non hermitian single}.
	Therefore, after the second ``evolve and collapse" process, we end up with the state:
	\begin{equation}
		\begin{aligned}
			\hat{b}^{(h)}_{out}[(G(t)^{\frac{1}{2}}\sqrt{2}(u(t)-\tilde{u}(t,0))(\beta_{0}(t+\tau,t)\ket{1;g_h}+\beta_{1}(t+\tau,t)\ket{0;e}) + \frac{(-\sqrt{2})}{\sqrt{2\gamma}}u(t)\tilde{u}(t,0)e^{-2\gamma \tau}\ket{0;e}]
		\end{aligned}
	\end{equation}
	After cleaning up, the above state is:
	\begin{equation}
		\begin{aligned}
			&[\sqrt{2}(u(t)-\tilde{u}(t,0))u(t+\tau)+\sqrt{2}(u(t)-\tilde{u}(t,0))(-\tilde{u}(t+\tau,t))+(-\sqrt{2}u(t)\tilde{u}(t,0)e^{-2\gamma \tau})]\ket{0;g_h}
		\end{aligned}
	\end{equation}
	Thus:
	\begin{equation}
		\begin{aligned}
			&\bra{0;g_h}\hat{b}^{(h)}_{out}(t+\tau)\hat{b}^{(h)}_{out}(t)\ket{2;g_h}\\
			&=[\sqrt{2}(u(t)-\tilde{u}(t,0))u(t+\tau)\\
			&+\sqrt{2}(u(t)-\tilde{u}(t,0))(-\tilde{u}(t+\tau,t))\\
			&+(-\sqrt{2}u(t)\tilde{u}(t,0)e^{-2\gamma \tau})]
		\end{aligned}
	\end{equation}
	and we can verify that $p_{fail}=\int_{0}^{\infty}dt\int_{0}^{\infty}d\tau |\bra{0;g_h}\hat{b}^{(h)}_{out}(t+\tau)\hat{b}^{(h)}_{out}(t)\ket{2;g_h}|^2$ from the correlation-wavefunction correspondence. Using this exact expression for $p_{fail}$, it is hard to prove rigorously that $p_{fail}$ can be made arbitrarily small given large enough $\gamma$, as we did for the single photon case. Therefore, below we would give an upper bound for $p_{fail}$ for any photon number $k$ that has a much simpler expression, and show that this upper bound can be made arbitrarily small for large enough $\gamma$. Nevertheless, we may still numerically find the actual value of $p_{fail}$ for $k=2$ based on its exact expression.

	\subsubsection{Generalization to arbitrary photon number}
	In order to estimate $p_{fail}$ For general $k$, we focus on $p_{succ}$ instead. This probability is defined in~\ref{eq: p_fail_p_succ_def}. Essentially, we can write $\ket{\psi'} = \ket{\psi'^{(1)}} + \ket{\psi'^{(2)}}$, where:
	\begin{equation}
		\begin{aligned}
			\ket{\psi'^{(1)}} = &\int_{-\infty}^{\infty}dz_{h,1}\cdots dz_{h,k} \psi'^{(1)}(z_{h,1},\cdots,z_{h,k})\hat{a}_{h,z_{h,1}}^{\dagger}\cdots\hat{a}_{h,z_{h,k}}^{\dagger}\ket{vac,g_h}\\
			\ket{\psi'^{(2)}}=&\int_{-\infty}^{\infty}dz_{v}dz_{h,1}\cdots dz_{h,k-1} \psi'^{(2)}(z_{v};z_{h,1},\cdots,z_{h,k-1})\hat{a}_{v,z_{v}}^{\dagger}\hat{a}_{h,z_{h,1}}^{\dagger}\cdots\hat{a}_{h,z_{h,k-1}}^{\dagger}\ket{vac,g_v}
		\end{aligned}
	\end{equation}
	Now we have $p_{succ}=\bra{\psi'^{(2)}}\ket{\psi'^{(2)}}$ and $p_{fail}=\bra{\psi'^{(1)}}\ket{\psi'^{(1)}}$. To estimate $p_{succ}$, we make use of the Cauchy-Schwartz inequality. In particular, we define $\ket{\psi'_{l.p.}} = \ket{\psi'^{(1)}_{l.p.}} + \ket{\psi'^{(2)}_{l.p.}}$, where:
	\begin{equation}
		\begin{aligned}
			&\psi'^{(1)}_{l.p.}(z_{h,1}\cdots,z_{h,k}) \equiv 0 \\
			\psi'^{(2)}_{l.p.}(z_{v};z_{h,1}\cdots,z_{h,k-1})  &= -\sqrt{\frac{k}{(k-1)!}}\psi(z_{v})\psi(z_{h,1})\cdots\psi(z_{h,k-1})\mathbf{1}_{z_{v} \geq max(z_{h,i})}
		\end{aligned}
	\end{equation}
	That is, the wavefunction corresponding to the component $\ket{\psi'^{(2)}_{l.p.}}$ can only takes non-zero value on the domain $z_{v,1} \geq max(z_{h,i})$. Then, using Cauchy-Schwartz inequality, $|\bra{\psi'^{(2)}}\ket{\psi'^{(2)}_{l.p.}}|^{2}\leq \bra{\psi'^{(2)}}\ket{\psi'^{(2)}}\bra{\psi'^{(2)}_{l.p.}}\ket{\psi'^{(2)}_{l.p.}}$. It is easy to verify that $\bra{\psi'^{(2)}_{l.p.}}\ket{\psi'^{(2)}_{l.p.}}=1$, so we have $p_{succ}\geq|\bra{\psi'^{(2)}}\ket{\psi'^{(2)}_{l.p.}}|^{2}$. We would then define the ``subtraction fidelity" $F_{sub}$ as:
	\begin{equation}
		F_{sub}=|\bra{\psi'^{(2)}}\ket{\psi'^{(2)}_{l.p.}}|^{2}
	\end{equation}
	which is distinguished from the gate fidelity $F_{gate}$. Note that to find $F_{sub}$, we only need to find the wavefunction $\psi'^{(2)}(z_{v};z_{h,1},\cdots,z_{h,k-1})$ on the domain $z_{v} \geq max(z_{h,i})$, since $\psi'^{(2)}_{l.p.}$ is $0$ otherwise. To obtain this wavefunction, we use the same virtual cavity method as above, and use the correlation-wavefunction correspondence given by:
	\begin{equation}
		\begin{aligned}
			&\bra{0;g_v}\hat{b}^{(h)}_{out}(t_{k-1})\cdots\hat{b}^{(h)}_{out}(t_2)\hat{b}^{(h)}_{out}(t_1)\hat{b}^{(v)}_{out}(t)\ket{k;g_h}\\
			&=(k-1)!\psi'^{(2)}(z_{v}=-t;z_{h,1}=-t_1,\cdots,z_{h,k-1}=-t_{k-1})\\
			& \ \ \ \ \ \ \ \ \ \ \ \ \ \ \ \ \ \ (t\leq t_1 \leq t_2 \leq \cdots \leq t_{k-1})
		\end{aligned}
	\end{equation}
	where the above correlation function can be obtained after $k$ rounds of ``evolve and collapse" process, which is examined below. Starting with initial state $\ket{k;g_h}$ at time $T=0$, and after evolving for time $t$ under the effective Hamiltonian $H_{eff}$, the state evolves into $\alpha_{0}(t)\ket{k;g_h}+\alpha_{1}(t)\ket{k-1;e}$, where $\alpha_{0},\alpha_{1}$ satisfies:
	
	\begin{equation}
		\begin{aligned}
			i\frac{d}{dT}\alpha_{0}(T) &= \left( -i k \frac{|g(T)|^2}{2} \right)\,\alpha_{0}(T)\\
			i\frac{d}{dT}\alpha_{1}(T) &= \left(-i\sqrt{k}\sqrt{2\gamma}g^{*}(T)\right)\,\alpha_{0}(T)+\left(-i (k-1)\frac{|g(T)|^2}{2}-2i\gamma \right)\,\alpha_{1}(T)\\
		\end{aligned}
	\end{equation}
	
	From the above equations, we obtain a closed-form solution for $\alpha_{0}$, given by:
	\begin{equation}
		\alpha_{0}(t)  = e^{-\frac{k}{2}(-ln(1-\int_{0}^{t}|u(t')|^2 dt'|))}
		= (G(t))^{\frac{k}{2}}
	\end{equation}
	
	For $\alpha_{1}(t)$, this equation can be written as:
	
	\begin{equation}
		\frac{d}{d t}\alpha_{1}(t)+((k-1)\frac{|g(t)|^2}{2} + 2\gamma)\alpha_{1}(t)= (-\sqrt{k}\sqrt{2\gamma}g^{*}(t))\alpha_{0}(t)
	\end{equation}
	
	Therefore, by plugging in $e^{\int_{0}^{t}dt'((k-1)\frac{|g(t')|^2}{2} + 2\gamma)}=e^{2\gamma t}(G(t))^{-\frac{k-1}{2}}$ on both sides, we get:
	
	\begin{equation}
		\begin{aligned}
			\frac{d}{d t}[\alpha_{1}(t)e^{\int_{0}^{t}dt'((k-1)\frac{|g(t')|^2}{2} + 2\gamma)}] &= -\sqrt{k}\sqrt{2\gamma}g^{*}(t)\alpha_{0}(t)e^{\int_{0}^{t}dt'((k-1)\frac{|g(t')|^2}{2} + 2\gamma)}\\
			&= -\sqrt{k}\sqrt{2\gamma}u(t)e^{2\gamma t}
		\end{aligned}
	\end{equation}
	
	Hence, we integrate from $0$ to $t$ on both sides:
	
	\begin{equation}
		\alpha_{1}(t)e^{2\gamma t}(G(t))^{-\frac{k-1}{2}} =\int_{0}^{t}(-\sqrt{k}\sqrt{2\gamma}u(t')e^{2\gamma t'})dt'
	\end{equation}
	
	giving:
	
	\begin{equation}
		\alpha_{1}(t) = (G(t))^{\frac{k-1}{2}}\int_{0}^{t}(-\sqrt{k}\sqrt{2\gamma}u(t')e^{-2\gamma (t-t')})dt'
		\label{single_no_approximation}
	\end{equation}
	
	
	
	
	We now consider the effect of the quantum jump, induced by applying the collapse operator $\hat{b}^{(v)}_{out}$ at time $t$ to the state 
	$\alpha_{0}(t)\ket{k;g_h}+\alpha_{1}(t)\ket{k-1;e}$. This gives $\sqrt{2\gamma}\alpha_{1}(t)\ket{k-1;g_v}$. This state then enters the next round of ``evolve and collapse" process. One can straightforwardly verify that after evolving for time $\tau_{1}$, we end up with the state $\sqrt{2\gamma}\alpha_{1}(t)\sigma_{1}(\tau_{1})\ket{k-1;g_v}$, where $\sigma_{1}(\tau_{1})$ satisfies:
	
	\begin{equation}
		\begin{aligned}
			\frac{d}{dT}\sigma_{1}(T) &= (- (k-1) \frac{|g(T)|^2}{2} ) \sigma_{1}(T)\\
			t & \leq T \leq t+\tau_1
		\end{aligned}
	\end{equation}
	
	Therefore, $\sigma_{1}(\tau_{1}) = (\frac{G(t+\tau_1)}{G(t)})^{\frac{k-1}{2}}$. Applying the collapse operator $\hat{b}^{(h)}_{out}=g^{*}(t)\hat{a}_{C}+\sqrt{2\gamma}\hat{\sigma}_{he}$ at time $t+\tau_1$ thus gives the state $\sqrt{2\gamma}\alpha_{1}(t)\sqrt{k-1}\frac{u(t+\tau_1)}{\sqrt{G(t+\tau_1)}}\sigma_{1}(\tau_{1})\ket{k-2;g_v}$. Plugging in the solution for $\alpha_{1}$ and $\sigma_{1}$, this state is $\sqrt{\frac{k!}{(k-2)!}}G(t+\tau_1)^{\frac{k-2}{2}}u(t)u(t+\tau_1)\ket{k-2;g_v}$. Continuing this ``evolve and collapse" process, we finally get:
	
	
	\begin{equation}
		\begin{aligned}
			\bra{0;g_v}&\hat{b}^{(h)}_{out}(t+\tau_1+\cdots+\tau_{k-1})\cdots\hat{b}^{(h)}_{out}(t+\tau_1+\tau_2)\hat{b}^{(h)}_{out}(t+\tau_1)\hat{b}^{(v)}_{out}(t)\ket{init}\\
			&=-\sqrt{k!}\,\tilde{u}(t)u(t_1)u(t_2)\cdots u(t_{k-1})\\
			&(t_1=t+\tau_1,\cdots,t_{k-1}=t+\tau_1+\cdots+\tau_{k-1})
		\end{aligned}
	\end{equation}
	where $\tilde{u}(t)$ is defined in equation~\ref{eq: u_tilde_u_0_define}.
	
	By plugging in the above expression, one can verify that $F_{sub}$ is given by:
	\begin{equation}
		\begin{aligned}
			F_{sub} & = |k!\int_{0}^{\infty}\tilde{u}(t)u^{*}(t)\int_{t}^{\infty}dt_1|u(t_1)|^2\int_{t_1}^{\infty}dt_2|u(t_2)|^2\cdots\int_{t_{k-2}}^{\infty}dt_{k-1}|u(t_{k-1})|^2 |^2    \\
			& = |k\int_{0}^{\infty}dt  \tilde{u}(t)u^{*}(t) [G(t)]^{k-1}|^2   = |k\int_{0}^{1}dt  \tilde{u}(t)u^{*}(t) [G(t)]^{k-1}|^2 \\
		\end{aligned}
	\end{equation}
	where we adopt the assumption that $u(t)$ is only supported on the interval $[0,1]$, and define $G(t)$ as in equation~\ref{eq: G define}. Here $u^{*}(t)$ is the Hermitian conjugate of $u(t)$, and since we have taken $u(t)$ to be a positive, smooth and real, $u^{*}(t)=u(t)$. Since $p_{succ}\geq F_{sub}$ and $p_{succ}+p_{fail}=1$, we have $p_{fail}\leq(1-F_{sub})$. Thus, we just need to show that $F_{sub}$ can approach arbitrarily close to $1$ for sufficiently large $\gamma$.
	
	Consider the integration $|k\int_{0}^{1} dt \ \tilde{u}(t)u^{*}(t) [G(t)]^{k-1}|$. We have:
	\begin{equation}
		\begin{aligned}
			&1 - |k\int_{0}^{1} dt \ \tilde{u}(t)u(t) [G(t)]^{k-1}|\\
			&\leq |1 - k\int_{0}^{1} dt \ \tilde{u}(t)u(t) [G(t)]^{k-1}| = |k\int_{0}^{1} dt (u(t) - \tilde{u}(t)) u(t) [G(t)]^{k-1}|\\
			&\leq k\int_{0}^{1} dt |u(t) - \tilde{u}(t)| u(t) [G(t)]^{k-1}\\
			&\leq k\int_{0}^{1} dt |u(t) - u^{(0)}(t)| u(t) [G(t)]^{k-1} + k\int_{0}^{1} dt |u^{(0)}(t) - \tilde{u}(t)| u(t) [G(t)]^{k-1}\\
		\end{aligned}
	\end{equation}
	where $u^{(0)}(t)$ and $\tilde{u}(t)$ are defined in~\ref{eq: u_tilde_u_0_define}. For the first term, we have:
	\begin{equation}
		\begin{aligned}
			&k\int_{0}^{1} dt |u(t) - u^{(0)}(t)| u(t) [G(t)]^{k-1} \\
			&=k\int_{0}^{1} dt e^{-2\gamma t}|u(t)|^2  [G(t)]^{k-1}\leq \frac{kM^2}{2\gamma}
		\end{aligned}
	\end{equation}
	where $M$ is the upper bound for $u(t)$. Then, the first term is less than $\epsilon Mk$ if $\gamma>\frac{M}{2\epsilon}$. For the second term, note that in the single photon analysis, we have already shown that for arbitrarily small $\epsilon$, taking $\gamma>\frac{1}{2\eta \epsilon}ln(\frac{M}{\epsilon})$ yields $|u^{(0)}(t) - \tilde{u}(t)|<2\epsilon$. So we have:
	\begin{equation}
		\begin{aligned}
			&k\int_{0}^{1} dt |u^{(0)}(t) - \tilde{u}(t)| u(t) [G(t)]^{k-1}\leq 2\epsilon Mk
		\end{aligned}
	\end{equation}
	Therefore, for $\gamma>max\{\frac{1}{2\eta \epsilon}ln(\frac{M}{\epsilon}) , \frac{M}{2\epsilon}\}$: 
	\begin{equation}
		\begin{aligned}
			1 - |k\int_{0}^{1} dt \ \tilde{u}(t)u(t) [G(t)]^{k-1}| \leq 3\epsilon kM 
		\end{aligned}
	\end{equation}
	for arbitrarily small $\epsilon$. Therefore, $p_{fail}=(1-F_{sub})$ can also me made arbitrarily small, since:
	\begin{equation}
		p_{fail}\leq1 - F_{sub} = 1 - |k\int_{0}^{1} dt \ \tilde{u}(t)u(t) [G(t)]^{k-1}|^2
	\end{equation}
	Importantly, the above upper bound grows linearly with the input photon number $k$ when $\gamma$ is held fixed. We can appreciate this fact in the following subsection, where we numerically examine $(1-F_{sub})$ for different $u(t)$, $\gamma$ and $k$. The above bound also suggest that, for some target $\epsilon'$ and input photon number $k$, we have $\gamma_{\epsilon',k}=\Theta(\frac{k}{\epsilon'}ln(\frac{k}{\epsilon'}))$ for the required $\gamma_{\epsilon',k}$.

	\subsection{III. Numerical results: Single subtraction case}
	
	\begin{figure*}[h]
		\centering
		\includegraphics[width=0.95\textwidth]{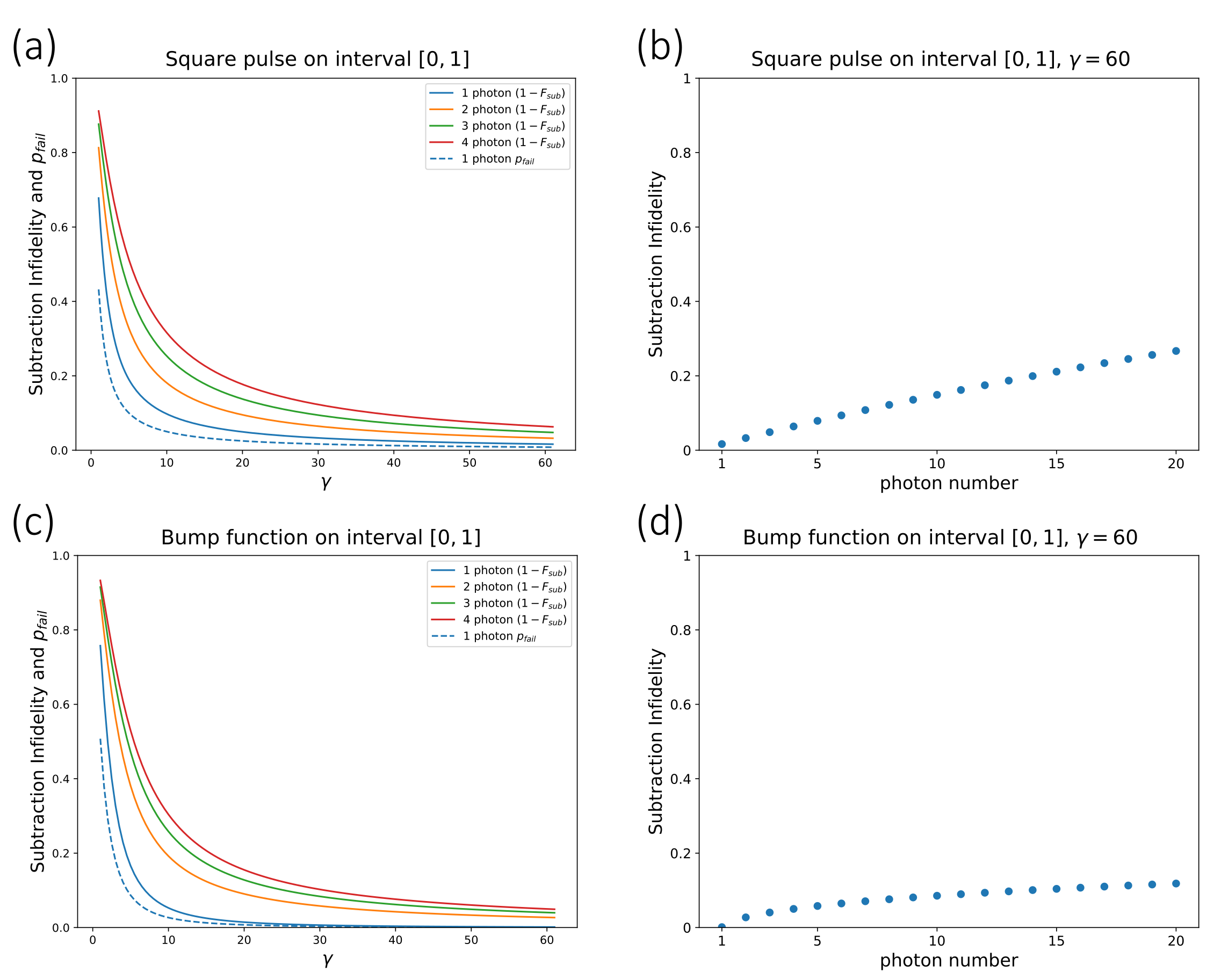}
		\caption{Subtraction infidelity $1-F_{sub}$ for different $u(t)$, $\gamma$ and input photon number $k$, which serves as an upper bound for $p_{fail}$. (a) Infidelity for $u(t)$ being a normalized square pulse on interval $[0,1]$, plotted against $\gamma$. (b) Infidelity plotted as a function of the input photon number $k$. We take the same pulse shape as in (a), but with fixed $\gamma=60$. (c-d) Infidelity plotted against $\gamma$ or photon number for $u(t)$ being a bump function on interval $[0,1]$.}
		\label{Fig:infidelity numeric single subtract}
	\end{figure*}
	In this part we focus primarily on the subtraction infidelity $(1-F_{sub})$, which serves as an upper bound for $p_{fail}$. The subtraction infidelity would depend on the temporal profile of the input photon pulse $u(t)$, together with the input photon number $k$ and atom decay rate $\gamma$. In particular, we considered two different $u(t)$: (i) Normalized square pulse on the interval $[0,1]$ and (ii) Normalized bump function $bump(t) \propto e^{-\frac{0.5^2}{0.5^2-(t-0.5)^2}}$ on the interval $[0,1]$. Equivalently, $\gamma$ is measured in units of $\frac{1}{L}$ where $L$ is the length of the input photon pulse (i.e. size of the time bin). 
	In Figure~\ref{Fig:infidelity numeric single subtract}(a)(c), we plot the subtraction infidelity against $\gamma$, after fixing the input photon number $k$ to different values. For all $k$, the infidelity converges to $0$ as $\frac{1}{\gamma}\rightarrow 0$ for both the square pulse (a) and bump function (c). We numerically observe that for small $\frac{1}{\gamma}$, the scaling of the subtraction infidelity is $O(\frac{C(k)}{\gamma})$ where $C(k)$ is a function that only depend on the photon number. The function $C(k)$ can then be visualized in Figure~\ref{Fig:infidelity numeric single subtract}(b)(d), where we fix $\gamma$ and plot the subtraction infidelity against the input photon number $k$.
	Moreover, for the square pulse, we can plug in the formula for $F_{sub}$ to get a simple solution for the subtraction infidelity. In particular, for photon number $k=1,2$, we have:
	\begin{equation}
		\begin{aligned}
			1-F_{sub}(k=1,\gamma ) &= \frac{1}{\gamma }\left( 1-e^{-2\gamma } \right) \left( 1- \frac{1-e^{-2\gamma }}{4\gamma }\right)\\
			1-F_{sub}(k=2,\gamma ) &= \frac{2}{\gamma }\left(1 -\frac{1 - e^{-2\gamma }}{2\gamma } \right) \left( 1 - \frac{1}{2\gamma } + \frac{1-e^{-2\gamma }}{4\gamma^2 }\right)
		\end{aligned}
	\end{equation}
	And thus by plugging these equations in, we find that $\gamma  = 4000$ yields $(1-F_{sub})=0.00025$ for a single photon Fock input state, and $(1-F_{sub})=0.0005$ for a two-photon Fock input state, where the input photon states are encapsulated in a square pulse of length $1$. Notably, the subtraction infidelity $1-F_{sub}$ only serve as an upper bound for $p_{fail}$, and the gate infidelity $INF_{gate}$ is ultimately determined by $p_{fail}$ via $INF_{gate}=2 p_{fail}(1-p_{fail})(1-cos(\delta \phi))$, as shown in equation~\ref{eq: infidel subtract failure relation}. 
	Figure~\ref{Fig:infidelity numeric single subtract}(a)(c) compare the subtraction infidelity (solid line) with $p_{fail}$ (dashed line) for input photon number $k=1$, indicating that this upper bound for $p_{fail}$ might only be a coarse estimate. Indeed, for the same $\gamma=4000$ and for the square pulse, the actual value of $p_{fail}$ is $p_{fail}=0.00012$, for both single and two photon input state.
	Finally, as shown in Figure~\ref{Fig:infidelity numeric single subtract}(c-d), the bump function yields even smaller infidelity than a square pulse with the same pulse width, indicating that further optimization of the temporal profile may yield even better fidelity.

	\subsection{IV. Fidelity analysis: Two layer subtraction and beyond}
	The logic of computing and estimating the gate fidelity $F_{gate}$ for the architecture in Figure~\ref{Fig:phase gate}(c) follows closely to that of Figure~\ref{Fig:phase gate}(b). Here we have one ``h" waveguide and two ``v" waveguides, which we label ``$v_1$" and ``$v_2$". Then, after two layers of subtraction, the output state may generally be written as:
	\begin{equation}
		\begin{aligned}
			\ket{\psi'} &= \ket{\psi'^{(1)}} + \ket{\psi'^{(2)}} + \ket{\psi'^{(3)}} + \ket{\psi'^{(4)}}\\
			\ket{\psi'^{(1)}}&=\int_{-\infty}^{\infty}dz_{h,1}\cdots dz_{h,k} \psi'^{(1)}\hat{a}_{h,z_{h,1}}^{\dagger}\cdots\hat{a}_{h,z_{h,k}}^{\dagger}\ket{vac,g_h,g_h}\\
			\ket{\psi'^{(2)}}&=\int_{-\infty}^{\infty}dz_{v_1}dz_{h,1}\cdots dz_{h,k-1} \psi'^{(2)}\hat{a}_{v_1,z_{v_1}}^{\dagger}\hat{a}_{h,z_{h,1}}^{\dagger}\cdots\hat{a}_{h,z_{h,k}}^{\dagger}\ket{vac,g_v,g_h}\\
			\ket{\psi'^{(3)}}&=\int_{-\infty}^{\infty}dz_{v_2} dz_{h,1}\cdots dz_{h,k-1} \psi'^{(3)}\hat{a}_{v_2,z_{v_2}}^{\dagger}\cdots\hat{a}_{h,z_{h,1}}^{\dagger}\hat{a}_{h,z_{h,k-1}}^{\dagger}\ket{vac,g_h,g_v}\\
			\ket{\psi'^{(4)}}&=\int_{-\infty}^{\infty}dz_{v_1}dz_{v_2} dz_{h,1}\cdots dz_{h,k-2} \psi'^{(4)}\hat{a}_{v_1,z_{v_1}}^{\dagger}\hat{a}_{v_2,z_{v_2}}^{\dagger}\hat{a}_{h,z_{h,1}}^{\dagger}\cdots\hat{a}_{h,z_{h,k-2}}^{\dagger}\ket{vac,g_v,g_v}\\
		\end{aligned}
	\end{equation}
	In the above equations we have suppress the $z$ variables for the wavefunctions $\psi'^{(1,2,3,4)}$. Essentially, the state after two rounds of subtraction is a superposition of four components, with $\ket{\psi'^{(1)}}$ representing the case where no photon gets subtracted, $\ket{\psi'^{(2)}}$ representing the case where the first subtraction succeed, $\ket{\psi'^{(3)}}$ representing the case where the first subtraction fails but the second subtraction succeed and $\ket{\psi'^{(4)}}$ representing the case where both subtraction succeed. Defining $p_{i}=\bra{\psi'^{(i)}}\ket{\psi'^{(i)}}$, and following the same argument as in the single subtraction case, the gate fidelity is now given by:
	\begin{equation}
		F_{gate} = |p_1 e^{i[(\phi_1-\phi_3)+(\phi_2-\phi_3)]}+ p_2 e^{i(\phi_2-\phi_3)}+ p_3 e^{i(\phi_1-\phi_3)} + p_4|^2
	\end{equation}
	From which we obtain:
	\begin{equation}
		INF_{gate} = 1-F_{gate} \leq 4 p_{fail}(1-p_{fail})
	\end{equation}
	Here, we have defined $p_{fail}=p_1+p_2+p_3$ and $p_{succ}=p_4$, to be consistent with the single subtraction architecture in Figure~\ref{Fig:phase gate}(b). Then, to show that $INF_{gate}\rightarrow0$ when $\gamma\rightarrow\infty$, we can just prove that $p_{succ}\rightarrow1$.
	
	In order to find a lower bound for $p_{succ}=p_4=\bra{\psi'^{(4)}}\ket{\psi'^{(4)}}$, we use the same Cauchy-Schwarz technique as before. In particular, we define the wavefunction:
	\begin{equation}
		\begin{aligned}
			\ket{\psi'^{(4)}_{l.p.}}&=\int_{-\infty}^{\infty}dz_{v_1}dz_{v_2} dz_{h,1}\cdots dz_{h,k-2} \psi'^{(4)}_{l.p.}\hat{a}_{v_1,z_{v_1}}^{\dagger}\hat{a}_{v_2,z_{v_2}}^{\dagger}\hat{a}_{h,z_{h,1}}^{\dagger}\cdots\hat{a}_{h,z_{h,k-2}}^{\dagger}\ket{vac,g_v,g_v}\\
			\psi'^{(4)}_{l.p.}  &= \sqrt{\frac{k(k-1)}{(k-2)!}}\psi(z_{v_1})\psi(z_{v_2})\psi(z_{h,1})\cdots\psi(z_{h,k-2})\mathbf{1}_{z_{v_1} \geq z_{v_2} \geq max(z_{h,i})}
		\end{aligned}
	\end{equation}
	where $\psi(z)$ is the wavefunction of the input $k$-photon Fock state. Then, we have 
	$\bra{\psi'^{(4)}}\ket{\psi'^{(4)}}\bra{\psi'^{(4)}_{l.p.}}\ket{\psi'^{(4)}_{l.p.}}\geq|\bra{\psi'^{(4)}_{l.p.}}\ket{\psi'^{(4)}}|^2$, and since $\bra{\psi'^{(4)}_{l.p.}}\ket{\psi'^{(4)}_{l.p.}}=1$, we get $p_{succ}\geq|\bra{\psi'^{(4)}_{l.p.}}\ket{\psi'^{(4)}}|^2$. We can therefore define the subtraction fidelity $F_{sub}$ for the two layer subtraction case as:
	\begin{equation}
		F_{sub}=|\bra{\psi'^{(4)}_{l.p.}}\ket{\psi'^{(4)}}|^2 \leq p_{succ}
	\end{equation}
	To solve for $F_{sub}$, we only need to find the wavefunction $\psi'^{(4)}$ subject to the domain $\{z_{v_1} \geq z_{v_2} \geq max(z_{h,i})\}$. As shown below, this wavefunction can be solved using the virtual cavity method, similar to the single subtraction case.
	
\begin{figure*}[h]
		\centering
		\includegraphics[width=0.8\textwidth]{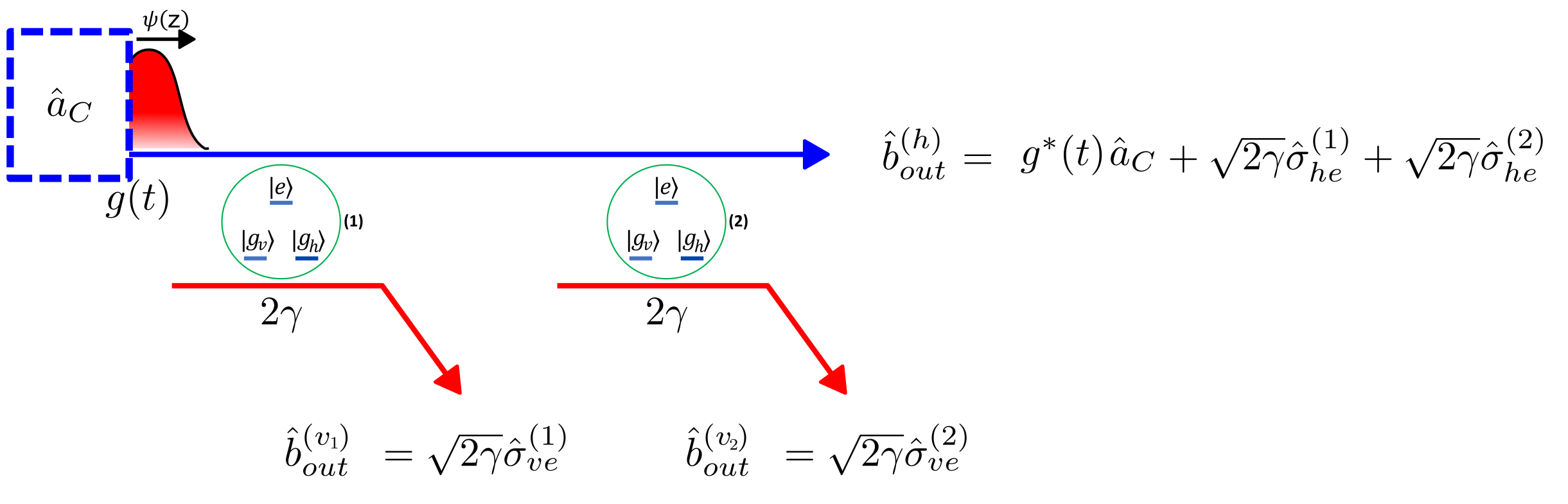}
		\caption{The virtual cavity approach for two layers of photon subtraction. Each green circle represents a three level atom coupled to two waveguides. The collapse operators are $\hat{b}_{out}^{(h)},\hat{b}_{out}^{(v_1)}$ and $\hat{b}_{out}^{(v_2)}$.}
		\label{Fig:virtual cavity double}
	\end{figure*}

	Figure~\ref{Fig:virtual cavity double} shows the virtual cavity and atom system to be considered. The effective Hamiltonian is given by:
	\begin{equation}
		\begin{aligned}
			\hat{H}_{eff} &= -i\frac{|g(t)|^2}{2} \hat{a}_{C}^{\dagger} \hat{a}_{C} - 2i\gamma \hat{\sigma}^{(1)}_{ee} - 2i\gamma \hat{\sigma}^{(2)}_{ee}\\
			& -i\sqrt{2\gamma}g^{*}(t)\hat{a}_{C}\hat{\sigma}^{(1)}_{eh} - i\sqrt{2\gamma}g^{*}(t)\hat{a}_{C}\hat{\sigma}^{(2)}_{eh}-i 2\gamma\hat{\sigma}^{(1)}_{he} \hat{\sigma}^{(2)}_{eh} 
			\label{non hermitian double}
		\end{aligned}
	\end{equation}
	with three collapse operators $\hat{b}_{out}^{(h)},\hat{b}_{out}^{(v_1)}$ and $\hat{b}_{out}^{(v_2)}$ as shown in the Figure~\ref{Fig:virtual cavity double}. We have defined $\hat{\sigma}^{(2)}_{eh}=\ket{e^{(2)}}\bra{g^{(2)}_{h}}$ where $\ket{e^{(2)}}$ and $\ket{g^{(2)}_{h}}$ are the excited state and $\ket{g_h}$ ground state of the second atom. Other operators are defined similarly.
	
	To obtain $\psi'^{(4)}$ on the target domain, we just need to find the following time-ordered correlation function:
	\begin{equation}
		\begin{aligned}
			&\bra{0;g_v;g_v}\hat{b}^{(h)}_{out}(t_{h,k-2})\cdots\hat{b}^{(h)}_{out}(t_{h,1})\hat{b}^{(v_2)}_{out}(t_{v_2})\hat{b}^{(v_1)}_{out}(t_{v_1})\ket{k;g_h;g_h}\\
			&\ \ \ \ \ \ \ \ \ \ \ \ \ \ \ \ \ \ \ \ \ \ (t_{v_1}\leq t_{v_2} \leq t_{h,1}\leq \cdots \leq t_{h,k-2})
		\end{aligned}
	\end{equation}
	where $\ket{k;g_h;g_h}$ denotes the initial state at $T=0$, with $k$ photons in the virtual cavity and both the atoms are in $\ket{g_h}$. The correlation-wavefunction correspondence is:
	\begin{equation}
		\begin{aligned}
			&\bra{0;g_v;g_v}\hat{b}^{(h)}_{out}(t_{h,k-2})\cdots\hat{b}^{(h)}_{out}(t_{h,1})\hat{b}^{(v_2)}_{out}(t_{v_2})\hat{b}^{(v_1)}_{out}(t_{v_1})\ket{k;g_h;g_h}\\
			& = (k-2)!\psi'^{(4)}(z_{v_1}=-t_{v_1},z_{v_2}=-t_{v_2};z_{h,1}=-t_{h,1},\cdots,z_{h,k-2}=-t_{h,k-2})
		\end{aligned}
	\end{equation}
	
	To find the above correlation function, we consider the process in which the initial state undergoes $k$ rounds of ``evolve and collapse" process, before ending up in a state proportional to $\ket{0;g_v;g_v}$. In particular, the first and second collapse are induced by the collapse operators $\hat{b}_{out}^{(v_1)}$ and $\hat{b}_{out}^{(v_2)}$ at time $t_{v_1}$ and $t_{v_2}$, respectively. The rest $(k-2)$ are induced by $\hat{b}_{out}^{(h)}$.
	
	Starting with the initial state $\ket{init} = \ket{k;g_h;g_h}$, from time $0$ to $t$ the state evolves under the effective Hamiltonian into 
	$\alpha_{0}(t)\ket{k;g_h;g_h}+\alpha_{1}(t)\ket{k-1;e;g_h}+\alpha_{2}(t)\ket{k-1;g_h;e}+\alpha_{3}(t)\ket{k-2;e;e}$. The corresponding differential equations are:
	\begin{equation}
		\begin{aligned}
			i\frac{d}{dt}\alpha_{0}(t) &= (-i \frac{|g(t)|^2}{2} k )\alpha_{0}(t)\\
			i\frac{d}{dt}\alpha_{1}(t) &= (-i\sqrt{k}g^{*}(t)\sqrt{2\gamma})\alpha_{0}(t)+(-i (k-1)\frac{|g(t)|^2}{2}-2i\gamma )\alpha_{1}(t)\\
			i\frac{d}{dt}\alpha_{2}(t) &= (-i\sqrt{k}g^{*}(t)\sqrt{2\gamma})\alpha_{0}(t)+(-i2\gamma)\alpha_{1}(t)+(-i (k-1)\frac{|g(t)|^2}{2}-2i\gamma )\alpha_{2}(t)\\
			i\frac{d}{dt}\alpha_{3}(t) &= (-i\sqrt{k-1}g^{*}(t)\sqrt{2\gamma})\alpha_{2}(t)+(-i\sqrt{k-1}g^{*}(t)\sqrt{2\gamma})\alpha_{1}(t)+(-i (k-2)\frac{|g(t)|^2}{2}-4i\gamma )\alpha_{3}(t)\\
		\end{aligned}
	\end{equation}
	
	For this family of equations, $\alpha_{0}(t)$ and $\alpha_{1}(t)$ are exactly the same as in the previous section, and are given by:
	
	\begin{equation}
		\begin{aligned}
			\alpha_{0}(t) &= [G(t)]^{\frac{k}{2}}\\
			\alpha_{1}(t) &= [G(t)]^{\frac{k-1}{2}}\times (-\frac{\sqrt{k}}{\sqrt{2\gamma}}\tilde{u}(t))\\
		\end{aligned}
	\end{equation}
	Where $\tilde{u}(t)$ is defined in~\ref{eq: u_tilde_u_0_define}. We further obtain $\alpha_{2}(t)$, giving:
	\begin{equation}
		\begin{aligned}
			\alpha_{2}(t) &= \frac{\sqrt{k}}{\sqrt{2\gamma}}[G(t)]^{\frac{k-1}{2}}\tilde{\Theta}(t)\\
			\tilde{\Theta}(t) &= \int_{0}^{t} 2\gamma(\tilde{u}(t')-u(t'))e^{-2\gamma(t-t')} dt'  
		\end{aligned}
	\end{equation}
	Finally, the differential equation for $\alpha_{3}(t)$ can be written as:
	\begin{equation}
		\begin{aligned}
			&\frac{d}{dt}\alpha_{3}(t) + [(k-2)\frac{|g(t)|^2}{2}+4\gamma] \alpha_{3}(t)\\
			&=\sqrt{k(k-1)}[G(t)]^{\frac{k-2}{2}}[u(t)\tilde{u}(t)-u(t)\tilde{\Theta}(t)]
		\end{aligned}
	\end{equation}
	which gives:
	\begin{equation}
		\begin{aligned}
			&\alpha_{3}(t) = [G(t)]^{\frac{k-2}{2}}\frac{\sqrt{k(k-1)}}{4\gamma} \tilde{w}(t)\\
			&\tilde{w}(t) = \int_{0}^{t} 4\gamma [u(t')\tilde{u}(t')-u(t')\tilde{\Theta}(t')] e^{-4\gamma(t-t')}dt'
		\end{aligned}
	\end{equation}
	
	After applying the collapse operator $\hat{b}_{out}^{(v_1)}$ at time $t$, we end up in the state: $\sqrt{2\gamma}\alpha_{1}(t)\ket{k-1;g_v;g_h} + \sqrt{2\gamma}\alpha_{3}(t)\ket{k-2;g_v;e}$. For $\ket{k-1;g_v;g_h}$, it evolves into $\beta_{0}(t+\tau_0,t)\ket{k-1;g_v;g_h}+\beta_{1}(t+\tau_0,t)\ket{k-2;g_v;e}$ from time $t$ to $t+\tau_0$. We have:
	\begin{equation}
		\begin{aligned}
			i\frac{d}{dT}\beta_{0}(T,t) &= (-i (k-1)\frac{|g(T)|^2}{2} )\beta_{0}(T,t)\\
			i\frac{d}{dT}\beta_{1}(T,t) &= (-i\sqrt{k-1}\sqrt{2\gamma}g^{*}(T))\beta_{0}(T,t)+(-i (k-2)\frac{|g(T)|^2}{2}-2i\gamma )\beta_{1}(T,t)\\
		\end{aligned}
	\end{equation}
	From which we obtain:
	\begin{equation}
		\begin{aligned}
			\beta_{0}(t+\tau_0,t) &= [\frac{G(t+\tau_0)}{G(t)}]^{\frac{k-1}{2}}\\
			\beta_{1}(t+\tau_0,t) &= \frac{G(t+\tau_0)^{\frac{k-2}{2}}}{G(t)^{\frac{k-1}{2}}}(-\frac{\sqrt{k-1}}{\sqrt{2\gamma}})\tilde{u}(t+\tau_0,t)\\
		\end{aligned}
	\end{equation}
	where the function $\tilde{u}(t+\tau_0,t)$ is defined in equation~\ref{eq: general u tilde define}.
	
	For $\ket{k-2;g_v;e}$, it evolves into $\beta_{2}(t+\tau_0,t)\ket{k-2;g_v;e}$ from time $t$ to $t+\tau_0$, where $\beta_{2}(t+\tau_0,t)=e^{-2\gamma \tau_{0}}[\frac{G(t+\tau_{0})}{G(t)}]^{\frac{k-2}{2}}$.
	
	Therefore, after applying the second collapse operator $\hat{b}_{out}^{(v_2)}$ at time $t+\tau_0$, the state becomes $2\gamma[\alpha_{1}(t)\beta_{1}(t+\tau_0,t) + \alpha_{3}(t)\beta_{2}(t+\tau_0,t)]\ket{k-2;g_v;g_v}$. Plugging in all the previous expression, we get:
	\begin{equation}
		\begin{aligned}
			&2\gamma[\alpha_{1}(t)\beta_{1}(t+\tau_0,t) + \alpha_{3}(t)\beta_{2}(t+\tau_0,t)]\ket{k-2;g_v;g_v}\\
			=&[G(t+\tau_0)]^{\frac{k-2}{2}}\sqrt{k(k-1)}[\tilde{u}(t)\tilde{u}(t+\tau_0,t)+\frac{1}{2}\tilde{w}(t)e^{-2\gamma \tau_0}]\ket{k-2;g_v;g_v}
		\end{aligned}
	\end{equation}
	The residue state proportional to $\ket{k-2;g_v;g_v}$ then undergoes a trivial evolve and collapse process as discussed in the previous section. By taking $t=t_{v_1}$ and $t+\tau_0=t_{v_2}$, we finally get:
	\begin{equation}
		\begin{aligned}
			&\bra{0;g_v;g_v}\hat{b}^{(h)}_{out}(t_{h,k-2})\cdots\hat{b}^{(h)}_{out}(t_{h,1})\hat{b}^{(v_2)}_{out}(t_{v_2})\hat{b}^{(v_1)}_{out}(t_{v_1})\ket{k;g_h;g_h}\\
			&=\sqrt{k!}[\tilde{u}(t_{v_1})\tilde{u}(t_{v_2},t_{v_1})+\frac{1}{2}\tilde{w}(t_{v_1})e^{-2\gamma (t_{v_2}-t_{v_1})}]u(t_{h,1})\cdots u(t_{h,k-2})\\
			&\ \ \ \ \ \ \ \ \ \ \ \ \ \ \ \ \ \ \ \ \ \ (t_{v_1}\leq t_{v_2} \leq t_{h,1}\leq \cdots \leq t_{h,k-2})
		\end{aligned}
	\end{equation}
	
	Plugging in the definition of subtraction infidelity $F_{sub}$ as well as the correlation-wavefunction correspondence, we find:
	\begin{equation}
		F_{sub}=|k(k-1)\int_{0}^{1}dt_{v_1}\int_{t_{v_1}}^{1}dt_{v_2}[\tilde{u}(t_{v_1})\tilde{u}(t_{v_2},t_{v_1})+\frac{1}{2}\tilde{w}(t_{v_1})e^{-2\gamma (t_{v_2}-t_{v_1})}][u(t_{v_1})u(t_{v_2})][G(t_{v_2})]^{k-2}|^2
	\end{equation}
	where we have already used the fact that $u(t)$ is a positive, smooth and real function supported on the interval $[0,1]$. To show that $\lim\limits_{\gamma\rightarrow1}F_{gate}= 1$, we just need to show $\lim\limits_{\gamma\rightarrow1}p_{succ}= 1$. Since $p_{succ}\geq F_{sub}$, we then need to show that $\lim\limits_{\gamma\rightarrow1}\sqrt{F_{sub}}= 1$.
	
	Consider $1-\sqrt{F_{sub}}$. We have:
	\begin{equation}
		|1-\sqrt{F_{sub}}| \leq I_1 + I_2 + I_3
	\end{equation}
	where:
	\begin{equation}
		\begin{aligned}
			I_1 = k(k-1)\int_{0}^{1}dt_{v_1}\int_{t_{v_1}}^{1}dt_{v_2}|\tilde{u}(t_{v_2},t_{v_1})-u(t_{v_2})|\tilde{u}(t_{v_1})u(t_{v_1})u(t_{v_2})G(t_{v_2})^{k-2}\\
			I_2 = k(k-1)\int_{0}^{1}dt_{v_1}\int_{t_{v_1}}^{1}dt_{v_2}|\tilde{u}(t_{v_1})-u(t_{v_1})|u(t_{v_2})u(t_{v_1})u(t_{v_2})G(t_{v_2})^{k-2}\\
			I_3 = k(k-1)\int_{0}^{1}dt_{v_1}\int_{t_{v_1}}^{1}dt_{v_2}\frac{1}{2}\tilde{w}(t_{v_1})e^{-2\gamma (t_{v_2}-t_{v_1})}u(t_{v_1})u(t_{v_2})G(t_{v_2})^{k-2}
		\end{aligned}
	\end{equation}
	For the integration $I_2$, we have already shown that $I_2\leq3\epsilon kM$ for $\gamma>max\{\frac{1}{2\eta \epsilon}ln(\frac{M}{\epsilon}) , \frac{M}{2\epsilon}\}$. For $I_3$, it is easy to show that $\max\limits_{t\in[0,1]}|w(t)|\leq M^2$ based on the definition of $w(t)$. Then we have $I_3\leq\frac{k(k-1)M^{4}}{4\gamma}$. 
	
	We now examine the integration $I_1$. Recall in equation~\ref{eq: general u tilde define} we have defined $\tilde{u}(t_2,t_1)=\int_{t_1}^{t_2}2\gamma e^{-2\gamma(t_2-T)}u(T)dT$. We further define:
	\begin{equation}
		u^{(0)}(t_2,t_1)=\int_{t_1}^{t_2}2\gamma e^{-2\gamma(t_2-T)}u(t_2)dT = u(t_2) (1-e^{-2\gamma(t_2 - t_1)}) 
	\end{equation}
	So that we have $I_1\leq I_{1}^{(1)}+I_{1}^{(2)}$, with:
	\begin{equation}
		\begin{aligned}
			I_{1}^{(1)} = k(k-1)\int_{0}^{1}dt_{v_1}\int_{t_{v_1}}^{1}dt_{v_2}|u^{(0)}(t_{v_2},t_{v_1})-u(t_{v_2})|\tilde{u}(t_{v_1})u(t_{v_1})u(t_{v_2})G(t_{v_2})^{k-2}\\
			= k(k-1)\int_{0}^{1}dt_{v_1}\int_{t_{v_1}}^{1}dt_{v_2}e^{-2\gamma(t_{v_2} - t_{v_1})} u(t_{v_2})\tilde{u}(t_{v_1})u(t_{v_1})u(t_{v_2})G(t_{v_2})^{k-2}\\
			I_{1}^{(2)} = k(k-1)\int_{0}^{1}dt_{v_1}\int_{t_{v_1}}^{1}dt_{v_2}|u^{(0)}(t_{v_2},t_{v_1})-\tilde{u}(t_{v_2},t_{v_1})|\tilde{u}(t_{v_1})u(t_{v_1})u(t_{v_2})G(t_{v_2})^{k-2}
		\end{aligned}
	\end{equation}
	Using the same trick as in the single subtraction case, we can prove that $|u^{(0)}(t_{v_2},t_{v_1})-\tilde{u}(t_{v_2},t_{v_1})|\leq 2\epsilon$ for $\gamma>\frac{1}{2\eta \epsilon}ln(\frac{M}{\epsilon})$. Therefore, $I_{1}^{(2)}\leq k(k-1)\epsilon M^{3}$. Finally, from the definition of $I_{1}^{(1)}$ we see that $I_{1}^{(1)}\leq\frac{k(k-1)M^{4}}{2\gamma}$. Wrapping up, we get:
	\begin{equation}
		|1-\sqrt{F_{sub}}|\leq 3\epsilon kM + \frac{5}{2}k(k-1)\epsilon M^{3}  
	\end{equation}
	for $\gamma>max\{\frac{1}{2\eta \epsilon}ln(\frac{M}{\epsilon}) , \frac{M}{2\epsilon}\}$. Since $M=\max\limits_{t\in[0,1]}u(t)$ and input photon number $k$ are fixed values, we see that $|1-\sqrt{F_{sub}}|$ can be made arbitrarily small and so is the subtraction infidelity $(1-F_{sub})$.
	
	Finally, we present numerical results for the subtraction infidelity $(1-F_{sub})$ for $u(t)$ being a square pulse on the interval $[0,1]$. Similar to the single subtraction case, the subtraction infidelity shows a $\propto \frac{1}{\gamma}$ scaling for fixed input photon number $k$ for large $\gamma$.

	\begin{figure*}[h]
		\centering
		\includegraphics[width=0.95\textwidth]{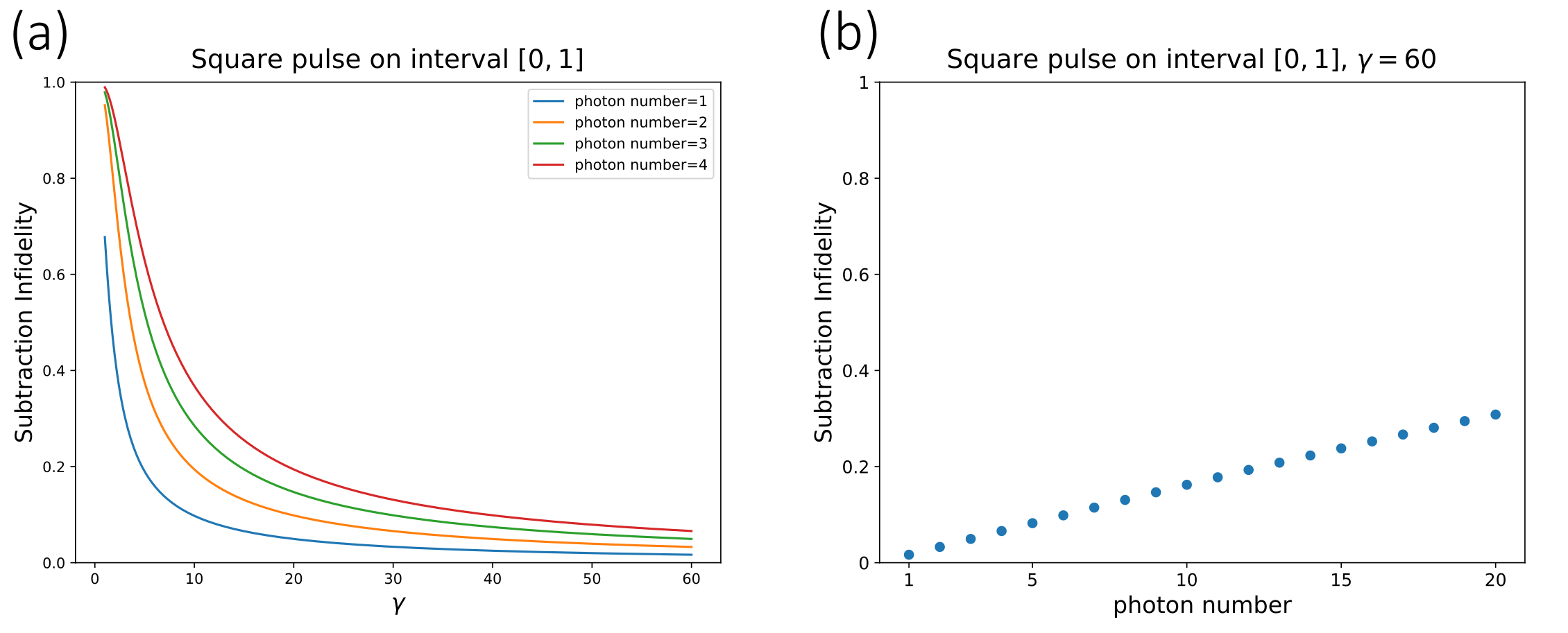}
		\caption{Subtraction infidelity for two-layer subtraction. (a) $(1-F_{sub})$ as a function of $\gamma$. (b) $(1-F_{sub})$ for fixed $\gamma=60$ plotted as a function of photon number. In both cases we have taken $u(t)$ being a normalized square pulse on interval $[0,1]$.}
		\label{Fig:infidelity numeric two layer subtract}
	\end{figure*}

	

	\subsection{V. Analytical solution for the FQH ground state}
\begin{figure*}[h]
		\centering
		\includegraphics[width=0.6\textwidth]{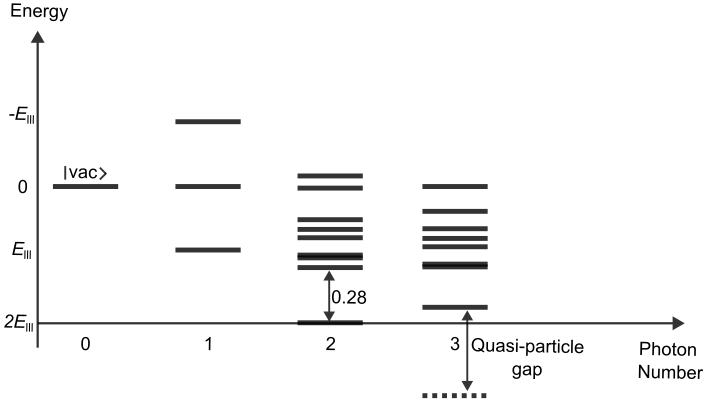}
		\caption{Sketch of the Energy level diagram of the FQH Hamiltonian considered in this work. For each photon number sector with photon number $k\leq 2$, the ground state energy is $kE_{lll}$ where $E_{lll}$ is the energy of the lowest Landau level.}
		\label{Fig:lll}
	\end{figure*}

	In the main article, we considered the FQH model on a square lattice with periodic boundary conditions. For such a model with on-site interaction term $\sum_{i}\frac{U}{2}(\hat{n}_{i}-1)\hat{n}_{i}$, the energy level structure can be schematically described as shown in Figure~\ref{Fig:lll}.
	
	The degenerate ground states of the FQH Hamiltonian are Hamiltonian eigenstates with the lowest energy (i.e. eigenvalue). Thus, the ground states should have a fixed photon number. In particular, it has been shown~\cite{kapit2014induced} that any ground state should have photon number $\phi_{entire}/2$. Here $\phi_{entire}$ is the flux through the entire lattice and is equal to the flux per plaquette times the number of lattice plaquettes $N_x\times N_y$.
	
	As for the minimal model discussed in our work, $\phi_{entire}=16\times\frac{1}{4}=4$, and thus the ground states should have two photons, i.e. they reside in the two-photon sector of Figure~\ref{Fig:lll}. The analytical expression of the ground state is given by:
	
	\begin{equation}
		\Psi^{(l)}(r_1,r_2) \propto F_{CM}^{(l)}(R)e^{-\pi\phi_{plaq} \sum_{i}y_{i}^2}\times(\vartheta\begin{bmatrix}\frac{1}{2}\\\frac{1}{2}\end{bmatrix}\begin{pmatrix}\frac{r_1-r_2}{N_x} & i\end{pmatrix})^2
	\end{equation}
	
	In the above equation, $l=1,2$ labels the two orthogonal two-photon wavefunctions that span the ground state, and correspond to the $\Psi_{ana}^{(1)}$ and $\Psi_{ana}^{(2)}$.
	
	
	\subsection{VI. Simulating Coherent drive and dissipation: single bosonic mode}
	
	Here, we confirm that the channel in Figure~\ref{Fig:FQH coherent drive}(a) in the main text simulates the coherent drive and dissipation combined for one lattice site. We define the superoperator $\mathbb{L}_{DS}$ as $\mathbb{L}_{DS}=-i[(F\hat{b}+h.c.),\rho]+\gamma( \hat{b}\rho\hat{b}^{\dagger}-\frac{1}{2}\{\hat{b}^{\dagger}\hat{b},\rho\} ) $. As mentioned in the main text, the channel that we wish to implement is given by: $e^{\mathbb{L}_{DS}\delta t}$. Since $\mathbb{L}_{DS}$ is time-invariant, we can formally expand the channel as:
	
	\begin{equation}
		e^{\mathbb{L}_{DS}\delta t} = \mathbb{I} + \mathbb{L}_{DS}\delta t + \frac{1}{2!}\delta t^2 (\mathbb{L}_{DS})^2 + ...
	\end{equation}
	
	Hence, up to the first order of $\delta t$, we have:
	
	\begin{equation}
		e^{\mathbb{L}_{DS}\delta t}[\rho] \simeq \rho + (-i[(F\delta t\hat{b}+h.c.),\rho]) + \gamma \delta t( \hat{b}\rho\hat{b}^{\dagger}-\frac{1}{2}\{\hat{b}^{\dagger}\hat{b},\rho\} ) 
		\label{DS channel expan}
	\end{equation}
	
	On the other hand, the actual channel used in our circuit is given by: $\epsilon[\rho]=Tr_{\hat{c}}[U_{\hat{b}-\hat{c}}(\rho\otimes\ket{\alpha}\bra{\alpha})U_{\hat{b}-\hat{c}}^{\dagger}]$, where $U_{\hat{b}-\hat{c}}=e^{-iK \delta t(\hat{b}^{\dagger}\hat{c}+\hat{c}^{\dagger}\hat{b}) }$. We now make use of the operator identity: $e^{A} B e^{-A}= B + [A,B] + \frac{1}{2!}[A,[A,B]]+...$ and expand $U_{\hat{b}-\hat{c}}(\rho\otimes\ket{\alpha}\bra{\alpha})U_{\hat{b}-\hat{c}}^{\dagger}$. We take $A=-iK \delta t(\hat{b}^{\dagger}\hat{c}+\hat{c}^{\dagger}\hat{b})$ and $B=\rho\otimes \ket{\alpha}\bra{\alpha}$. Note that:
	
	\begin{equation}
		\begin{aligned}
			[A,B] & = (-iK\delta t) \hat{b}\rho\otimes \hat{c}^{\dagger}\ket{\alpha}\bra{\alpha}\\
			& + (-iK\delta t) \hat{b}^{\dagger}\rho\otimes \hat{c}\ket{\alpha}\bra{\alpha}\\
			& - (-iK\delta t) \rho\hat{b}^{\dagger}\otimes \ket{\alpha}\bra{\alpha}\hat{c}\\
			& - (-iK\delta t) \rho\hat{b}\otimes \ket{\alpha}\bra{\alpha}\hat{c}^{\dagger}\\
		\end{aligned}
	\end{equation}
	
	Also, since $\ket{\alpha}$ is the coherent state for bosonic mode $\hat{c}$, we have $Tr[\hat{c}\ket{\alpha}\bra{\alpha}]=Tr[\ket{\alpha}\bra{\alpha}]\hat{c}=\bra{\alpha}\hat{c}\ket{\alpha}=\alpha$, and $Tr[\hat{c}^{\dagger}\ket{\alpha}\bra{\alpha}]=Tr[\ket{\alpha}\bra{\alpha}]\hat{c}^{\dagger}=\bra{\alpha}\hat{c}^{\dagger}\ket{\alpha}=\alpha^{*}$. Plugging in, we have:
	
	\begin{equation}
		\begin{aligned}
			Tr_{\hat{c}}[[A,B]] & = (-iK\alpha^{*}\delta t) \hat{b}\rho \\
			& + (-iK\alpha\delta t) \hat{b}^{\dagger}\rho\\
			& - (-iK\alpha\delta t) \rho\hat{b}^{\dagger}\\
			& - (-iK\alpha^{*}\delta t) \rho\hat{b}\\
		\end{aligned}
	\end{equation}
	
	Hence we should take $F=K\alpha^{*}$. To show that $(K\delta t)^2=\gamma \delta t$, we consider the term $\frac{1}{2!}[A,[A,B]]$ in the expansion. Since $[A,[A,B]]=AAB-2ABA+BAA$, we consider the three terms separately. 
	The first term reads:
	
	\begin{equation}
		\begin{aligned}
			AAB & = (-iK\delta t)^2\hat{b}^{\dagger}\hat{b}^{\dagger}\rho\otimes \hat{c}\hat{c}\ket{\alpha}\bra{\alpha}\\
			& + (-iK\delta t)^2\hat{b}\hat{b}^{\dagger}\rho\otimes \hat{c}^{\dagger}\hat{c}\ket{\alpha}\bra{\alpha}\\
			& + (-iK\delta t)^2\hat{b}^{\dagger}\hat{b}\rho\otimes \hat{c}\hat{c}^{\dagger}\ket{\alpha}\bra{\alpha}\\
			& + (-iK\delta t)^2\hat{b}\hat{b}\rho\otimes \hat{c}^{\dagger}\hat{c}^{\dagger}\ket{\alpha}\bra{\alpha}\\
		\end{aligned}
	\end{equation}
	
	We have $Tr[\hat{c}\hat{c}\ket{\alpha}\bra{\alpha}]=\alpha^2$, $Tr[\hat{c}^{\dagger}\hat{c}\ket{\alpha}\bra{\alpha}]=|\alpha|^2$, $Tr[\hat{c}\hat{c}^{\dagger}\ket{\alpha}\bra{\alpha}]= 1 + Tr[\hat{c}^{\dagger}\hat{c}\ket{\alpha}\bra{\alpha}]=1+|\alpha|^2$ and $Tr[\hat{c}^{\dagger}\hat{c}^{\dagger}\ket{\alpha}\bra{\alpha}]=(\alpha^{*})^2$. In this work, we have $\alpha \ll K\delta t \ll 1$. Hence, neglecting all the $O(|\alpha|^2)$ terms, we get:
	
	\begin{equation}
		Tr_{\hat{c}}[AAB]= - (K\delta t)^2 \hat{b}^{\dagger}\hat{b}\rho
	\end{equation}
	
	Following the same procedure for $-2ABA$ and $BAA$, and neglecting $O(|\alpha|^2)$ terms, we obtain:
	
	\begin{equation}
		\begin{aligned}
			Tr_{\hat{c}}[AAB] & = - (K\delta t)^2 \hat{b}^{\dagger}\hat{b}\rho\\
			Tr_{\hat{c}}[BAA] & = - (K\delta t)^2 \rho\hat{b}^{\dagger}\hat{b}\\
			Tr_{\hat{c}}[-2ABA] & = 2(K\delta t)^2 \hat{b}\rho\hat{b}^{\dagger}\\
		\end{aligned}
	\end{equation}
	
	Therefore, $\frac{1}{2!}[A,[A,B]]\simeq (K\delta t)^2 (\hat{b}\rho\hat{b}^{\dagger}- \frac{1}{2}\{\hat{b}^{\dagger}\hat{b},\rho\} )$.
	By comparing this to equation~\ref{DS channel expan}, we get $(K\delta t)^2=\gamma \delta t$.

	\subsection{VII. Additional result for FQH under coherent drive and dissipation}
	Here we present additional observables computed from the fixed point $\rho_{fix}$ in the main text.
	\begin{figure*}[h]
		\centering
		\includegraphics[width=0.8\textwidth]{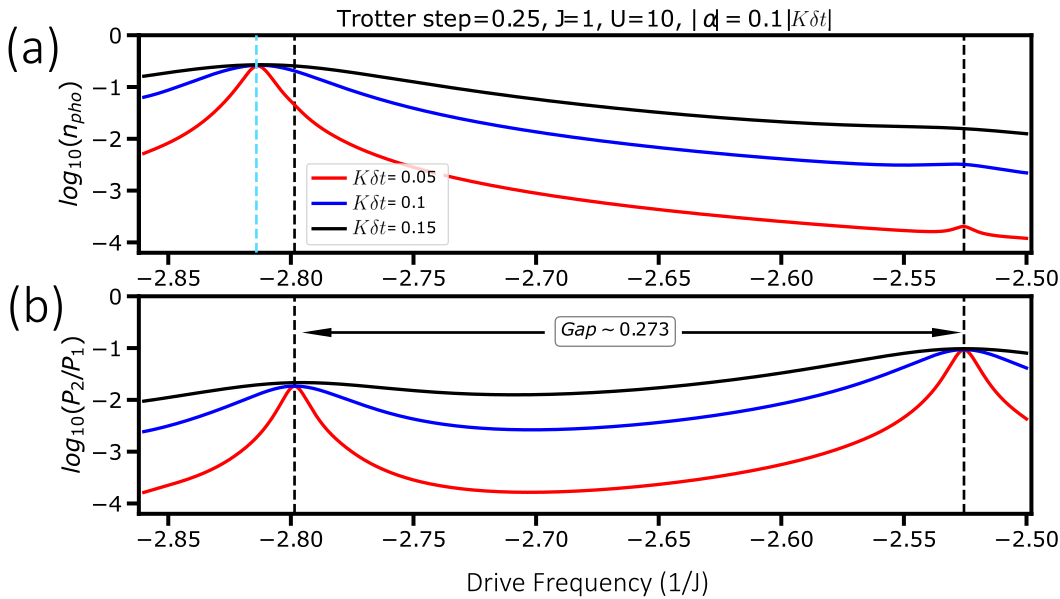}
		\caption{(a) Average photon number of the steady state $\rho_{fix}$ of the circuit shown in Figure 4(b) in the main text. (b) Two-photon population $P_2$ v.s. one photon population $P_1$ of the steady state. Both panels are plotted as a function of driving frequency.}
		\label{Fig:supp_coherent_drive}
	\end{figure*}
	We first calculate the average photon number for the steady state, with the same parameters chosen as in the main text.
	The observable is given by $n_{photon}=Tr[\rho_{fix}\sum_{i}\hat{b}_{i}^{\dagger}\hat{b}_{i}]$. As shown in Figure~\ref{Fig:supp_coherent_drive}(a), all three curves show a peak at $-2.82$ (dashed light-blue line in Figure~\ref{Fig:supp_coherent_drive}(a)). This is because, at this frequency, we are resonantly pumping the system from the vacuum to the lowest single photon eigenstate (i.e. the lowest landau level). However, even at this frequency where $n_{photon}$ is maximized, the average photon number never exceeds $0.269$. Thus, the steady-state $\rho_{fix}$ is mostly in the vacuum state with no photons, and the lowest energy eigenstate in each photon number sector is only sparsely occupied.
	
	To solidify this observation, we further investigate the ratio between the two-photon population and the one-photon population of the steady state, shown in Figure~\ref{Fig:supp_coherent_drive}(b). 
	The one photon population is defined by $P_1=Tr[\Pi_{1}\rho_{fix}\Pi_{1}]$, where $\Pi_{1}$ is the projector onto the Hilbert space spanned by all single photon states $\hat{b}_{i}^{\dagger}\ket{vac}$ (i.e. single photon sector). Similarly, we define two-photon population $P_2=Tr[\Pi_{2}\rho_{fix}\Pi_{2}]$ where $\Pi_2$ is the projector onto the Hilbert space spanned by all two-photon states $\hat{b}_{i}^{\dagger}\hat{b}_{j}^{\dagger}\ket{vac}$(i.e. the two-photon sector). Within the range of driving frequency, we observe two peaks for the curve $P_2/P_1$, for all $K\delta t$. The first peak occurs at $\Omega_{drive}=-2.7985$.
	At this frequency, we are resonantly pumping the system from the ground state of the one-photon sector to the ground state of the two-photon sector, hence achieving a local optimum for $P2/P1$. The other peak occurs at frequency $\Omega_{drive}=-2.5255$, for which we are resonantly pumping the system from the ground state of the one-photon sector to the first excited states of the two-photon sector. The energy difference between the ground and the first excited state in the two-photon sector is $~0.28$, as shown in Figure~\ref{Fig:FQH Hamiltonian simulation}(c), and this gap is manifestly shown from the distance between the two peaks in Figure~\ref{Fig:supp_coherent_drive}(b). However, for any frequency, the maximum value for $P2/P1$ never exceeds $0.0973$, indicating the weakness of the pump.

	\subsection{VIII. A more efficient protocol for FQH ground state preparation}
	As shown in Figure~\ref{Fig:supp_coherent_drive}(a-b), for the protocol in Figure~\ref{Fig:FQH coherent drive}(b), the population in the correct photon number sector remains small despite the high post-selected overlap between $\rho_{fix}$ and FQH ground space. Here we propose an alternative protocol that circumvents post-selection and sweeps the majority of the population into the correct photon number sector.
	
	\begin{figure*}[h]
		\centering
		\includegraphics[width=0.8\textwidth]{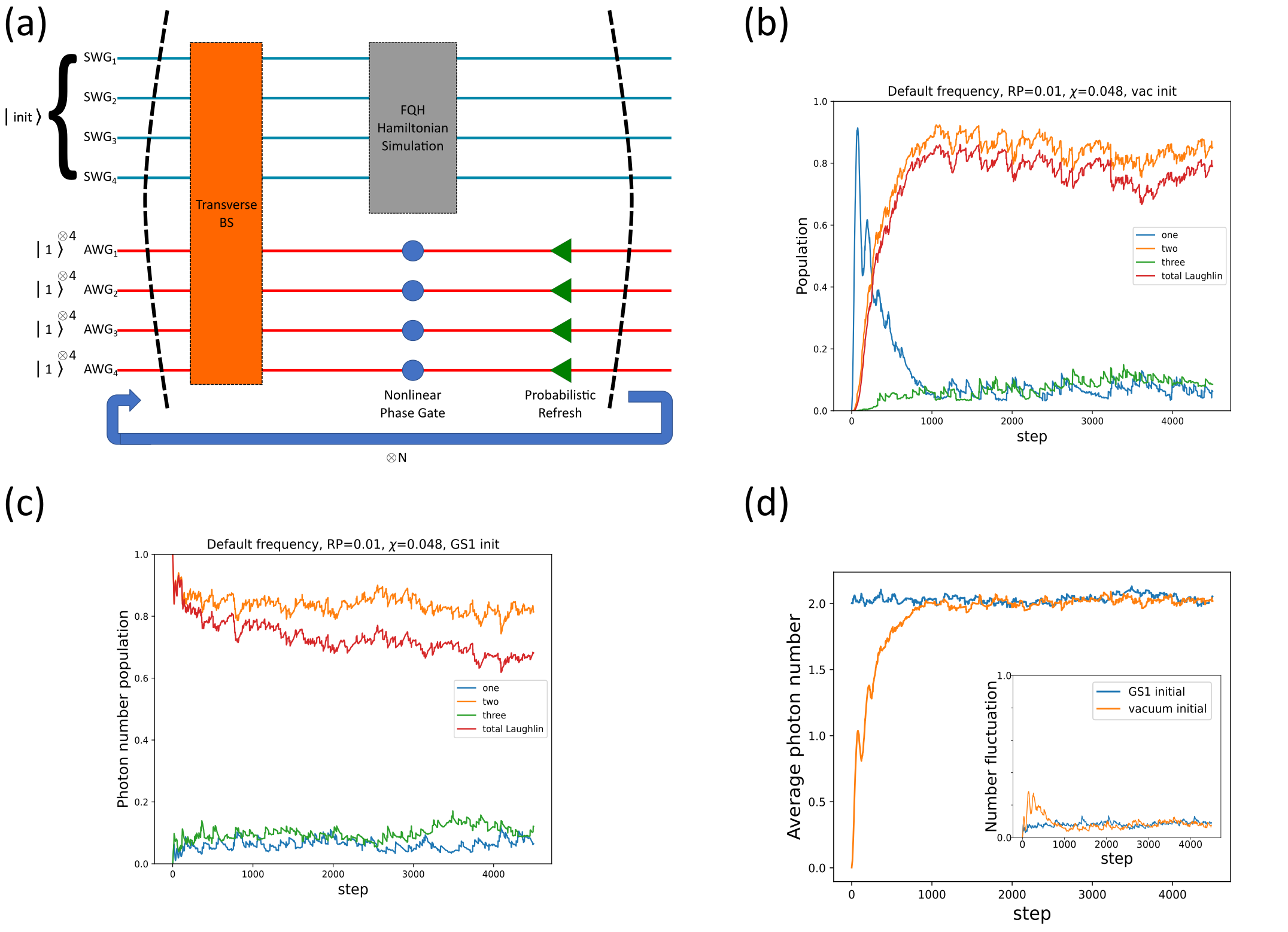}
		\caption{(a) The new protocol for FQH ground state preparation, using ancilla initialized in the single-photon state combined with probabilistic refresh operation. (b) The evolution of the population in each photon number sector and the FQH ground space with the system initialized in the vacuum state. (c) Same as (b), but with the system initialized in an FQH ground state. (d) Evolution of average photon number $\langle N \rangle$ and photon number fluctuation $\langle N^{2} \rangle - \langle N \rangle^2$ for the initial state in (b) and (c).}
		\label{Fig:incoherent FQH}
	\end{figure*}
	
	As shown in Figure~\ref{Fig:incoherent FQH}(a), the new protocol associates one ancilla Bosonic mode with one system Bosonic mode. We encode the system and ancilla bosonic modes into time bins in waveguides in the same way as the coherent driving scheme in Figure~\ref{Fig:FQH coherent drive}, and thus we have four system waveguides (SWG) and four ancilla waveguides (AWG). We initialized each ancilla bosonic modes into a single photon state and apply the channel in the dashed parentheses repeatedly. The first layer of the channel is a transverse beamsplitter between the SWGs and AWGs, with transmission coefficient $t=isin(\chi \delta t)$ and reflection coefficient $r=cos(\chi \delta t)$. The second layer includes the FQH Hamiltonian simulation circuit for the SWG and transverse nonlinear phase gates for the AWGs. In particular, the nonlinear phase gate for the AWGs applies a phase $\phi_1 = \phi_{g}^{(2)}-\phi_{g}^{(1)}$ and $\phi_2 = \phi_{g}^{(2)}-\phi_{g}^{(3)}+\phi_1$, where $\phi_{g}^{(k)}$ is the ground state eigen phase of the $k$-th photon sector. Finally, the last layer of the channel is a probabilistic refresh operation. For this operation, we flip a coin with head probability $p_{ref}$ for each time bin in each waveguide. If the coin turns out head, we dump any photon in the time bin and refresh it with a new single photon state.
	
	Figure~\ref{Fig:incoherent FQH}(b-c) shows the evolution of various observables after initializing the system into the vacuum (b) and FQH ground state (c). These observables include the population in the one, two, and three-photon number sectors (blue, orange, and green curves) along with the population in the FQH ground space (red curve). We have chosen parameters $\chi=0.048,\delta t=0.25$ and $p_{ref}=0.01$. As shown in (b-c), the population in the two-photon sector all converges to the same value $~85\%$, whereas the population in the one-photon and three-photon sector both converges to a transient $8\%$, indicating the effectiveness of the new protocol compared to the previous one in Figure~\ref{Fig:FQH coherent drive}. Moreover, we observe an approximate $75\%$ population in the FQH ground space. Finally, (d) shows the average photon number $\langle N \rangle$ and photon number fluctuation $\langle N^{2} \rangle - \langle N \rangle^2$ for the initial state in (b) and (c).

	\subsection{IX. Implementing periodic boundary condition}
	Here we present the complete waveguide circuit that simulates 1D and 2D lattices with periodic boundary conditions.
	Here we are only using the 1D Bose-Hubbard and 2D FQH model as examples, but the generalization to arbitrary 1D and 2D lattice with nearest-neighbor coupling is straightforward. 
	\begin{figure*}[h]
		\centering
		\includegraphics[width=0.8\textwidth]{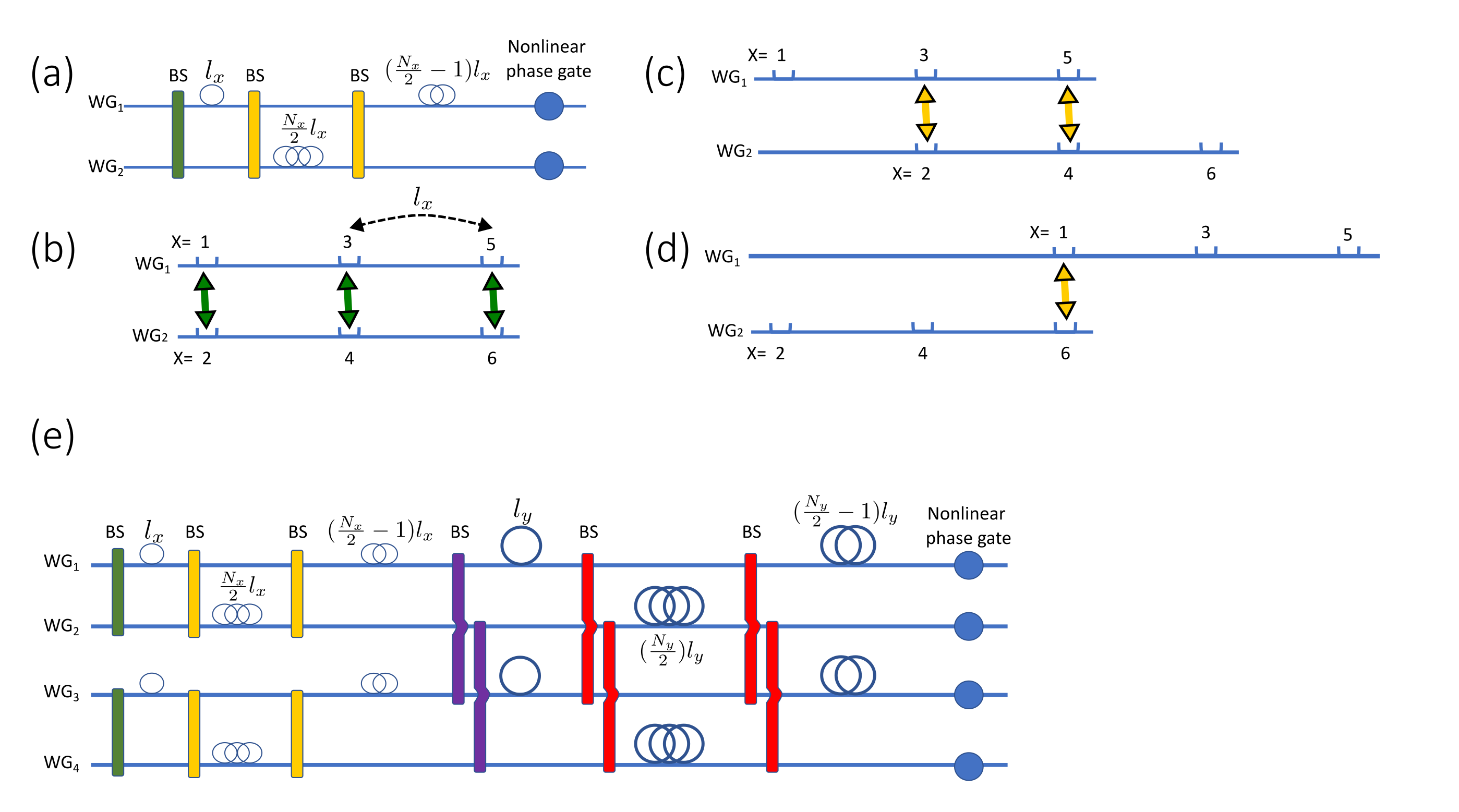}
		\caption{Complete waveguide circuit that implements periodic boundary conditions for the 1D Bose-Hubbard model (a) and 2D FQH model (e). (b) The configuration of time bins when they fly through the first green beamsplitter. The green arrows indicated the beamsplitter unitary applied to the corresponding time bins. (c) Same as (b) but for the first yellow beamsplitter. (d) Same as (b) and (c) but for the second yellow beamsplitter.}
		\label{Fig:periodic}
	\end{figure*}
	
	Figure~\ref{Fig:periodic}(a-d) shows the circuit for the 1D lattice. Initially, the time bins encoding the lattice sites are aligned according to the configuration in (b), and thus the green beamsplitter in (a) applies the beamsplitter operations in (b), indicated as green arrows. After the first $l_x$ delay, we obtain the configuration in (c). Thus we can apply the beamsplitter operations indicated as yellow arrows in (c) with the first yellow beamsplitter in (a). Importantly, the first yellow beamsplitter in (a) should be time-dependent, such that it only turns on when the desired pairs of time bins fly by. When time bin $x=1$ and time bin $x=6$ flies by, the two waveguides are uncoupled. Immediately after the first yellow beamsplitter, we use a $\frac{N_x}{2}l_x$ delay to align the time bins in a configuration shown in (d). The second time-dependent yellow beamsplitter then applies the beamsplitter operation indicated as the yellow arrow in (d). Finally, a fiber delay of length $(\frac{N_x}{2}-1)l_x$ is used to align the time bins to their initial configuration. The waveguide circuit in (a) thus implements all the beamsplitter operations needed for a Trotterized 1D lattice Hamiltonian evolution with periodic boundary conditions.
	
	Moreover, for a 1D lattice with an even number of sites $N_x$, a much simpler architecture can be applied to realize periodic boundary conditions, which eliminates the need for time-dependent beamsplitters. We use the same encoding scheme and circuit architecture as shown in Figure~\ref{Fig:bose hubbard}, while taking the total length of each fiber to be $\frac{N_x}{2}l_x$ (i.e. the time bins are fully packed in each fiber with no extra space left). This would naturally allow the first time bin in WG1 to align with the last time bin in WG2 after the first fiber delay of $l_x$, so that all the beamsplitter operations needed for periodic boundary condition can be implemented with the two time-independent beamsplitters in Figure~\ref{Fig:bose hubbard}(b).

	Similarly, we can generalize the circuit in Figure~\ref{Fig:periodic}(a) to the circuit in Figure~\ref{Fig:periodic}(e), which implements 2D periodic boundary conditions. 
	We use the same scheme as shown in Figure~\ref{Fig:FQH Hamiltonian simulation}(b) to encode the 2D lattice into time bins in four waveguides. That is, all the time bins representing lattice sites with the same $y$ and different $x$s are grouped into a ``coarse" time bin, where the separation between the center of two adjacent ``coarse" time bins is $l_y$. Within each ``coarse" time bins, we have the equally spaced ``fine" time bins (with $l_x$ spacing) representing different $x$s. For this encoding scheme, it should be noted that one would not get the desired periodic boundary condition in both $x$ and $y$ by simply taking $\frac{N_x}{2}l_x = l_y$ and the total length of each fiber loop to be $\frac{N_y}{2}l_y$. Therefore, the circuit architecture in Figure~\ref{Fig:periodic}(e) is necessary for 2D periodicity. For Figure~\ref{Fig:periodic}(e), the first three layers of beamsplitters and delays are chosen to be exactly the same as in (a), and implements all the beamsplitter operations between lattice sites with the same $y$ but different $x$s. The next three layers of beamsplitters and delays (i.e. red and purple beamsplitter and delays proportional to $l_y$) then implements the beamsplitter operations between lattice sites with the same $x$ but different $y$s. Essentially, the next three layers implements the same operation for the ``coarse" time bins in the same way (a) implements the operations for the ``fine" time bins.

	
	\section{Acknowledgements}
	
	The authors wish to acknowledge fruitful discussions with Mohammad Hafezi, Alex Ma, Sunil Mittal, and Eliot Kapit. This research was supported by The Office of Naval Research ONR-MURI grant N00014-20-1-2325, AFOSR FA95502010223, NSF OMA1936314, NSF PHY1820938, IMOD NSF DMR-2019444, and ARL W911NF1920181.
	
	
	
	\section{Competing interests}
	The authors declare no competing interests.
	
	
	\section{Data availability}
	All of the data that support the findings of this study are reported in the main text and Supplementary Information. Source data are available from the corresponding authors on reasonable request.
	
	
	\bibliography{Main.bib}

\begin{thebibliography}{57}%
\makeatletter
\providecommand \@ifxundefined [1]{%
 \@ifx{#1\undefined}
}%
\providecommand \@ifnum [1]{%
 \ifnum #1\expandafter \@firstoftwo
 \else \expandafter \@secondoftwo
 \fi
}%
\providecommand \@ifx [1]{%
 \ifx #1\expandafter \@firstoftwo
 \else \expandafter \@secondoftwo
 \fi
}%
\providecommand \natexlab [1]{#1}%
\providecommand \enquote  [1]{``#1''}%
\providecommand \bibnamefont  [1]{#1}%
\providecommand \bibfnamefont [1]{#1}%
\providecommand \citenamefont [1]{#1}%
\providecommand \href@noop [0]{\@secondoftwo}%
\providecommand \href [0]{\begingroup \@sanitize@url \@href}%
\providecommand \@href[1]{\@@startlink{#1}\@@href}%
\providecommand \@@href[1]{\endgroup#1\@@endlink}%
\providecommand \@sanitize@url [0]{\catcode `\\12\catcode `\$12\catcode
  `\&12\catcode `\#12\catcode `\^12\catcode `\_12\catcode `\%12\relax}%
\providecommand \@@startlink[1]{}%
\providecommand \@@endlink[0]{}%
\providecommand \url  [0]{\begingroup\@sanitize@url \@url }%
\providecommand \@url [1]{\endgroup\@href {#1}{\urlprefix }}%
\providecommand \urlprefix  [0]{URL }%
\providecommand \Eprint [0]{\href }%
\providecommand \doibase [0]{https://doi.org/}%
\providecommand \selectlanguage [0]{\@gobble}%
\providecommand \bibinfo  [0]{\@secondoftwo}%
\providecommand \bibfield  [0]{\@secondoftwo}%
\providecommand \translation [1]{[#1]}%
\providecommand \BibitemOpen [0]{}%
\providecommand \bibitemStop [0]{}%
\providecommand \bibitemNoStop [0]{.\EOS\space}%
\providecommand \EOS [0]{\spacefactor3000\relax}%
\providecommand \BibitemShut  [1]{\csname bibitem#1\endcsname}%
\let\auto@bib@innerbib\@empty
\bibitem [{\citenamefont {Noh}\ and\ \citenamefont
  {Angelakis}(2016)}]{noh2016quantum}%
  \BibitemOpen
  \bibfield  {author} {\bibinfo {author} {\bibfnamefont {C.}~\bibnamefont
  {Noh}}\ and\ \bibinfo {author} {\bibfnamefont {D.~G.}\ \bibnamefont
  {Angelakis}},\ }\bibfield  {title} {\bibinfo {title} {Quantum simulations and
  many-body physics with light},\ }\href@noop {} {\bibfield  {journal}
  {\bibinfo  {journal} {Reports on Progress in Physics}\ }\textbf {\bibinfo
  {volume} {80}},\ \bibinfo {pages} {016401} (\bibinfo {year}
  {2016})}\BibitemShut {NoStop}%
\bibitem [{\citenamefont {Hartmann}(2016)}]{hartmann2016quantum}%
  \BibitemOpen
  \bibfield  {author} {\bibinfo {author} {\bibfnamefont {M.~J.}\ \bibnamefont
  {Hartmann}},\ }\bibfield  {title} {\bibinfo {title} {Quantum simulation with
  interacting photons},\ }\href@noop {} {\bibfield  {journal} {\bibinfo
  {journal} {Journal of Optics}\ }\textbf {\bibinfo {volume} {18}},\ \bibinfo
  {pages} {104005} (\bibinfo {year} {2016})}\BibitemShut {NoStop}%
\bibitem [{\citenamefont {Yuan}\ \emph {et~al.}(2018)\citenamefont {Yuan},
  \citenamefont {Lin}, \citenamefont {Xiao},\ and\ \citenamefont
  {Fan}}]{yuan2018synthetic}%
  \BibitemOpen
  \bibfield  {author} {\bibinfo {author} {\bibfnamefont {L.}~\bibnamefont
  {Yuan}}, \bibinfo {author} {\bibfnamefont {Q.}~\bibnamefont {Lin}}, \bibinfo
  {author} {\bibfnamefont {M.}~\bibnamefont {Xiao}},\ and\ \bibinfo {author}
  {\bibfnamefont {S.}~\bibnamefont {Fan}},\ }\bibfield  {title} {\bibinfo
  {title} {Synthetic dimension in photonics},\ }\href@noop {} {\bibfield
  {journal} {\bibinfo  {journal} {Optica}\ }\textbf {\bibinfo {volume} {5}},\
  \bibinfo {pages} {1396} (\bibinfo {year} {2018})}\BibitemShut {NoStop}%
\bibitem [{\citenamefont {Aspuru-Guzik}\ and\ \citenamefont
  {Walther}(2012)}]{aspuru2012photonic}%
  \BibitemOpen
  \bibfield  {author} {\bibinfo {author} {\bibfnamefont {A.}~\bibnamefont
  {Aspuru-Guzik}}\ and\ \bibinfo {author} {\bibfnamefont {P.}~\bibnamefont
  {Walther}},\ }\bibfield  {title} {\bibinfo {title} {Photonic quantum
  simulators},\ }\href@noop {} {\bibfield  {journal} {\bibinfo  {journal}
  {Nature physics}\ }\textbf {\bibinfo {volume} {8}},\ \bibinfo {pages} {285}
  (\bibinfo {year} {2012})}\BibitemShut {NoStop}%
\bibitem [{\citenamefont {Underwood}\ and\ \citenamefont
  {Feder}(2012)}]{underwood2012bose}%
  \BibitemOpen
  \bibfield  {author} {\bibinfo {author} {\bibfnamefont {M.~S.}\ \bibnamefont
  {Underwood}}\ and\ \bibinfo {author} {\bibfnamefont {D.~L.}\ \bibnamefont
  {Feder}},\ }\bibfield  {title} {\bibinfo {title} {Bose-hubbard model for
  universal quantum-walk-based computation},\ }\href@noop {} {\bibfield
  {journal} {\bibinfo  {journal} {Physical Review A—Atomic, Molecular, and
  Optical Physics}\ }\textbf {\bibinfo {volume} {85}},\ \bibinfo {pages}
  {052314} (\bibinfo {year} {2012})}\BibitemShut {NoStop}%
\bibitem [{\citenamefont {Childs}\ \emph {et~al.}(2014)\citenamefont {Childs},
  \citenamefont {Gosset},\ and\ \citenamefont {Webb}}]{childs2014bose}%
  \BibitemOpen
  \bibfield  {author} {\bibinfo {author} {\bibfnamefont {A.~M.}\ \bibnamefont
  {Childs}}, \bibinfo {author} {\bibfnamefont {D.}~\bibnamefont {Gosset}},\
  and\ \bibinfo {author} {\bibfnamefont {Z.}~\bibnamefont {Webb}},\ }\bibfield
  {title} {\bibinfo {title} {The bose-hubbard model is qma-complete},\ }in\
  \href@noop {} {\emph {\bibinfo {booktitle} {Automata, Languages, and
  Programming: 41st International Colloquium, ICALP 2014, Copenhagen, Denmark,
  July 8-11, 2014, Proceedings, Part I 41}}}\ (\bibinfo {organization}
  {Springer},\ \bibinfo {year} {2014})\ pp.\ \bibinfo {pages}
  {308--319}\BibitemShut {NoStop}%
\bibitem [{\citenamefont {Knill}\ \emph {et~al.}(2001)\citenamefont {Knill},
  \citenamefont {Laflamme},\ and\ \citenamefont {Milburn}}]{knill2001scheme}%
  \BibitemOpen
  \bibfield  {author} {\bibinfo {author} {\bibfnamefont {E.}~\bibnamefont
  {Knill}}, \bibinfo {author} {\bibfnamefont {R.}~\bibnamefont {Laflamme}},\
  and\ \bibinfo {author} {\bibfnamefont {G.~J.}\ \bibnamefont {Milburn}},\
  }\bibfield  {title} {\bibinfo {title} {A scheme for efficient quantum
  computation with linear optics},\ }\href@noop {} {\bibfield  {journal}
  {\bibinfo  {journal} {nature}\ }\textbf {\bibinfo {volume} {409}},\ \bibinfo
  {pages} {46} (\bibinfo {year} {2001})}\BibitemShut {NoStop}%
\bibitem [{\citenamefont {Umucal{\i}lar}\ and\ \citenamefont
  {Carusotto}(2012)}]{umucalilar2012fractional}%
  \BibitemOpen
  \bibfield  {author} {\bibinfo {author} {\bibfnamefont {R.}~\bibnamefont
  {Umucal{\i}lar}}\ and\ \bibinfo {author} {\bibfnamefont {I.}~\bibnamefont
  {Carusotto}},\ }\bibfield  {title} {\bibinfo {title} {Fractional quantum hall
  states of photons in an array of dissipative coupled cavities},\ }\href@noop
  {} {\bibfield  {journal} {\bibinfo  {journal} {Physical Review Letters}\
  }\textbf {\bibinfo {volume} {108}},\ \bibinfo {pages} {206809} (\bibinfo
  {year} {2012})}\BibitemShut {NoStop}%
\bibitem [{\citenamefont {Cho}\ \emph {et~al.}(2008)\citenamefont {Cho},
  \citenamefont {Angelakis},\ and\ \citenamefont {Bose}}]{cho_fractional_2008}%
  \BibitemOpen
  \bibfield  {author} {\bibinfo {author} {\bibfnamefont {J.}~\bibnamefont
  {Cho}}, \bibinfo {author} {\bibfnamefont {D.~G.}\ \bibnamefont {Angelakis}},\
  and\ \bibinfo {author} {\bibfnamefont {S.}~\bibnamefont {Bose}},\ }\bibfield
  {title} {\bibinfo {title} {Fractional {Quantum} {Hall} {State} in {Coupled}
  {Cavities}},\ }\href {https://doi.org/10.1103/PhysRevLett.101.246809}
  {\bibfield  {journal} {\bibinfo  {journal} {Phys. Rev. Lett.}\ }\textbf
  {\bibinfo {volume} {101}},\ \bibinfo {pages} {246809} (\bibinfo {year}
  {2008})},\ \bibinfo {note} {publisher: American Physical Society}\BibitemShut
  {NoStop}%
\bibitem [{\citenamefont {Birnbaum}\ \emph {et~al.}(2005)\citenamefont
  {Birnbaum}, \citenamefont {Boca}, \citenamefont {Miller}, \citenamefont
  {Boozer}, \citenamefont {Northup},\ and\ \citenamefont
  {Kimble}}]{birnbaum2005photon}%
  \BibitemOpen
  \bibfield  {author} {\bibinfo {author} {\bibfnamefont {K.~M.}\ \bibnamefont
  {Birnbaum}}, \bibinfo {author} {\bibfnamefont {A.}~\bibnamefont {Boca}},
  \bibinfo {author} {\bibfnamefont {R.}~\bibnamefont {Miller}}, \bibinfo
  {author} {\bibfnamefont {A.~D.}\ \bibnamefont {Boozer}}, \bibinfo {author}
  {\bibfnamefont {T.~E.}\ \bibnamefont {Northup}},\ and\ \bibinfo {author}
  {\bibfnamefont {H.~J.}\ \bibnamefont {Kimble}},\ }\bibfield  {title}
  {\bibinfo {title} {Photon blockade in an optical cavity with one trapped
  atom},\ }\href@noop {} {\bibfield  {journal} {\bibinfo  {journal} {Nature}\
  }\textbf {\bibinfo {volume} {436}},\ \bibinfo {pages} {87} (\bibinfo {year}
  {2005})}\BibitemShut {NoStop}%
\bibitem [{\citenamefont {Reinhard}\ \emph {et~al.}(2012)\citenamefont
  {Reinhard}, \citenamefont {Volz}, \citenamefont {Winger}, \citenamefont
  {Badolato}, \citenamefont {Hennessy}, \citenamefont {Hu},\ and\ \citenamefont
  {Imamo{\u{g}}lu}}]{reinhard2012strongly}%
  \BibitemOpen
  \bibfield  {author} {\bibinfo {author} {\bibfnamefont {A.}~\bibnamefont
  {Reinhard}}, \bibinfo {author} {\bibfnamefont {T.}~\bibnamefont {Volz}},
  \bibinfo {author} {\bibfnamefont {M.}~\bibnamefont {Winger}}, \bibinfo
  {author} {\bibfnamefont {A.}~\bibnamefont {Badolato}}, \bibinfo {author}
  {\bibfnamefont {K.~J.}\ \bibnamefont {Hennessy}}, \bibinfo {author}
  {\bibfnamefont {E.~L.}\ \bibnamefont {Hu}},\ and\ \bibinfo {author}
  {\bibfnamefont {A.}~\bibnamefont {Imamo{\u{g}}lu}},\ }\bibfield  {title}
  {\bibinfo {title} {Strongly correlated photons on a chip},\ }\href@noop {}
  {\bibfield  {journal} {\bibinfo  {journal} {Nature Photonics}\ }\textbf
  {\bibinfo {volume} {6}},\ \bibinfo {pages} {93} (\bibinfo {year}
  {2012})}\BibitemShut {NoStop}%
\bibitem [{\citenamefont {Wang}\ \emph {et~al.}(2024)\citenamefont {Wang},
  \citenamefont {Liu}, \citenamefont {Chen}, \citenamefont {Chen},
  \citenamefont {Zhao}, \citenamefont {Ying}, \citenamefont {Shang},
  \citenamefont {Wang}, \citenamefont {Huo}, \citenamefont {Peng} \emph
  {et~al.}}]{wang2024realization}%
  \BibitemOpen
  \bibfield  {author} {\bibinfo {author} {\bibfnamefont {C.}~\bibnamefont
  {Wang}}, \bibinfo {author} {\bibfnamefont {F.-M.}\ \bibnamefont {Liu}},
  \bibinfo {author} {\bibfnamefont {M.-C.}\ \bibnamefont {Chen}}, \bibinfo
  {author} {\bibfnamefont {H.}~\bibnamefont {Chen}}, \bibinfo {author}
  {\bibfnamefont {X.-H.}\ \bibnamefont {Zhao}}, \bibinfo {author}
  {\bibfnamefont {C.}~\bibnamefont {Ying}}, \bibinfo {author} {\bibfnamefont
  {Z.-X.}\ \bibnamefont {Shang}}, \bibinfo {author} {\bibfnamefont {J.-W.}\
  \bibnamefont {Wang}}, \bibinfo {author} {\bibfnamefont {Y.-H.}\ \bibnamefont
  {Huo}}, \bibinfo {author} {\bibfnamefont {C.-Z.}\ \bibnamefont {Peng}}, \emph
  {et~al.},\ }\bibfield  {title} {\bibinfo {title} {Realization of fractional
  quantum hall state with interacting photons},\ }\href@noop {} {\bibfield
  {journal} {\bibinfo  {journal} {Science}\ }\textbf {\bibinfo {volume}
  {384}},\ \bibinfo {pages} {579} (\bibinfo {year} {2024})}\BibitemShut
  {NoStop}%
\bibitem [{\citenamefont {Karamlou}\ \emph {et~al.}(2024)\citenamefont
  {Karamlou}, \citenamefont {Rosen}, \citenamefont {Muschinske}, \citenamefont
  {Barrett}, \citenamefont {Di~Paolo}, \citenamefont {Ding}, \citenamefont
  {Harrington}, \citenamefont {Hays}, \citenamefont {Das}, \citenamefont {Kim}
  \emph {et~al.}}]{karamlou2024probing}%
  \BibitemOpen
  \bibfield  {author} {\bibinfo {author} {\bibfnamefont {A.~H.}\ \bibnamefont
  {Karamlou}}, \bibinfo {author} {\bibfnamefont {I.~T.}\ \bibnamefont {Rosen}},
  \bibinfo {author} {\bibfnamefont {S.~E.}\ \bibnamefont {Muschinske}},
  \bibinfo {author} {\bibfnamefont {C.~N.}\ \bibnamefont {Barrett}}, \bibinfo
  {author} {\bibfnamefont {A.}~\bibnamefont {Di~Paolo}}, \bibinfo {author}
  {\bibfnamefont {L.}~\bibnamefont {Ding}}, \bibinfo {author} {\bibfnamefont
  {P.~M.}\ \bibnamefont {Harrington}}, \bibinfo {author} {\bibfnamefont
  {M.}~\bibnamefont {Hays}}, \bibinfo {author} {\bibfnamefont {R.}~\bibnamefont
  {Das}}, \bibinfo {author} {\bibfnamefont {D.~K.}\ \bibnamefont {Kim}}, \emph
  {et~al.},\ }\bibfield  {title} {\bibinfo {title} {Probing entanglement in a
  2d hard-core bose--hubbard lattice},\ }\href@noop {} {\bibfield  {journal}
  {\bibinfo  {journal} {Nature}\ ,\ \bibinfo {pages} {1}} (\bibinfo {year}
  {2024})}\BibitemShut {NoStop}%
\bibitem [{\citenamefont {Kapit}\ \emph {et~al.}(2014)\citenamefont {Kapit},
  \citenamefont {Hafezi},\ and\ \citenamefont {Simon}}]{kapit2014induced}%
  \BibitemOpen
  \bibfield  {author} {\bibinfo {author} {\bibfnamefont {E.}~\bibnamefont
  {Kapit}}, \bibinfo {author} {\bibfnamefont {M.}~\bibnamefont {Hafezi}},\ and\
  \bibinfo {author} {\bibfnamefont {S.~H.}\ \bibnamefont {Simon}},\ }\bibfield
  {title} {\bibinfo {title} {Induced self-stabilization in fractional quantum
  hall states of light},\ }\href@noop {} {\bibfield  {journal} {\bibinfo
  {journal} {Physical Review X}\ }\textbf {\bibinfo {volume} {4}},\ \bibinfo
  {pages} {031039} (\bibinfo {year} {2014})}\BibitemShut {NoStop}%
\bibitem [{\citenamefont {Schreiber}\ \emph {et~al.}(2012)\citenamefont
  {Schreiber}, \citenamefont {G{\'a}bris}, \citenamefont {Rohde}, \citenamefont
  {Laiho}, \citenamefont {{\v{S}}tefa{\v{n}}{\'a}k}, \citenamefont
  {Poto{\v{c}}ek}, \citenamefont {Hamilton}, \citenamefont {Jex},\ and\
  \citenamefont {Silberhorn}}]{schreiber20122d}%
  \BibitemOpen
  \bibfield  {author} {\bibinfo {author} {\bibfnamefont {A.}~\bibnamefont
  {Schreiber}}, \bibinfo {author} {\bibfnamefont {A.}~\bibnamefont
  {G{\'a}bris}}, \bibinfo {author} {\bibfnamefont {P.~P.}\ \bibnamefont
  {Rohde}}, \bibinfo {author} {\bibfnamefont {K.}~\bibnamefont {Laiho}},
  \bibinfo {author} {\bibfnamefont {M.}~\bibnamefont
  {{\v{S}}tefa{\v{n}}{\'a}k}}, \bibinfo {author} {\bibfnamefont
  {V.}~\bibnamefont {Poto{\v{c}}ek}}, \bibinfo {author} {\bibfnamefont
  {C.}~\bibnamefont {Hamilton}}, \bibinfo {author} {\bibfnamefont
  {I.}~\bibnamefont {Jex}},\ and\ \bibinfo {author} {\bibfnamefont
  {C.}~\bibnamefont {Silberhorn}},\ }\bibfield  {title} {\bibinfo {title} {A 2d
  quantum walk simulation of two-particle dynamics},\ }\href@noop {} {\bibfield
   {journal} {\bibinfo  {journal} {Science}\ }\textbf {\bibinfo {volume}
  {336}},\ \bibinfo {pages} {55} (\bibinfo {year} {2012})}\BibitemShut
  {NoStop}%
\bibitem [{\citenamefont {Marques~Muniz}\ \emph {et~al.}(2023)\citenamefont
  {Marques~Muniz}, \citenamefont {Wu}, \citenamefont {Jung}, \citenamefont
  {Khajavikhan}, \citenamefont {Christodoulides},\ and\ \citenamefont
  {Peschel}}]{marques2023observation}%
  \BibitemOpen
  \bibfield  {author} {\bibinfo {author} {\bibfnamefont {A.}~\bibnamefont
  {Marques~Muniz}}, \bibinfo {author} {\bibfnamefont {F.}~\bibnamefont {Wu}},
  \bibinfo {author} {\bibfnamefont {P.}~\bibnamefont {Jung}}, \bibinfo {author}
  {\bibfnamefont {M.}~\bibnamefont {Khajavikhan}}, \bibinfo {author}
  {\bibfnamefont {D.}~\bibnamefont {Christodoulides}},\ and\ \bibinfo {author}
  {\bibfnamefont {U.}~\bibnamefont {Peschel}},\ }\bibfield  {title} {\bibinfo
  {title} {Observation of photon-photon thermodynamic processes under negative
  optical temperature conditions},\ }\href@noop {} {\bibfield  {journal}
  {\bibinfo  {journal} {Science}\ }\textbf {\bibinfo {volume} {379}},\ \bibinfo
  {pages} {1019} (\bibinfo {year} {2023})}\BibitemShut {NoStop}%
\bibitem [{\citenamefont {Bogaerts}\ \emph {et~al.}(2020)\citenamefont
  {Bogaerts}, \citenamefont {P{\'e}rez}, \citenamefont {Capmany}, \citenamefont
  {Miller}, \citenamefont {Poon}, \citenamefont {Englund}, \citenamefont
  {Morichetti},\ and\ \citenamefont {Melloni}}]{bogaerts2020programmable}%
  \BibitemOpen
  \bibfield  {author} {\bibinfo {author} {\bibfnamefont {W.}~\bibnamefont
  {Bogaerts}}, \bibinfo {author} {\bibfnamefont {D.}~\bibnamefont {P{\'e}rez}},
  \bibinfo {author} {\bibfnamefont {J.}~\bibnamefont {Capmany}}, \bibinfo
  {author} {\bibfnamefont {D.~A.}\ \bibnamefont {Miller}}, \bibinfo {author}
  {\bibfnamefont {J.}~\bibnamefont {Poon}}, \bibinfo {author} {\bibfnamefont
  {D.}~\bibnamefont {Englund}}, \bibinfo {author} {\bibfnamefont
  {F.}~\bibnamefont {Morichetti}},\ and\ \bibinfo {author} {\bibfnamefont
  {A.}~\bibnamefont {Melloni}},\ }\bibfield  {title} {\bibinfo {title}
  {Programmable photonic circuits},\ }\href@noop {} {\bibfield  {journal}
  {\bibinfo  {journal} {Nature}\ }\textbf {\bibinfo {volume} {586}},\ \bibinfo
  {pages} {207} (\bibinfo {year} {2020})}\BibitemShut {NoStop}%
\bibitem [{\citenamefont {Monika}\ \emph {et~al.}(2024)\citenamefont {Monika},
  \citenamefont {Nosrati}, \citenamefont {George}, \citenamefont {Sciara},
  \citenamefont {Fazili}, \citenamefont {Marques~Muniz}, \citenamefont
  {Bisianov}, \citenamefont {Lo~Franco}, \citenamefont {Munro}, \citenamefont
  {Chemnitz} \emph {et~al.}}]{monika2024quantum}%
  \BibitemOpen
  \bibfield  {author} {\bibinfo {author} {\bibfnamefont {M.}~\bibnamefont
  {Monika}}, \bibinfo {author} {\bibfnamefont {F.}~\bibnamefont {Nosrati}},
  \bibinfo {author} {\bibfnamefont {A.}~\bibnamefont {George}}, \bibinfo
  {author} {\bibfnamefont {S.}~\bibnamefont {Sciara}}, \bibinfo {author}
  {\bibfnamefont {R.}~\bibnamefont {Fazili}}, \bibinfo {author} {\bibfnamefont
  {A.~L.}\ \bibnamefont {Marques~Muniz}}, \bibinfo {author} {\bibfnamefont
  {A.}~\bibnamefont {Bisianov}}, \bibinfo {author} {\bibfnamefont
  {R.}~\bibnamefont {Lo~Franco}}, \bibinfo {author} {\bibfnamefont {W.~J.}\
  \bibnamefont {Munro}}, \bibinfo {author} {\bibfnamefont {M.}~\bibnamefont
  {Chemnitz}}, \emph {et~al.},\ }\bibfield  {title} {\bibinfo {title} {Quantum
  state processing through controllable synthetic temporal photonic lattices},\
  }\href@noop {} {\bibfield  {journal} {\bibinfo  {journal} {Nature Photonics}\
  ,\ \bibinfo {pages} {1}} (\bibinfo {year} {2024})}\BibitemShut {NoStop}%
\bibitem [{\citenamefont {Zheng}\ \emph {et~al.}(2024)\citenamefont {Zheng},
  \citenamefont {Jalali~Mehrabad}, \citenamefont {Vannucci}, \citenamefont
  {Li}, \citenamefont {Dutt}, \citenamefont {Hafezi}, \citenamefont {Mittal},\
  and\ \citenamefont {Waks}}]{zheng2024dynamic}%
  \BibitemOpen
  \bibfield  {author} {\bibinfo {author} {\bibfnamefont {X.}~\bibnamefont
  {Zheng}}, \bibinfo {author} {\bibfnamefont {M.}~\bibnamefont
  {Jalali~Mehrabad}}, \bibinfo {author} {\bibfnamefont {J.}~\bibnamefont
  {Vannucci}}, \bibinfo {author} {\bibfnamefont {K.}~\bibnamefont {Li}},
  \bibinfo {author} {\bibfnamefont {A.}~\bibnamefont {Dutt}}, \bibinfo {author}
  {\bibfnamefont {M.}~\bibnamefont {Hafezi}}, \bibinfo {author} {\bibfnamefont
  {S.}~\bibnamefont {Mittal}},\ and\ \bibinfo {author} {\bibfnamefont
  {E.}~\bibnamefont {Waks}},\ }\bibfield  {title} {\bibinfo {title} {Dynamic
  control of 2d non-hermitian photonic corner skin modes in synthetic
  dimensions},\ }\href@noop {} {\bibfield  {journal} {\bibinfo  {journal}
  {Nature Communications}\ }\textbf {\bibinfo {volume} {15}},\ \bibinfo {pages}
  {10881} (\bibinfo {year} {2024})}\BibitemShut {NoStop}%
\bibitem [{\citenamefont {Chalabi}\ \emph {et~al.}(2019)\citenamefont
  {Chalabi}, \citenamefont {Barik}, \citenamefont {Mittal}, \citenamefont
  {Murphy}, \citenamefont {Hafezi},\ and\ \citenamefont
  {Waks}}]{chalabi2019synthetic}%
  \BibitemOpen
  \bibfield  {author} {\bibinfo {author} {\bibfnamefont {H.}~\bibnamefont
  {Chalabi}}, \bibinfo {author} {\bibfnamefont {S.}~\bibnamefont {Barik}},
  \bibinfo {author} {\bibfnamefont {S.}~\bibnamefont {Mittal}}, \bibinfo
  {author} {\bibfnamefont {T.~E.}\ \bibnamefont {Murphy}}, \bibinfo {author}
  {\bibfnamefont {M.}~\bibnamefont {Hafezi}},\ and\ \bibinfo {author}
  {\bibfnamefont {E.}~\bibnamefont {Waks}},\ }\bibfield  {title} {\bibinfo
  {title} {Synthetic gauge field for two-dimensional time-multiplexed quantum
  random walks},\ }\href@noop {} {\bibfield  {journal} {\bibinfo  {journal}
  {Physical Review Letters}\ }\textbf {\bibinfo {volume} {123}},\ \bibinfo
  {pages} {150503} (\bibinfo {year} {2019})}\BibitemShut {NoStop}%
\bibitem [{\citenamefont {Chalabi}\ \emph {et~al.}(2020)\citenamefont
  {Chalabi}, \citenamefont {Barik}, \citenamefont {Mittal}, \citenamefont
  {Murphy}, \citenamefont {Hafezi},\ and\ \citenamefont
  {Waks}}]{chalabi2020guiding}%
  \BibitemOpen
  \bibfield  {author} {\bibinfo {author} {\bibfnamefont {H.}~\bibnamefont
  {Chalabi}}, \bibinfo {author} {\bibfnamefont {S.}~\bibnamefont {Barik}},
  \bibinfo {author} {\bibfnamefont {S.}~\bibnamefont {Mittal}}, \bibinfo
  {author} {\bibfnamefont {T.~E.}\ \bibnamefont {Murphy}}, \bibinfo {author}
  {\bibfnamefont {M.}~\bibnamefont {Hafezi}},\ and\ \bibinfo {author}
  {\bibfnamefont {E.}~\bibnamefont {Waks}},\ }\bibfield  {title} {\bibinfo
  {title} {Guiding and confining of light in a two-dimensional synthetic space
  using electric fields},\ }\href@noop {} {\bibfield  {journal} {\bibinfo
  {journal} {Optica}\ }\textbf {\bibinfo {volume} {7}},\ \bibinfo {pages} {506}
  (\bibinfo {year} {2020})}\BibitemShut {NoStop}%
\bibitem [{\citenamefont {Leefmans}\ \emph {et~al.}(2022)\citenamefont
  {Leefmans}, \citenamefont {Dutt}, \citenamefont {Williams}, \citenamefont
  {Yuan}, \citenamefont {Parto}, \citenamefont {Nori}, \citenamefont {Fan},\
  and\ \citenamefont {Marandi}}]{leefmans2022topological}%
  \BibitemOpen
  \bibfield  {author} {\bibinfo {author} {\bibfnamefont {C.}~\bibnamefont
  {Leefmans}}, \bibinfo {author} {\bibfnamefont {A.}~\bibnamefont {Dutt}},
  \bibinfo {author} {\bibfnamefont {J.}~\bibnamefont {Williams}}, \bibinfo
  {author} {\bibfnamefont {L.}~\bibnamefont {Yuan}}, \bibinfo {author}
  {\bibfnamefont {M.}~\bibnamefont {Parto}}, \bibinfo {author} {\bibfnamefont
  {F.}~\bibnamefont {Nori}}, \bibinfo {author} {\bibfnamefont {S.}~\bibnamefont
  {Fan}},\ and\ \bibinfo {author} {\bibfnamefont {A.}~\bibnamefont {Marandi}},\
  }\bibfield  {title} {\bibinfo {title} {Topological dissipation in a
  time-multiplexed photonic resonator network},\ }\href@noop {} {\bibfield
  {journal} {\bibinfo  {journal} {Nature Physics}\ }\textbf {\bibinfo {volume}
  {18}},\ \bibinfo {pages} {442} (\bibinfo {year} {2022})}\BibitemShut
  {NoStop}%
\bibitem [{\citenamefont {Parto}\ \emph {et~al.}(2023)\citenamefont {Parto},
  \citenamefont {Leefmans}, \citenamefont {Williams}, \citenamefont {Nori},\
  and\ \citenamefont {Marandi}}]{parto2023non}%
  \BibitemOpen
  \bibfield  {author} {\bibinfo {author} {\bibfnamefont {M.}~\bibnamefont
  {Parto}}, \bibinfo {author} {\bibfnamefont {C.}~\bibnamefont {Leefmans}},
  \bibinfo {author} {\bibfnamefont {J.}~\bibnamefont {Williams}}, \bibinfo
  {author} {\bibfnamefont {F.}~\bibnamefont {Nori}},\ and\ \bibinfo {author}
  {\bibfnamefont {A.}~\bibnamefont {Marandi}},\ }\bibfield  {title} {\bibinfo
  {title} {Non-abelian effects in dissipative photonic topological lattices},\
  }\href@noop {} {\bibfield  {journal} {\bibinfo  {journal} {Nature
  Communications}\ }\textbf {\bibinfo {volume} {14}},\ \bibinfo {pages} {1440}
  (\bibinfo {year} {2023})}\BibitemShut {NoStop}%
\bibitem [{\citenamefont {Weidemann}\ \emph {et~al.}(2020)\citenamefont
  {Weidemann}, \citenamefont {Kremer}, \citenamefont {Helbig}, \citenamefont
  {Hofmann}, \citenamefont {Stegmaier}, \citenamefont {Greiter}, \citenamefont
  {Thomale},\ and\ \citenamefont {Szameit}}]{weidemann2020topological}%
  \BibitemOpen
  \bibfield  {author} {\bibinfo {author} {\bibfnamefont {S.}~\bibnamefont
  {Weidemann}}, \bibinfo {author} {\bibfnamefont {M.}~\bibnamefont {Kremer}},
  \bibinfo {author} {\bibfnamefont {T.}~\bibnamefont {Helbig}}, \bibinfo
  {author} {\bibfnamefont {T.}~\bibnamefont {Hofmann}}, \bibinfo {author}
  {\bibfnamefont {A.}~\bibnamefont {Stegmaier}}, \bibinfo {author}
  {\bibfnamefont {M.}~\bibnamefont {Greiter}}, \bibinfo {author} {\bibfnamefont
  {R.}~\bibnamefont {Thomale}},\ and\ \bibinfo {author} {\bibfnamefont
  {A.}~\bibnamefont {Szameit}},\ }\bibfield  {title} {\bibinfo {title}
  {Topological funneling of light},\ }\href@noop {} {\bibfield  {journal}
  {\bibinfo  {journal} {Science}\ }\textbf {\bibinfo {volume} {368}},\ \bibinfo
  {pages} {311} (\bibinfo {year} {2020})}\BibitemShut {NoStop}%
\bibitem [{\citenamefont {Weidemann}\ \emph {et~al.}(2022)\citenamefont
  {Weidemann}, \citenamefont {Kremer}, \citenamefont {Longhi},\ and\
  \citenamefont {Szameit}}]{weidemann2022topological}%
  \BibitemOpen
  \bibfield  {author} {\bibinfo {author} {\bibfnamefont {S.}~\bibnamefont
  {Weidemann}}, \bibinfo {author} {\bibfnamefont {M.}~\bibnamefont {Kremer}},
  \bibinfo {author} {\bibfnamefont {S.}~\bibnamefont {Longhi}},\ and\ \bibinfo
  {author} {\bibfnamefont {A.}~\bibnamefont {Szameit}},\ }\bibfield  {title}
  {\bibinfo {title} {Topological triple phase transition in non-hermitian
  floquet quasicrystals},\ }\href@noop {} {\bibfield  {journal} {\bibinfo
  {journal} {Nature}\ }\textbf {\bibinfo {volume} {601}},\ \bibinfo {pages}
  {354} (\bibinfo {year} {2022})}\BibitemShut {NoStop}%
\bibitem [{\citenamefont {Lin}\ \emph {et~al.}(2023)\citenamefont {Lin},
  \citenamefont {Yi},\ and\ \citenamefont {Xue}}]{lin2023manipulating}%
  \BibitemOpen
  \bibfield  {author} {\bibinfo {author} {\bibfnamefont {Q.}~\bibnamefont
  {Lin}}, \bibinfo {author} {\bibfnamefont {W.}~\bibnamefont {Yi}},\ and\
  \bibinfo {author} {\bibfnamefont {P.}~\bibnamefont {Xue}},\ }\bibfield
  {title} {\bibinfo {title} {Manipulating directional flow in a two-dimensional
  photonic quantum walk under a synthetic magnetic field},\ }\href@noop {}
  {\bibfield  {journal} {\bibinfo  {journal} {Nature Communications}\ }\textbf
  {\bibinfo {volume} {14}},\ \bibinfo {pages} {6283} (\bibinfo {year}
  {2023})}\BibitemShut {NoStop}%
\bibitem [{\citenamefont {J{\"u}rgensen}\ \emph {et~al.}(2021)\citenamefont
  {J{\"u}rgensen}, \citenamefont {Mukherjee},\ and\ \citenamefont
  {Rechtsman}}]{jurgensen2021quantized}%
  \BibitemOpen
  \bibfield  {author} {\bibinfo {author} {\bibfnamefont {M.}~\bibnamefont
  {J{\"u}rgensen}}, \bibinfo {author} {\bibfnamefont {S.}~\bibnamefont
  {Mukherjee}},\ and\ \bibinfo {author} {\bibfnamefont {M.~C.}\ \bibnamefont
  {Rechtsman}},\ }\bibfield  {title} {\bibinfo {title} {Quantized nonlinear
  thouless pumping},\ }\href@noop {} {\bibfield  {journal} {\bibinfo  {journal}
  {Nature}\ }\textbf {\bibinfo {volume} {596}},\ \bibinfo {pages} {63}
  (\bibinfo {year} {2021})}\BibitemShut {NoStop}%
\bibitem [{\citenamefont {Leefmans}\ \emph {et~al.}(2024)\citenamefont
  {Leefmans}, \citenamefont {Parto}, \citenamefont {Williams}, \citenamefont
  {Li}, \citenamefont {Dutt}, \citenamefont {Nori},\ and\ \citenamefont
  {Marandi}}]{leefmans2024topological}%
  \BibitemOpen
  \bibfield  {author} {\bibinfo {author} {\bibfnamefont {C.~R.}\ \bibnamefont
  {Leefmans}}, \bibinfo {author} {\bibfnamefont {M.}~\bibnamefont {Parto}},
  \bibinfo {author} {\bibfnamefont {J.}~\bibnamefont {Williams}}, \bibinfo
  {author} {\bibfnamefont {G.~H.}\ \bibnamefont {Li}}, \bibinfo {author}
  {\bibfnamefont {A.}~\bibnamefont {Dutt}}, \bibinfo {author} {\bibfnamefont
  {F.}~\bibnamefont {Nori}},\ and\ \bibinfo {author} {\bibfnamefont
  {A.}~\bibnamefont {Marandi}},\ }\bibfield  {title} {\bibinfo {title}
  {Topological temporally mode-locked laser},\ }\href@noop {} {\bibfield
  {journal} {\bibinfo  {journal} {Nature Physics}\ ,\ \bibinfo {pages} {1}}
  (\bibinfo {year} {2024})}\BibitemShut {NoStop}%
\bibitem [{\citenamefont {Flower}\ \emph {et~al.}(2024)\citenamefont {Flower},
  \citenamefont {Jalali~Mehrabad}, \citenamefont {Xu}, \citenamefont {Moille},
  \citenamefont {Suarez-Forero}, \citenamefont {{\"O}rsel}, \citenamefont
  {Bahl}, \citenamefont {Chembo}, \citenamefont {Srinivasan}, \citenamefont
  {Mittal} \emph {et~al.}}]{flower2024observation}%
  \BibitemOpen
  \bibfield  {author} {\bibinfo {author} {\bibfnamefont {C.~J.}\ \bibnamefont
  {Flower}}, \bibinfo {author} {\bibfnamefont {M.}~\bibnamefont
  {Jalali~Mehrabad}}, \bibinfo {author} {\bibfnamefont {L.}~\bibnamefont {Xu}},
  \bibinfo {author} {\bibfnamefont {G.}~\bibnamefont {Moille}}, \bibinfo
  {author} {\bibfnamefont {D.~G.}\ \bibnamefont {Suarez-Forero}}, \bibinfo
  {author} {\bibfnamefont {O.}~\bibnamefont {{\"O}rsel}}, \bibinfo {author}
  {\bibfnamefont {G.}~\bibnamefont {Bahl}}, \bibinfo {author} {\bibfnamefont
  {Y.}~\bibnamefont {Chembo}}, \bibinfo {author} {\bibfnamefont
  {K.}~\bibnamefont {Srinivasan}}, \bibinfo {author} {\bibfnamefont
  {S.}~\bibnamefont {Mittal}}, \emph {et~al.},\ }\bibfield  {title} {\bibinfo
  {title} {Observation of topological frequency combs},\ }\href@noop {}
  {\bibfield  {journal} {\bibinfo  {journal} {Science}\ }\textbf {\bibinfo
  {volume} {384}},\ \bibinfo {pages} {1356} (\bibinfo {year}
  {2024})}\BibitemShut {NoStop}%
\bibitem [{\citenamefont {Jalali~Mehrabad}\ and\ \citenamefont
  {Hafezi}(2024)}]{jalali2024strain}%
  \BibitemOpen
  \bibfield  {author} {\bibinfo {author} {\bibfnamefont {M.}~\bibnamefont
  {Jalali~Mehrabad}}\ and\ \bibinfo {author} {\bibfnamefont {M.}~\bibnamefont
  {Hafezi}},\ }\bibfield  {title} {\bibinfo {title} {Strain-induced landau
  levels in photonic crystals},\ }\href@noop {} {\bibfield  {journal} {\bibinfo
   {journal} {Nature Photonics}\ }\textbf {\bibinfo {volume} {18}},\ \bibinfo
  {pages} {527} (\bibinfo {year} {2024})}\BibitemShut {NoStop}%
\bibitem [{\citenamefont {Yang}\ \emph {et~al.}(2022)\citenamefont {Yang},
  \citenamefont {Lund}, \citenamefont {Pohl}, \citenamefont {Lodahl},\ and\
  \citenamefont {M{\o}lmer}}]{yang2022deterministic}%
  \BibitemOpen
  \bibfield  {author} {\bibinfo {author} {\bibfnamefont {F.}~\bibnamefont
  {Yang}}, \bibinfo {author} {\bibfnamefont {M.~M.}\ \bibnamefont {Lund}},
  \bibinfo {author} {\bibfnamefont {T.}~\bibnamefont {Pohl}}, \bibinfo {author}
  {\bibfnamefont {P.}~\bibnamefont {Lodahl}},\ and\ \bibinfo {author}
  {\bibfnamefont {K.}~\bibnamefont {M{\o}lmer}},\ }\bibfield  {title} {\bibinfo
  {title} {Deterministic photon sorting in waveguide qed systems},\ }\href@noop
  {} {\bibfield  {journal} {\bibinfo  {journal} {Physical Review Letters}\
  }\textbf {\bibinfo {volume} {128}},\ \bibinfo {pages} {213603} (\bibinfo
  {year} {2022})}\BibitemShut {NoStop}%
\bibitem [{\citenamefont {Lund}\ \emph {et~al.}(2024)\citenamefont {Lund},
  \citenamefont {Yang}, \citenamefont {Christiansen}, \citenamefont
  {Kornovan},\ and\ \citenamefont {M{\o}lmer}}]{lund2024subtraction}%
  \BibitemOpen
  \bibfield  {author} {\bibinfo {author} {\bibfnamefont {M.~M.}\ \bibnamefont
  {Lund}}, \bibinfo {author} {\bibfnamefont {F.}~\bibnamefont {Yang}}, \bibinfo
  {author} {\bibfnamefont {V.~R.}\ \bibnamefont {Christiansen}}, \bibinfo
  {author} {\bibfnamefont {D.}~\bibnamefont {Kornovan}},\ and\ \bibinfo
  {author} {\bibfnamefont {K.}~\bibnamefont {M{\o}lmer}},\ }\bibfield  {title}
  {\bibinfo {title} {Subtraction and addition of propagating photons by
  two-level emitters},\ }\href@noop {} {\bibfield  {journal} {\bibinfo
  {journal} {Physical Review Letters}\ }\textbf {\bibinfo {volume} {133}},\
  \bibinfo {pages} {103601} (\bibinfo {year} {2024})}\BibitemShut {NoStop}%
\bibitem [{\citenamefont {Tomm}\ \emph {et~al.}(2023)\citenamefont {Tomm},
  \citenamefont {Mahmoodian}, \citenamefont {Antoniadis}, \citenamefont
  {Schott}, \citenamefont {Valentin}, \citenamefont {Wieck}, \citenamefont
  {Ludwig}, \citenamefont {Javadi},\ and\ \citenamefont
  {Warburton}}]{tomm2023photon}%
  \BibitemOpen
  \bibfield  {author} {\bibinfo {author} {\bibfnamefont {N.}~\bibnamefont
  {Tomm}}, \bibinfo {author} {\bibfnamefont {S.}~\bibnamefont {Mahmoodian}},
  \bibinfo {author} {\bibfnamefont {N.~O.}\ \bibnamefont {Antoniadis}},
  \bibinfo {author} {\bibfnamefont {R.}~\bibnamefont {Schott}}, \bibinfo
  {author} {\bibfnamefont {S.~R.}\ \bibnamefont {Valentin}}, \bibinfo {author}
  {\bibfnamefont {A.~D.}\ \bibnamefont {Wieck}}, \bibinfo {author}
  {\bibfnamefont {A.}~\bibnamefont {Ludwig}}, \bibinfo {author} {\bibfnamefont
  {A.}~\bibnamefont {Javadi}},\ and\ \bibinfo {author} {\bibfnamefont {R.~J.}\
  \bibnamefont {Warburton}},\ }\bibfield  {title} {\bibinfo {title} {Photon
  bound state dynamics from a single artificial atom},\ }\href@noop {}
  {\bibfield  {journal} {\bibinfo  {journal} {Nature Physics}\ }\textbf
  {\bibinfo {volume} {19}},\ \bibinfo {pages} {857} (\bibinfo {year}
  {2023})}\BibitemShut {NoStop}%
\bibitem [{\citenamefont {Babin}\ \emph {et~al.}(2022)\citenamefont {Babin},
  \citenamefont {St{\"o}hr}, \citenamefont {Morioka}, \citenamefont
  {Linkewitz}, \citenamefont {Steidl}, \citenamefont {W{\"o}rnle},
  \citenamefont {Liu}, \citenamefont {Hesselmeier}, \citenamefont {Vorobyov},
  \citenamefont {Denisenko} \emph {et~al.}}]{babin2022fabrication}%
  \BibitemOpen
  \bibfield  {author} {\bibinfo {author} {\bibfnamefont {C.}~\bibnamefont
  {Babin}}, \bibinfo {author} {\bibfnamefont {R.}~\bibnamefont {St{\"o}hr}},
  \bibinfo {author} {\bibfnamefont {N.}~\bibnamefont {Morioka}}, \bibinfo
  {author} {\bibfnamefont {T.}~\bibnamefont {Linkewitz}}, \bibinfo {author}
  {\bibfnamefont {T.}~\bibnamefont {Steidl}}, \bibinfo {author} {\bibfnamefont
  {R.}~\bibnamefont {W{\"o}rnle}}, \bibinfo {author} {\bibfnamefont
  {D.}~\bibnamefont {Liu}}, \bibinfo {author} {\bibfnamefont {E.}~\bibnamefont
  {Hesselmeier}}, \bibinfo {author} {\bibfnamefont {V.}~\bibnamefont
  {Vorobyov}}, \bibinfo {author} {\bibfnamefont {A.}~\bibnamefont {Denisenko}},
  \emph {et~al.},\ }\bibfield  {title} {\bibinfo {title} {Fabrication and
  nanophotonic waveguide integration of silicon carbide colour centres with
  preserved spin-optical coherence},\ }\href@noop {} {\bibfield  {journal}
  {\bibinfo  {journal} {Nature materials}\ }\textbf {\bibinfo {volume} {21}},\
  \bibinfo {pages} {67} (\bibinfo {year} {2022})}\BibitemShut {NoStop}%
\bibitem [{\citenamefont {Ferreira}\ \emph {et~al.}(2024)\citenamefont
  {Ferreira}, \citenamefont {Kim}, \citenamefont {Butler}, \citenamefont
  {Pichler},\ and\ \citenamefont {Painter}}]{ferreira2024deterministic}%
  \BibitemOpen
  \bibfield  {author} {\bibinfo {author} {\bibfnamefont {V.~S.}\ \bibnamefont
  {Ferreira}}, \bibinfo {author} {\bibfnamefont {G.}~\bibnamefont {Kim}},
  \bibinfo {author} {\bibfnamefont {A.}~\bibnamefont {Butler}}, \bibinfo
  {author} {\bibfnamefont {H.}~\bibnamefont {Pichler}},\ and\ \bibinfo {author}
  {\bibfnamefont {O.}~\bibnamefont {Painter}},\ }\bibfield  {title} {\bibinfo
  {title} {Deterministic generation of multidimensional photonic cluster states
  with a single quantum emitter},\ }\href@noop {} {\bibfield  {journal}
  {\bibinfo  {journal} {Nature Physics}\ ,\ \bibinfo {pages} {1}} (\bibinfo
  {year} {2024})}\BibitemShut {NoStop}%
\bibitem [{\citenamefont {Shen}\ and\ \citenamefont
  {Fan}(2007)}]{shen2007strongly}%
  \BibitemOpen
  \bibfield  {author} {\bibinfo {author} {\bibfnamefont {J.-T.}\ \bibnamefont
  {Shen}}\ and\ \bibinfo {author} {\bibfnamefont {S.}~\bibnamefont {Fan}},\
  }\bibfield  {title} {\bibinfo {title} {Strongly correlated multiparticle
  transport in one dimension through a quantum impurity},\ }\href@noop {}
  {\bibfield  {journal} {\bibinfo  {journal} {Physical Review A—Atomic,
  Molecular, and Optical Physics}\ }\textbf {\bibinfo {volume} {76}},\ \bibinfo
  {pages} {062709} (\bibinfo {year} {2007})}\BibitemShut {NoStop}%
\bibitem [{\citenamefont {Mahmoodian}\ \emph {et~al.}(2020)\citenamefont
  {Mahmoodian}, \citenamefont {Calaj{\'o}}, \citenamefont {Chang},
  \citenamefont {Hammerer},\ and\ \citenamefont
  {S{\o}rensen}}]{mahmoodian2020dynamics}%
  \BibitemOpen
  \bibfield  {author} {\bibinfo {author} {\bibfnamefont {S.}~\bibnamefont
  {Mahmoodian}}, \bibinfo {author} {\bibfnamefont {G.}~\bibnamefont
  {Calaj{\'o}}}, \bibinfo {author} {\bibfnamefont {D.~E.}\ \bibnamefont
  {Chang}}, \bibinfo {author} {\bibfnamefont {K.}~\bibnamefont {Hammerer}},\
  and\ \bibinfo {author} {\bibfnamefont {A.~S.}\ \bibnamefont {S{\o}rensen}},\
  }\bibfield  {title} {\bibinfo {title} {Dynamics of many-body photon bound
  states in chiral waveguide qed},\ }\href@noop {} {\bibfield  {journal}
  {\bibinfo  {journal} {Physical Review X}\ }\textbf {\bibinfo {volume} {10}},\
  \bibinfo {pages} {031011} (\bibinfo {year} {2020})}\BibitemShut {NoStop}%
\bibitem [{\citenamefont {Peyronel}\ \emph {et~al.}(2012)\citenamefont
  {Peyronel}, \citenamefont {Firstenberg}, \citenamefont {Liang}, \citenamefont
  {Hofferberth}, \citenamefont {Gorshkov}, \citenamefont {Pohl}, \citenamefont
  {Lukin},\ and\ \citenamefont {Vuleti{\'c}}}]{peyronel2012quantum}%
  \BibitemOpen
  \bibfield  {author} {\bibinfo {author} {\bibfnamefont {T.}~\bibnamefont
  {Peyronel}}, \bibinfo {author} {\bibfnamefont {O.}~\bibnamefont
  {Firstenberg}}, \bibinfo {author} {\bibfnamefont {Q.-Y.}\ \bibnamefont
  {Liang}}, \bibinfo {author} {\bibfnamefont {S.}~\bibnamefont {Hofferberth}},
  \bibinfo {author} {\bibfnamefont {A.~V.}\ \bibnamefont {Gorshkov}}, \bibinfo
  {author} {\bibfnamefont {T.}~\bibnamefont {Pohl}}, \bibinfo {author}
  {\bibfnamefont {M.~D.}\ \bibnamefont {Lukin}},\ and\ \bibinfo {author}
  {\bibfnamefont {V.}~\bibnamefont {Vuleti{\'c}}},\ }\bibfield  {title}
  {\bibinfo {title} {Quantum nonlinear optics with single photons enabled by
  strongly interacting atoms},\ }\href@noop {} {\bibfield  {journal} {\bibinfo
  {journal} {Nature}\ }\textbf {\bibinfo {volume} {488}},\ \bibinfo {pages}
  {57} (\bibinfo {year} {2012})}\BibitemShut {NoStop}%
\bibitem [{\citenamefont {Chang}\ \emph {et~al.}(2014)\citenamefont {Chang},
  \citenamefont {Vuleti{\'c}},\ and\ \citenamefont {Lukin}}]{chang2014quantum}%
  \BibitemOpen
  \bibfield  {author} {\bibinfo {author} {\bibfnamefont {D.~E.}\ \bibnamefont
  {Chang}}, \bibinfo {author} {\bibfnamefont {V.}~\bibnamefont {Vuleti{\'c}}},\
  and\ \bibinfo {author} {\bibfnamefont {M.~D.}\ \bibnamefont {Lukin}},\
  }\bibfield  {title} {\bibinfo {title} {Quantum nonlinear optics—photon by
  photon},\ }\href@noop {} {\bibfield  {journal} {\bibinfo  {journal} {Nature
  Photonics}\ }\textbf {\bibinfo {volume} {8}},\ \bibinfo {pages} {685}
  (\bibinfo {year} {2014})}\BibitemShut {NoStop}%
\bibitem [{\citenamefont {Rosen}\ \emph {et~al.}(2024)\citenamefont {Rosen},
  \citenamefont {Muschinske}, \citenamefont {Barrett}, \citenamefont
  {Chatterjee}, \citenamefont {Hays}, \citenamefont {DeMarco}, \citenamefont
  {Karamlou}, \citenamefont {Rower}, \citenamefont {Das}, \citenamefont {Kim}
  \emph {et~al.}}]{rosen2024synthetic}%
  \BibitemOpen
  \bibfield  {author} {\bibinfo {author} {\bibfnamefont {I.~T.}\ \bibnamefont
  {Rosen}}, \bibinfo {author} {\bibfnamefont {S.}~\bibnamefont {Muschinske}},
  \bibinfo {author} {\bibfnamefont {C.~N.}\ \bibnamefont {Barrett}}, \bibinfo
  {author} {\bibfnamefont {A.}~\bibnamefont {Chatterjee}}, \bibinfo {author}
  {\bibfnamefont {M.}~\bibnamefont {Hays}}, \bibinfo {author} {\bibfnamefont
  {M.~A.}\ \bibnamefont {DeMarco}}, \bibinfo {author} {\bibfnamefont {A.~H.}\
  \bibnamefont {Karamlou}}, \bibinfo {author} {\bibfnamefont {D.~A.}\
  \bibnamefont {Rower}}, \bibinfo {author} {\bibfnamefont {R.}~\bibnamefont
  {Das}}, \bibinfo {author} {\bibfnamefont {D.~K.}\ \bibnamefont {Kim}}, \emph
  {et~al.},\ }\bibfield  {title} {\bibinfo {title} {A synthetic magnetic vector
  potential in a 2d superconducting qubit array},\ }\href@noop {} {\bibfield
  {journal} {\bibinfo  {journal} {Nature Physics}\ ,\ \bibinfo {pages} {1}}
  (\bibinfo {year} {2024})}\BibitemShut {NoStop}%
\bibitem [{\citenamefont {Cheng}\ \emph {et~al.}(2023)\citenamefont {Cheng},
  \citenamefont {Wang},\ and\ \citenamefont {Fan}}]{cheng2023artificial}%
  \BibitemOpen
  \bibfield  {author} {\bibinfo {author} {\bibfnamefont {D.}~\bibnamefont
  {Cheng}}, \bibinfo {author} {\bibfnamefont {K.}~\bibnamefont {Wang}},\ and\
  \bibinfo {author} {\bibfnamefont {S.}~\bibnamefont {Fan}},\ }\bibfield
  {title} {\bibinfo {title} {Artificial non-abelian lattice gauge fields for
  photons in the synthetic frequency dimension},\ }\href@noop {} {\bibfield
  {journal} {\bibinfo  {journal} {Physical Review Letters}\ }\textbf {\bibinfo
  {volume} {130}},\ \bibinfo {pages} {083601} (\bibinfo {year}
  {2023})}\BibitemShut {NoStop}%
\bibitem [{\citenamefont {Albert}\ \emph {et~al.}(2016)\citenamefont {Albert},
  \citenamefont {Bradlyn}, \citenamefont {Fraas},\ and\ \citenamefont
  {Jiang}}]{albert2016geometry}%
  \BibitemOpen
  \bibfield  {author} {\bibinfo {author} {\bibfnamefont {V.~V.}\ \bibnamefont
  {Albert}}, \bibinfo {author} {\bibfnamefont {B.}~\bibnamefont {Bradlyn}},
  \bibinfo {author} {\bibfnamefont {M.}~\bibnamefont {Fraas}},\ and\ \bibinfo
  {author} {\bibfnamefont {L.}~\bibnamefont {Jiang}},\ }\bibfield  {title}
  {\bibinfo {title} {Geometry and response of lindbladians},\ }\href@noop {}
  {\bibfield  {journal} {\bibinfo  {journal} {Physical Review X}\ }\textbf
  {\bibinfo {volume} {6}},\ \bibinfo {pages} {041031} (\bibinfo {year}
  {2016})}\BibitemShut {NoStop}%
\bibitem [{\citenamefont {Lloyd}(1996)}]{lloyd1996universal}%
  \BibitemOpen
  \bibfield  {author} {\bibinfo {author} {\bibfnamefont {S.}~\bibnamefont
  {Lloyd}},\ }\bibfield  {title} {\bibinfo {title} {Universal quantum
  simulators},\ }\href@noop {} {\bibfield  {journal} {\bibinfo  {journal}
  {Science}\ }\textbf {\bibinfo {volume} {273}},\ \bibinfo {pages} {1073}
  (\bibinfo {year} {1996})}\BibitemShut {NoStop}%
\bibitem [{\citenamefont {Duan}\ and\ \citenamefont
  {Kimble}(2004)}]{duan2004scalable}%
  \BibitemOpen
  \bibfield  {author} {\bibinfo {author} {\bibfnamefont {L.-M.}\ \bibnamefont
  {Duan}}\ and\ \bibinfo {author} {\bibfnamefont {H.}~\bibnamefont {Kimble}},\
  }\bibfield  {title} {\bibinfo {title} {Scalable photonic quantum computation
  through cavity-assisted interactions},\ }\href@noop {} {\bibfield  {journal}
  {\bibinfo  {journal} {Physical review letters}\ }\textbf {\bibinfo {volume}
  {92}},\ \bibinfo {pages} {127902} (\bibinfo {year} {2004})}\BibitemShut
  {NoStop}%
\bibitem [{\citenamefont {Hacker}\ \emph {et~al.}(2016)\citenamefont {Hacker},
  \citenamefont {Welte}, \citenamefont {Rempe},\ and\ \citenamefont
  {Ritter}}]{hacker2016photon}%
  \BibitemOpen
  \bibfield  {author} {\bibinfo {author} {\bibfnamefont {B.}~\bibnamefont
  {Hacker}}, \bibinfo {author} {\bibfnamefont {S.}~\bibnamefont {Welte}},
  \bibinfo {author} {\bibfnamefont {G.}~\bibnamefont {Rempe}},\ and\ \bibinfo
  {author} {\bibfnamefont {S.}~\bibnamefont {Ritter}},\ }\bibfield  {title}
  {\bibinfo {title} {A photon--photon quantum gate based on a single atom in an
  optical resonator},\ }\href@noop {} {\bibfield  {journal} {\bibinfo
  {journal} {Nature}\ }\textbf {\bibinfo {volume} {536}},\ \bibinfo {pages}
  {193} (\bibinfo {year} {2016})}\BibitemShut {NoStop}%
\bibitem [{\citenamefont {Hafezi}\ \emph {et~al.}(2012)\citenamefont {Hafezi},
  \citenamefont {Chang}, \citenamefont {Gritsev}, \citenamefont {Demler},\ and\
  \citenamefont {Lukin}}]{hafezi2012quantum}%
  \BibitemOpen
  \bibfield  {author} {\bibinfo {author} {\bibfnamefont {M.}~\bibnamefont
  {Hafezi}}, \bibinfo {author} {\bibfnamefont {D.~E.}\ \bibnamefont {Chang}},
  \bibinfo {author} {\bibfnamefont {V.}~\bibnamefont {Gritsev}}, \bibinfo
  {author} {\bibfnamefont {E.}~\bibnamefont {Demler}},\ and\ \bibinfo {author}
  {\bibfnamefont {M.~D.}\ \bibnamefont {Lukin}},\ }\bibfield  {title} {\bibinfo
  {title} {Quantum transport of strongly interacting photons in a
  one-dimensional nonlinear waveguide},\ }\href@noop {} {\bibfield  {journal}
  {\bibinfo  {journal} {Physical Review A—Atomic, Molecular, and Optical
  Physics}\ }\textbf {\bibinfo {volume} {85}},\ \bibinfo {pages} {013822}
  (\bibinfo {year} {2012})}\BibitemShut {NoStop}%
\bibitem [{\citenamefont {Shomroni}\ \emph {et~al.}(2014)\citenamefont
  {Shomroni}, \citenamefont {Rosenblum}, \citenamefont {Lovsky}, \citenamefont
  {Bechler}, \citenamefont {Guendelman},\ and\ \citenamefont
  {Dayan}}]{shomroni2014all}%
  \BibitemOpen
  \bibfield  {author} {\bibinfo {author} {\bibfnamefont {I.}~\bibnamefont
  {Shomroni}}, \bibinfo {author} {\bibfnamefont {S.}~\bibnamefont {Rosenblum}},
  \bibinfo {author} {\bibfnamefont {Y.}~\bibnamefont {Lovsky}}, \bibinfo
  {author} {\bibfnamefont {O.}~\bibnamefont {Bechler}}, \bibinfo {author}
  {\bibfnamefont {G.}~\bibnamefont {Guendelman}},\ and\ \bibinfo {author}
  {\bibfnamefont {B.}~\bibnamefont {Dayan}},\ }\bibfield  {title} {\bibinfo
  {title} {All-optical routing of single photons by a one-atom switch
  controlled by a single photon},\ }\href@noop {} {\bibfield  {journal}
  {\bibinfo  {journal} {Science}\ }\textbf {\bibinfo {volume} {345}},\ \bibinfo
  {pages} {903} (\bibinfo {year} {2014})}\BibitemShut {NoStop}%
\bibitem [{\citenamefont {Rosenblum}\ \emph {et~al.}(2011)\citenamefont
  {Rosenblum}, \citenamefont {Parkins},\ and\ \citenamefont
  {Dayan}}]{rosenblum2011photon}%
  \BibitemOpen
  \bibfield  {author} {\bibinfo {author} {\bibfnamefont {S.}~\bibnamefont
  {Rosenblum}}, \bibinfo {author} {\bibfnamefont {S.}~\bibnamefont {Parkins}},\
  and\ \bibinfo {author} {\bibfnamefont {B.}~\bibnamefont {Dayan}},\ }\bibfield
   {title} {\bibinfo {title} {Photon routing in cavity qed: Beyond the
  fundamental limit of photon blockade},\ }\href@noop {} {\bibfield  {journal}
  {\bibinfo  {journal} {Physical Review A—Atomic, Molecular, and Optical
  Physics}\ }\textbf {\bibinfo {volume} {84}},\ \bibinfo {pages} {033854}
  (\bibinfo {year} {2011})}\BibitemShut {NoStop}%
\bibitem [{\citenamefont {Sivan}\ and\ \citenamefont
  {Pendry}(2011)}]{sivan2011time}%
  \BibitemOpen
  \bibfield  {author} {\bibinfo {author} {\bibfnamefont {Y.}~\bibnamefont
  {Sivan}}\ and\ \bibinfo {author} {\bibfnamefont {J.~B.}\ \bibnamefont
  {Pendry}},\ }\bibfield  {title} {\bibinfo {title} {Time reversal in
  dynamically tuned zero-gap periodic systems},\ }\href@noop {} {\bibfield
  {journal} {\bibinfo  {journal} {Physical review letters}\ }\textbf {\bibinfo
  {volume} {106}},\ \bibinfo {pages} {193902} (\bibinfo {year}
  {2011})}\BibitemShut {NoStop}%
\bibitem [{\citenamefont {Chumak}\ \emph {et~al.}(2010)\citenamefont {Chumak},
  \citenamefont {Tiberkevich}, \citenamefont {Karenowska}, \citenamefont
  {Serga}, \citenamefont {Gregg}, \citenamefont {Slavin},\ and\ \citenamefont
  {Hillebrands}}]{chumak2010all}%
  \BibitemOpen
  \bibfield  {author} {\bibinfo {author} {\bibfnamefont {A.~V.}\ \bibnamefont
  {Chumak}}, \bibinfo {author} {\bibfnamefont {V.~S.}\ \bibnamefont
  {Tiberkevich}}, \bibinfo {author} {\bibfnamefont {A.~D.}\ \bibnamefont
  {Karenowska}}, \bibinfo {author} {\bibfnamefont {A.~A.}\ \bibnamefont
  {Serga}}, \bibinfo {author} {\bibfnamefont {J.~F.}\ \bibnamefont {Gregg}},
  \bibinfo {author} {\bibfnamefont {A.~N.}\ \bibnamefont {Slavin}},\ and\
  \bibinfo {author} {\bibfnamefont {B.}~\bibnamefont {Hillebrands}},\
  }\bibfield  {title} {\bibinfo {title} {All-linear time reversal by a dynamic
  artificial crystal},\ }\href@noop {} {\bibfield  {journal} {\bibinfo
  {journal} {Nature communications}\ }\textbf {\bibinfo {volume} {1}},\
  \bibinfo {pages} {141} (\bibinfo {year} {2010})}\BibitemShut {NoStop}%
\bibitem [{\citenamefont {Yanik}\ and\ \citenamefont
  {Fan}(2004)}]{yanik2004time}%
  \BibitemOpen
  \bibfield  {author} {\bibinfo {author} {\bibfnamefont {M.~F.}\ \bibnamefont
  {Yanik}}\ and\ \bibinfo {author} {\bibfnamefont {S.}~\bibnamefont {Fan}},\
  }\bibfield  {title} {\bibinfo {title} {Time reversal of light with linear
  optics and modulators},\ }\href@noop {} {\bibfield  {journal} {\bibinfo
  {journal} {Physical review letters}\ }\textbf {\bibinfo {volume} {93}},\
  \bibinfo {pages} {173903} (\bibinfo {year} {2004})}\BibitemShut {NoStop}%
\bibitem [{\citenamefont {Preiss}\ \emph {et~al.}(2015)\citenamefont {Preiss},
  \citenamefont {Ma}, \citenamefont {Tai}, \citenamefont {Lukin}, \citenamefont
  {Rispoli}, \citenamefont {Zupancic}, \citenamefont {Lahini}, \citenamefont
  {Islam},\ and\ \citenamefont {Greiner}}]{preiss2015strongly}%
  \BibitemOpen
  \bibfield  {author} {\bibinfo {author} {\bibfnamefont {P.~M.}\ \bibnamefont
  {Preiss}}, \bibinfo {author} {\bibfnamefont {R.}~\bibnamefont {Ma}}, \bibinfo
  {author} {\bibfnamefont {M.~E.}\ \bibnamefont {Tai}}, \bibinfo {author}
  {\bibfnamefont {A.}~\bibnamefont {Lukin}}, \bibinfo {author} {\bibfnamefont
  {M.}~\bibnamefont {Rispoli}}, \bibinfo {author} {\bibfnamefont
  {P.}~\bibnamefont {Zupancic}}, \bibinfo {author} {\bibfnamefont
  {Y.}~\bibnamefont {Lahini}}, \bibinfo {author} {\bibfnamefont
  {R.}~\bibnamefont {Islam}},\ and\ \bibinfo {author} {\bibfnamefont
  {M.}~\bibnamefont {Greiner}},\ }\bibfield  {title} {\bibinfo {title}
  {Strongly correlated quantum walks in optical lattices},\ }\href@noop {}
  {\bibfield  {journal} {\bibinfo  {journal} {Science}\ }\textbf {\bibinfo
  {volume} {347}},\ \bibinfo {pages} {1229} (\bibinfo {year}
  {2015})}\BibitemShut {NoStop}%
\bibitem [{\citenamefont {Su{\'a}rez-Forero}\ \emph {et~al.}(2025)\citenamefont
  {Su{\'a}rez-Forero}, \citenamefont {Jalali~Mehrabad}, \citenamefont {Vega},
  \citenamefont {Gonz{\'a}lez-Tudela},\ and\ \citenamefont
  {Hafezi}}]{suarez2025chiral}%
  \BibitemOpen
  \bibfield  {author} {\bibinfo {author} {\bibfnamefont {D.}~\bibnamefont
  {Su{\'a}rez-Forero}}, \bibinfo {author} {\bibfnamefont {M.}~\bibnamefont
  {Jalali~Mehrabad}}, \bibinfo {author} {\bibfnamefont {C.}~\bibnamefont
  {Vega}}, \bibinfo {author} {\bibfnamefont {A.}~\bibnamefont
  {Gonz{\'a}lez-Tudela}},\ and\ \bibinfo {author} {\bibfnamefont
  {M.}~\bibnamefont {Hafezi}},\ }\bibfield  {title} {\bibinfo {title} {Chiral
  quantum optics: Recent developments and future directions},\ }\href@noop {}
  {\bibfield  {journal} {\bibinfo  {journal} {PRX Quantum}\ }\textbf {\bibinfo
  {volume} {6}},\ \bibinfo {pages} {020101} (\bibinfo {year}
  {2025})}\BibitemShut {NoStop}%
\bibitem [{\citenamefont {Jalali~Mehrabad}\ \emph {et~al.}(2023)\citenamefont
  {Jalali~Mehrabad}, \citenamefont {Mittal},\ and\ \citenamefont
  {Hafezi}}]{jalali2023topological}%
  \BibitemOpen
  \bibfield  {author} {\bibinfo {author} {\bibfnamefont {M.}~\bibnamefont
  {Jalali~Mehrabad}}, \bibinfo {author} {\bibfnamefont {S.}~\bibnamefont
  {Mittal}},\ and\ \bibinfo {author} {\bibfnamefont {M.}~\bibnamefont
  {Hafezi}},\ }\bibfield  {title} {\bibinfo {title} {Topological photonics:
  Fundamental concepts, recent developments, and future directions},\
  }\href@noop {} {\bibfield  {journal} {\bibinfo  {journal} {Physical Review
  A}\ }\textbf {\bibinfo {volume} {108}},\ \bibinfo {pages} {040101} (\bibinfo
  {year} {2023})}\BibitemShut {NoStop}%
\bibitem [{\citenamefont {Skinner}\ \emph {et~al.}(2019)\citenamefont
  {Skinner}, \citenamefont {Ruhman},\ and\ \citenamefont
  {Nahum}}]{skinner2019measurement}%
  \BibitemOpen
  \bibfield  {author} {\bibinfo {author} {\bibfnamefont {B.}~\bibnamefont
  {Skinner}}, \bibinfo {author} {\bibfnamefont {J.}~\bibnamefont {Ruhman}},\
  and\ \bibinfo {author} {\bibfnamefont {A.}~\bibnamefont {Nahum}},\ }\bibfield
   {title} {\bibinfo {title} {Measurement-induced phase transitions in the
  dynamics of entanglement},\ }\href@noop {} {\bibfield  {journal} {\bibinfo
  {journal} {Physical Review X}\ }\textbf {\bibinfo {volume} {9}},\ \bibinfo
  {pages} {031009} (\bibinfo {year} {2019})}\BibitemShut {NoStop}%
\bibitem [{\citenamefont {Steinbrecher}\ \emph {et~al.}(2019)\citenamefont
  {Steinbrecher}, \citenamefont {Olson}, \citenamefont {Englund},\ and\
  \citenamefont {Carolan}}]{steinbrecher2019quantum}%
  \BibitemOpen
  \bibfield  {author} {\bibinfo {author} {\bibfnamefont {G.~R.}\ \bibnamefont
  {Steinbrecher}}, \bibinfo {author} {\bibfnamefont {J.~P.}\ \bibnamefont
  {Olson}}, \bibinfo {author} {\bibfnamefont {D.}~\bibnamefont {Englund}},\
  and\ \bibinfo {author} {\bibfnamefont {J.}~\bibnamefont {Carolan}},\
  }\bibfield  {title} {\bibinfo {title} {Quantum optical neural networks},\
  }\href@noop {} {\bibfield  {journal} {\bibinfo  {journal} {npj Quantum
  Information}\ }\textbf {\bibinfo {volume} {5}},\ \bibinfo {pages} {60}
  (\bibinfo {year} {2019})}\BibitemShut {NoStop}%
\bibitem [{\citenamefont {Kiilerich}\ and\ \citenamefont
  {M{\o}lmer}(2019)}]{kiilerich2019input}%
  \BibitemOpen
  \bibfield  {author} {\bibinfo {author} {\bibfnamefont {A.~H.}\ \bibnamefont
  {Kiilerich}}\ and\ \bibinfo {author} {\bibfnamefont {K.}~\bibnamefont
  {M{\o}lmer}},\ }\bibfield  {title} {\bibinfo {title} {Input-output theory
  with quantum pulses},\ }\href@noop {} {\bibfield  {journal} {\bibinfo
  {journal} {Physical review letters}\ }\textbf {\bibinfo {volume} {123}},\
  \bibinfo {pages} {123604} (\bibinfo {year} {2019})}\BibitemShut {NoStop}%
\end{thebibliography}%
	
\end{document}